\documentstyle[preprint,eqsecnum,aps,epsf,floats,tighten]{revtex} 

\def\vslash{v\!\!\!\slash}
\newcommand{\nn}{\nonumber}

\begin{document}

\preprint{\tighten \vbox{\hbox{CALT-68-2120} 
		\hbox{hep-ph/9705467}
		\hbox{\footnotesize DOE RESEARCH AND}
		\hbox{\footnotesize DEVELOPMENT REPORT} \hbox{} }}

\title{Semileptonic $B$ decays to excited charmed mesons}
 
\author{Adam K.\ Leibovich, Zoltan Ligeti, Iain W.\ Stewart, Mark B.\ Wise}

\address{California Institute of Technology, Pasadena, CA 91125}

\maketitle

{\tighten
\begin{abstract}%
Exclusive semileptonic $B$ decays into excited charmed mesons are investigated
at order $\Lambda_{\rm QCD}/m_Q$ in the heavy quark effective theory. 
Differential decay rates for each helicity state of the four lightest excited
$D$ mesons ($D_1$, $D_2^*$, $D_0^*$, and $D_1^*$) are examined.  At zero
recoil, $\Lambda_{\rm QCD}/m_Q$ corrections to the matrix elements of the weak
currents can be written in terms of the leading Isgur-Wise functions for the
corresponding transition and meson mass splittings.  A model independent
prediction is found for the slope parameter of the decay rate into helicity
zero $D_1$ at zero recoil.  The differential decay rates are predicted,
including $\Lambda_{\rm QCD}/m_Q$ corrections with some model dependence away
from zero recoil and including order $\alpha_s$ corrections.  Ratios of various
exclusive branching ratios are computed.  Matrix elements of the weak currents
between $B$ mesons and other excited charmed mesons are discussed at zero
recoil to order $\Lambda_{\rm QCD}/m_Q$.  These amplitudes vanish at leading
order, and can be written at order $\Lambda_{\rm QCD}/m_Q$ in terms of local
matrix elements.  Applications to $B$ decay sum rules and factorization are
presented.

\end{abstract}
}

\newpage

\section{Introduction}

Heavy quark symmetry \cite{HQS} implies that in the $m_Q\to\infty$
limit matrix elements of the weak currents between a $B$ meson and an
excited charmed meson vanish at zero recoil (where in the rest frame
of the $B$ the final state charmed meson is also at rest).  However, in
some cases at order $\Lambda_{\rm QCD}/m_Q$ these matrix elements are
not zero \cite{llsw}.  Since most of the phase space for semileptonic
$B$ decay to excited charmed mesons is near zero recoil, $\Lambda_{\rm
QCD}/m_Q$ corrections can be very important.  This paper is concerned
with rates for $B$ semileptonic decay to excited charmed mesons,
including the effects of $\Lambda_{\rm QCD}/m_Q$ corrections.

The use of heavy quark symmetry resulted in a dramatic improvement in our
understanding of the spectroscopy and weak decays of hadrons containing a
single heavy quark, $Q$.  In the limit where the heavy quark mass goes to
infinity, $m_Q\to\infty$, such hadrons are classified not only by their total
spin $J$, but also by the spin of their light degrees of freedom (i.e., light
quarks and gluons), $s_\ell$ \cite{IWprl}.  In this limit hadrons containing a
single heavy quark come in degenerate doublets with total spin, $J_\pm = s_\ell
\pm \frac12$, coming from combining the spin of the light degrees of freedom
with the spin of the heavy quark, $s_Q = \frac12$.  (An exception occurs for
baryons with $s_\ell = 0$, where there is only a single state with $J =
\frac12$.)  The ground state mesons with $Q\,\bar q$ flavor quantum numbers
contain light degrees of freedom with spin-parity
$s_\ell^{\pi_\ell}=\frac12^-$, giving a doublet containing a spin zero and spin
one meson.  For $Q=c$ these mesons are the $D$ and $D^*$, while $Q=b$ gives the
$B$ and $B^*$ mesons.  

Excited charmed mesons with $s_\ell^{\pi_\ell} = \frac32^+$ have been observed.
These are the $D_1$ and $D_2^*$ mesons with spin one and two, respectively. 
(There is also evidence for the analogous $Q=b$ heavy meson doublet.)  For
$q=u,d$, the $D_1$ and $D_2^*$ mesons have been observed to decay to
$D^{(*)}\,\pi$ and are narrow with widths around $20\,$MeV.  (The $D_{s1}$ and
$D_{s2}^*$ strange mesons decay to $D^{(*)}K$.)  In the nonrelativistic
constituent quark model these states correspond to $L=1$ orbital excitations. 
Combining the unit of orbital angular momentum with the spin of the light
antiquark leads to states with $s_\ell^{\pi_\ell}=\frac12^+$ and $\frac32^+$. 
The $\frac12^+$ doublet, $(D_0^*,D_1^*)$, has not been observed.  Presumably
this is because these states are much broader than those with
$s_\ell^{\pi_\ell}=\frac32^+$.  A vast discrepancy in widths is expected since
the members of the $\frac12^+$ doublet of charmed mesons decay to
$D^{(*)}\,\pi$ in an $S$-wave while the members of the $\frac32^+$ doublet of
charmed mesons decay to $D^{(*)}\,\pi$ in a $D$-wave.  (An $S$-wave $D_1\to
D^*\,\pi$ amplitude is allowed by total angular momentum conservation, but
forbidden in the $m_Q\to\infty$ limit by heavy quark spin symmetry
\cite{IWprl}.)

The heavy quark effective theory (HQET) is the limit of QCD where the
heavy quark mass goes to infinity with its four velocity, $v$, fixed.
The heavy quark field in QCD, $Q$, is related to its counterpart in HQET,
$h_v^{(Q)}$, by
\begin{equation}\label{HQETfield}
Q(x) = e^{-im_Qv\cdot x}\left[1 + \frac{iD\!\!\!\!\slash}{2m_Q} +
  \ldots \right] h_v^{(Q)} \,,
\end{equation}
where $\vslash h_v^{(Q)} = h_v^{(Q)}$ and the ellipses denote terms
suppressed by further powers of $\Lambda_{\rm QCD}/m_Q$.  Putting
Eq.~(\ref{HQETfield}) into the part of the QCD Lagrangian involving
the heavy quark field, ${\mathcal L}=\bar Q\,(iD\!\!\!\!\slash - m_Q)\,Q$, 
gives
\begin{equation}\label{fulllag}
{\mathcal L} = {\mathcal L}_{\mathrm HQET} + \delta {\mathcal L} + \ldots \,.
\end{equation}
The HQET Lagrangian \cite{eft}
\begin{equation}\label{HQETlag}
{\mathcal L}_{\mathrm HQET} = \bar h_v^{(Q)}\, iv\cdot D\, h_v^{(Q)} 
\end{equation}
is independent of the mass of the heavy quark and
its spin, and so for $N_Q$ heavy quarks with the same four velocity $v$ there
is a $U(2N_Q)$ spin-flavor symmetry.  This symmetry is broken by the order 
$\Lambda_{\rm QCD}/m_Q$ terms \cite{eft/m} in $\delta{\mathcal L}$,
\begin{equation}\label{lag}
\delta{\cal L} = \frac1{2 m_Q}\, \Big[ O_{{\rm kin},v}^{(Q)} +
  O_{{\rm mag},v}^{(Q)} \Big] \,,
\end{equation}  
where 
\begin{equation}
O_{{\rm kin},v}^{(Q)} = \bar h_v^{(Q)}\, (iD)^2\, h_v^{(Q)} \,, \qquad
O_{{\rm mag},v}^{(Q)} = \bar h_v^{(Q)}\,
  \frac{g_s}2\, \sigma_{\alpha\beta} G^{\alpha\beta}\, h_v^{(Q)} \,.
\end{equation}  
The first term in Eq.~(\ref{lag}) is the heavy quark kinetic energy.  It
breaks the flavor symmetry but leaves the spin symmetry intact.  The second is
the chromomagnetic term, which breaks both the spin and flavor symmetries.  
(In the rest frame, it is of the form 
$\vec{\mu}_Q\cdot\vec B_{\mathrm color}$, where $\vec{\mu}_Q$ is the heavy
quark color magnetic moment.)

The hadron masses give important information on some HQET matrix
elements.  The mass formula for a spin symmetry doublet of hadrons
$H_\pm$ with total spin $J_\pm = s_\ell \pm \frac12$ is
\begin{equation}\label{mass}
m_{H_\pm} = m_Q + \bar\Lambda^H - {\lambda_1^H \over 2 m_Q} 
  \pm {n_\mp\, \lambda_2^H \over 2m_Q} + \ldots \,,
\end{equation}
where the ellipsis denote terms suppressed by more powers of 
$\Lambda_{\rm QCD}/m_Q$ and $n_\pm = 2J_\pm+1$ is the number of spin states 
in the hadron $H_\pm$.  The parameter $\bar\Lambda$ is the energy of the light 
degrees of freedom in the $m_Q\to\infty$ limit, $\lambda_1$ determines the 
heavy quark kinetic energy\footnote{Hadron states labeled by their 
four-velocity, $v=p_H/m_H$, satisfy the standard covariant 
normalization $\langle H(p_H')\,|\,H(p_H)\rangle =
(2\pi)^3\,2E_H\,\delta^3(\vec{p}_H^{\,\prime} - \vec{p}_H)$.}
\begin{equation}\label{lambda1}
\lambda_1^H = \frac1{2v^0\,m_{H_\pm}}\, \langle H_\pm(v)|\,
  \bar h_v^{(Q)}\, (iD)^2\, h_v^{(Q)}\, |H_\pm(v)\rangle \,,
\end{equation}
and $\lambda_2$ determines the chromomagnetic energy
\begin{equation}\label{lambda2}
\lambda_2^H = {\mp1\over 2v^0\,m_{H_\pm}n_\mp}\, 
  \langle H_\pm(v)|\, \bar h_v^{(Q)}\, \frac{g_s}2\, \sigma_{\alpha\beta} 
  G^{\alpha\beta}\, h_v^{(Q)}\, |H_\pm(v)\rangle \,.
\end{equation}
$\bar\Lambda$ and $\lambda_1$ are independent of the heavy quark mass, while
$\lambda_2$ has a weak logarithmic dependence on $m_Q$.  Of course
they depend on the particular spin symmetry doublet to which $H_\pm$
belong.  In this paper, we consider heavy mesons in the ground state
$s_\ell^{\pi_\ell} = \frac12^-$ doublet and the excited 
$s_\ell^{\pi_\ell} = \frac32^+$ and $\frac12^+$ doublets.  We reserve
the notation $\bar\Lambda$, $\lambda_1$, $\lambda_2$ for the ground state
multiplet and use $\bar\Lambda'$, $\lambda_1'$, $\lambda_2'$
and $\bar\Lambda^*$, $\lambda_1^*$, $\lambda_2^*$ for the excited 
$s_\ell^{\pi_\ell} = \frac32^+$ and $\frac12^+$ doublets, respectively. 

The average mass $\overline{m}_H$, weighted by the number of helicity states
\begin{equation}\label{avemass}
  \overline{m}_H = \frac{n_-m_{H_-} + n_+m_{H_+}}{n_+ + n_-} \,,
\end{equation}
is independent of $\lambda_2$.  The spin average masses for the lowest lying
charmed mesons is given in Table~\ref{tab:charm}.  Identifying the $B^{(*)}\pi$
resonances observed at LEP with the bottom $s_\ell^{\pi_\ell}=\frac32^+$ meson
doublet we can use their average mass, $\overline{m}_B'=5.73\,$GeV
\cite{talks}, to determine the differences $\bar\Lambda'-\bar\Lambda$ and
$\lambda_1'-\lambda_1$:
\begin{eqnarray}\label{HQET_diff}
\bar\Lambda' - \bar\Lambda &=& {m_b\,(\overline{m}_B'-\overline{m}_B)
  - m_c\,(\overline{m}_D'-\overline{m}_D) \over m_b-m_c} 
  \simeq 0.39\, {\rm GeV} \,, \nn\\*
\lambda_1' - \lambda_1 &=& {2 m_c m_b\, [(\overline{m}_B'-\overline{m}_B)
  - (\overline{m}_D'-\overline{m}_D)]\over m_b-m_c} 
  \simeq -0.23\, {\rm GeV}^2 \,. 
\end{eqnarray}
The numerical values in Eq.~(\ref{HQET_diff}) follow from the choices
$m_b=4.8\,$GeV and $m_c=1.4\,$GeV.  To the order we are working, $m_b$ and
$m_c$ in Eq.~(\ref{HQET_diff}) can be replaced by $\overline{m}_B$ and
$\overline{m}_D$.  This changes the value of $\bar\Lambda'-\bar\Lambda$ only
slightly, but has a significant impact on the value of $\lambda_1'-\lambda_1$.
The value of $\bar\Lambda'-\bar\Lambda$ given in Eq.~(\ref{HQET_diff}) has
considerable uncertainty because the experimental error on $\overline{m}_B'$
is large, and because it is not clear that the peak of the $B^{(*)}\pi$ mass
distribution corresponds to the narrow $\frac32^+$ doublet.\footnote{The
$B_{s1}$ and $B_{s2}^*$ masses could also be used to determine 
$\bar\Lambda'-\bar\Lambda$ from the relation
\begin{eqnarray}
\bar\Lambda' - \bar\Lambda = \bar\Lambda_s' - \bar\Lambda +
  (\overline{m}_D' - \overline{m}_{D_s}') 
  + {\cal O}(\Lambda_{\rm QCD}\,m_s/m_c) \,, \nonumber
\end{eqnarray}
with the analog of Eq.~(\ref{HQET_diff}) used to fix
$\bar\Lambda_s'-\bar\Lambda$, and
$\overline{m}_D'-\overline{m}_{D_s}'=-114\,$MeV.  The $B_s^*$ has not been
observed, but its mass can be determined from
$(m_{B_s^*}-m_{B_s})-(m_{B^*}-m_B) =
(m_c/m_b)\,[(m_{D_s^*}-m_{D_s})-(m_{D^*}-m_D)]$.  However, because of
uncertainties in the $B_{s1}$ and $B_{s2}^*$ masses and the unknown order
$(\Lambda_{\rm QCD}\,m_s/m_c)$ term, this relation does not give a more
reliable determination of $\bar\Lambda'-\bar\Lambda$ than
Eq.~(\ref{HQET_diff}).}

\begin{table}[t]
\begin{tabular}{cccccc}  
& $s_l^{\pi_l}$ & Particles	  &  $J^P$  & $\overline{m}$ (GeV) \\ \hline 
& $\frac12^-$  & $D$, $D^*$       &  $0^-$, $1^-$ & $1.971$  \\
& $\frac12^+$  & $D_0^*$, $D_1^*$ &  $0^+$, $1^+$ & $\sim2.40$\\
& $\frac32^+$  & $D_1$, $D_2^*$   &  $1^+$, $2^+$ & $2.445$ \\
\end{tabular} \vspace{6pt}
\caption{Charmed meson spin multiplets ($q=u,d$).} \label{tab:charm}
\end{table}

At the present time, $\bar\Lambda$ and $\lambda_1$ are not well determined.  A
fit to the electron energy spectrum in semileptonic $B$ decay gives \cite{gklw}
$\bar\Lambda \simeq 0.4\,$GeV and $\lambda_1 \simeq -0.2\,{\mathrm GeV}^2$, but
the uncertainties are quite large \cite{AM}.  (A linear combination of
$\bar\Lambda$ and $\lambda_1$ is better determined than the individual values.)

The measured $D^*-D$ mass difference ($142\,$MeV) and the measured $D_2^*-D_1$
mass difference ($37\,$MeV) fix $\lambda_2 = 0.10\,{\mathrm GeV}^2$ and
$\lambda_2' = 0.013\,{\mathrm GeV}^2$.  Note that the matrix element of the
chromomagnetic operator is substantially smaller in the excited
$s_\ell^{\pi_\ell} = \frac32^+$ multiplet than in the ground state multiplet. 
This is consistent with expectations based on the nonrelativistic constituent
quark model.  In this phenomenological model, the splitting between members of
a $Q\,\bar q$ meson spin symmetry doublet arises mostly from matrix elements of
the operator $\vec s_Q\cdot\vec s_{\bar q}\,\delta^3(\vec r\,)$, and these
vanish for $Q\,\bar q$ mesons with orbital angular momentum.

Semileptonic $B$ meson decays have been studied extensively.  The semileptonic
decays $B\to D\,e\,\bar\nu_e$ and $B\to D^*\,e\,\bar\nu_e$ have branching
ratios of $(1.8\pm0.4)\%$ and $(4.6\pm0.3)\%$, respectively \cite{PDG}, and
comprise about 60\% of the semileptonic decays.  The differential decay rates
for these decays are determined by matrix elements of the weak $b\to c$
axial-vector and vector currents between the $B$ meson and the recoiling
$D^{(*)}$ meson.  These matrix elements are usually parameterized by a set of
Lorentz scalar form factors and the differential decay rate is expressed in
terms of these form factors.  For comparison with the predictions of HQET, it
is convenient to write the form factors as functions of the dot-product,
$w=v\cdot v'$, of the four-velocity of the $B$ meson, $v$, and that of the
recoiling $D^{(*)}$ meson, $v'$.  In the $m_Q\to\infty$ limit, heavy quark spin
symmetry implies that the six form factors that parameterize the $B\to D$ and
$B\to D^*$ matrix elements of the $b\to c$ axial-vector and vector currents can
be written in terms of a single function of $w$ \cite{HQS}.  Furthermore, heavy
quark flavor symmetry implies that this function is normalized to unity at zero
recoil, $w = 1$, where the $D^{(*)}$ is at rest in the rest frame of the $B$
\cite{NuWe,VoSi,HQS}.  The functions of $w$ that occur in predictions for weak
decay form factors based on HQET are usually called Isgur-Wise functions. 
There are perturbative $\alpha_s(m_Q)$ and nonperturbative $\Lambda_{\mathrm
QCD}/m_Q$ corrections to the predictions of the $m_Q\to\infty$ limit for the
$B\to D^{(*)}\,e\,\bar\nu_e$ semileptonic decay form factors.  The perturbative
QCD corrections do not cause any loss of predictive power.  They involve the
same Isgur-Wise function that occurs in the $m_Q\to\infty$ limit.  At order
$\Lambda_{\mathrm QCD}/m_Q$ several new Isgur-Wise functions occur; however, at
zero recoil, there are no $\Lambda_{\mathrm QCD}/m_Q$ corrections \cite{Luke}. 
Expectations for the $B\to D^{(*)}\,e\,\bar\nu_e$ differential decay rate based
on HQET are in agreement with experiment \cite{rich}.

Recently, semileptonic $B$ decay to an excited heavy meson has been observed
\cite{OPAL,ALEPH,CLEO}.  With some assumptions, CLEO \cite{CLEO} and ALEPH
\cite{ALEPH} find respectively the branching ratios ${\mathcal B}(B\to
D_1\,e\,\bar\nu_e) = (0.49 \pm 0.14)\%$ and ${\mathcal B}(B\to
D_1\,e\,\bar\nu_e) = (0.74 \pm 0.16)\%$, as well as the limits ${\mathcal
B}(B\to D_2^*\,e\,\bar\nu_e)<1\%$ and  ${\mathcal B}(B\to
D_2^*\,e\,\bar\nu_e)<0.2\%$.  In the future it should be possible to get
detailed experimental information on the $B\to D_1\,e\,\bar\nu_e$ and $B\to
D_2^*\,e\,\bar\nu_e$ differential decay rates.

In this paper we study the predictions of HQET for $B$ semileptonic decay to
excited charmed mesons.  This paper elaborates on the work in Ref.~\cite{llsw}
and contains some new results.  In the infinite mass limit the matrix elements
of the weak axial-vector and vector current between the $B$ meson and any
excited charmed meson vanish at zero recoil by heavy quark symmetry. 
Corrections to the infinite mass limit of order $\Lambda_{\mathrm QCD}/m_Q$ and
order $\alpha_s(m_Q)$ are discussed.  The corrections of order
$\Lambda_{\mathrm QCD}/m_Q$ are very important, particularly near zero recoil.

Section II discusses the differential decay rate ${\rm d}^2\Gamma/{\rm
d}w\,{\rm d}\!\cos\theta$ for $B\to(D_1,D_2^*)\,e\,\bar\nu_e$, where $\theta$
is the angle between the the charged lepton and the charmed meson in the rest
frame of the virtual $W$ boson.  Corrections of order $\Lambda_{\mathrm
QCD}/m_Q$ are included.  At order $\Lambda_{\mathrm QCD}/m_Q$ the $B\to D_1$
zero recoil matrix element does not vanish and is expressible in terms of the
leading $m_Q\to\infty$ Isgur-Wise function, $\tau$, and
$\bar\Lambda'-\bar\Lambda$ (which is known in terms of hadron mass splittings
from Eq.~(\ref{HQET_diff})).  Away from zero recoil new Isgur-Wise functions
occur, which are unknown.  These introduce a significant uncertainty.  The
$\Lambda_{\mathrm QCD}/m_Q$ corrections enhance considerably the $B$ 
semileptonic decay rate to the $D_1$ state, and for zero helicity the slope 
of ${\rm d}\Gamma(B\to D_1\,e\,\bar\nu_e)/{\rm d}w$ at $w=1$ is predicted.  
These corrections also reduce the ratio 
$R ={\cal B}(B\to D_2^*\,e\,\bar\nu_e)/{\cal B}(B\to D_1\,e\,\bar\nu_e)$ 
compared to its value in the $m_Q\to\infty$ limit.  The value of $\tau$ at
zero recoil is not fixed by heavy quark symmetry, and must be determined from 
experiment.  The measured $B\to D_1\,e\,\bar\nu_e$ branching ratio is used to 
determine (with some model dependent assumptions) $|\tau(1)|=0.71$. The
effects of perturbative QCD corrections are also discussed, with further 
details given in Appendix A.  

It is interesting to understand the composition of the inclusive $B$
semileptonic decay rate in terms of exclusive final states.  In Section III,
the HQET predictions for the differential decay rates for $B\to
D_0^*\,e\,\bar\nu_e$ and $B\to D_1^*\,e\,\bar\nu_e$ are investigated.  The
situation for the excited $s_\ell^{\pi_\ell}=\frac12^+$ multiplet is similar to
the $s_\ell^{\pi_\ell}=\frac32^+$ multiplet discussed in Section~II.  Using a
quark model relation between the leading $m_Q\to\infty$ Isgur-Wise functions
for $B$ decays to the $s_\ell^{\pi_\ell}=\frac32^+$ and
$s_\ell^{\pi_\ell}=\frac12^+$ charmed mesons (and some other model dependent
assumptions), the rates for $B\to D_0^*\,e\,\bar\nu_e$ and $B\to
D_1^*\,e\,\bar\nu_e$ are predicted.

Section IV discusses the contribution of other excited charmed mesons to the
matrix elements of the vector and axial-vector current at zero recoil.  Only
excited charmed hadrons with $s_\ell^{\pi_\ell}=\frac12^-$, $\frac32^-$ and
$\frac12^+$, $\frac32^+$ can contribute.  The $\frac32^+$ and $\frac12^+$
doublets are discussed in Sections~II and III.  This section deals with the
$\frac12^-$ and $\frac32^-$ cases, where the $\Lambda_{\mathrm QCD}/m_Q$
corrections to the states from $\delta{\mathcal L}$ give rise to non-vanishing
zero recoil matrix elements.

Section V examines other applications of our results.  Using factorization,
predictions are made for nonleptonic $B$ decay widths to $D_2^*\,\pi$,
$D_1\,\pi$ and to $D_1^*\,\pi$, $D_0^*\,\pi$.  The importance of our results
for $B$ decay sum rules is discussed.  Including the excited states
dramatically strengthens the Bjorken lower bound on the slope of the $B\to
D^{(*)}\,e\,\bar\nu_e$ Isgur-Wise function.

Concluding remarks and a summary of our most significant predictions are given
in Section~VI.

\section{$B\to D_1\,\lowercase{e}\,\bar\nu_{\lowercase{e}}$ and 
  $B\to D_2^*\,\lowercase{e}\,\bar\nu_{\lowercase{e}}$ decays}

The matrix elements of the vector and axial-vector currents 
($V^\mu=\bar c\,\gamma^\mu\,b$ and $A^\mu=\bar c\,\gamma^\mu\gamma_5\,b$) 
between $B$ mesons and $D_1$ or $D_2^*$ mesons can be parameterized as
\begin{eqnarray}\label{formf1}
{\langle D_1(v',\epsilon)|\, V^\mu\, |B(v)\rangle \over \sqrt{m_{D_1}\,m_B}}
  &=& f_{V_1}\, \epsilon^{*\mu} 
  + (f_{V_2} v^\mu + f_{V_3} v'^\mu)\, (\epsilon^*\cdot v) \,, \nn\\*
{\langle D_1(v',\epsilon)|\, A^\mu\, |B(v)\rangle \over \sqrt{m_{D_1}\,m_B}}
  &=& i\, f_A\, \varepsilon^{\mu\alpha\beta\gamma} 
  \epsilon^*_\alpha v_\beta v'_\gamma \,, \nn\\*
{\langle D^*_2(v',\epsilon)|\, A^\mu\, |B(v)\rangle \over\sqrt{m_{D_2^*}\,m_B}}
  &=& k_{A_1}\, \epsilon^{*\mu\alpha} v_\alpha 
  + (k_{A_2} v^\mu + k_{A_3} v'^\mu)\,
  \epsilon^*_{\alpha\beta}\, v^\alpha v^\beta \,, \nn\\*
{\langle D^*_2(v',\epsilon)|\, V^\mu\, |B(v)\rangle \over\sqrt{m_{D_2^*}\,m_B}}
  &=& i\,k_V\, \varepsilon^{\mu\alpha\beta\gamma} 
  \epsilon^*_{\alpha\sigma} v^\sigma v_\beta v'_\gamma \,,  \label{ff_nrwD}
\end{eqnarray}
where the form factors $f_i$ and $k_i$ are dimensionless functions of $w$.  
At zero recoil ($v=v'$) only the $f_{V_1}$ form factor can contribute, since 
$v'$ dotted into the polarization ($\epsilon^{*\mu}$ or 
$\epsilon^{*\mu\alpha}$) vanishes.

The differential decay rates can be written in terms of the form factors in
Eq.~(\ref{ff_nrwD}).  It is useful to separate the contributions to the
different helicities of the $D_1$ and $D_2^*$ mesons, since the
$\Lambda_{\rm QCD}/m_Q$ corrections effect these differently, and the decay
rates into different helicity states will probably be measurable.  We define
$\theta$ as the angle between the charged lepton and the charmed meson in the
rest frame of the virtual $W$ boson, i.e., in the center of momentum frame of
the lepton pair.  The different helicity amplitudes yield different
distributions in $\theta$.  In terms of $w=v\cdot v'$ and $\theta$, the double
differential decay rates are 
\begin{eqnarray}\label{rate1}
{{\rm d}^2\Gamma_{D_1}\over {\rm d}w\,{\rm d}\!\cos\theta} &=& 
  3\Gamma_0\, r_1^3\, \sqrt{w^2-1}\, \bigg\{ \sin^2\theta\,
  \Big[ (w-r_1) f_{V_1}+(w^2-1) (f_{V_3}+r_1 f_{V_2}) \Big]^2 \\*
&& + (1-2r_1w+r_1^2)\, \Big[ (1+\cos^2\theta)\, [f_{V_1}^2 + (w^2-1) f_A^2] 
  - 4\cos\theta\, \sqrt{w^2-1}\, f_{V_1}\, f_A \Big] \bigg\} \,,\nonumber\\*
{{\rm d}^2\Gamma_{D_2^*}\over {\rm d}w\,{\rm d}\!\cos\theta} &=& 
   \frac32\,\Gamma_0\, r_2^3\, (w^2-1)^{3/2}\, \bigg\{ \frac43\,\sin^2\theta\,
  \Big[ (w-r_2) k_{A_1}+(w^2-1) (k_{A_3}+r_2 k_{A_2}) \Big]^2 \nn\\*
&& + (1-2r_2w+r_2^2)\, \Big[ (1+\cos^2\theta)\, [k_{A_1}^2 + (w^2-1) k_V^2] 
  - 4\cos\theta\, \sqrt{w^2-1}\, k_{A_1}\, k_V \Big] \bigg\} \,,\nonumber
\end{eqnarray}
where $\Gamma_0 = {G_F^2\,|V_{cb}|^2\,m_B^5 /(192\pi^3)}$, $r_1=m_{D_1}/m_B$,
$r_2=m_{D_2^*}/m_B$.  The semileptonic $B$ decay rate into any $J\neq1$ state
involves an extra factor of $w^2-1$.  The $\sin^2\theta$ term is the helicity
zero rate, while the $1+\cos^2\theta$ and $\cos\theta$ terms determine the
helicity $\lambda=\pm1$ rates.  Since the weak current is $V-A$ in the standard
model, $B$ mesons can only decay into the helicity $|\lambda|=0,1$ components
of any excited charmed mesons.  The decay rate for
$|\lambda|=1$ vanishes at maximal recoil, $w_{\rm max}=(1+r^2)/(2r)$, as
implied by the $1-2rw+r^2$ factors above ($r=r_1$ or $r_2$).  From
Eq.~(\ref{rate1}) it is straightforward to obtain the double differential rate
${\rm d}^2\Gamma/{\rm d}w\,{\rm d}y$ using the relation 
\begin{equation}\label{yct}
y = 1 - rw - r\sqrt{w^2-1}\, \cos\theta \,,
\end{equation}
where $y=2E_e/m_B$ is the rescaled lepton energy.

The form factors $f_i$ and $k_i$ can be parameterized by a set of Isgur-Wise
functions at each order in $\Lambda_{\rm QCD}/m_Q$.  It is simplest to
calculate the matrix elements in Eq.~(\ref{formf1}) using the trace formalism
\cite{falk,trace}.  The fields $P_v$ and $P_v^{*\mu}$ that destroy members of
the $s_l^{\pi_l}=\frac12^-$ doublet with four-velocity $v$ are in 
the $4\times4$ matrix
\begin{equation}
H_v = \frac{1+\vslash}2\, \Big[ P_v^{*\mu} \gamma_\mu 
  - P_v\, \gamma_5 \Big] \,.  \label{Hdef}
\end{equation} 
while for $s_l^{\pi_l}=\frac32^+$ the fields $P_v^\nu$ and 
$P_v^{*\mu\nu}$ are in
\begin{equation}\label{Fdef}
F_v^\mu = \frac{1+\vslash}2\, \bigg\{ P_v^{*\mu\nu} \gamma_\nu 
  - \sqrt{\frac32}\, P_v^\nu \gamma_5 \bigg[ g^\mu_\nu - 
  \frac13 \gamma_\nu (\gamma^\mu-v^\mu) \bigg] \bigg\} \,.
\end{equation}  
The matrices $H$ and $F$ satisfy the properties $\vslash H_v=H_v=-H_v\vslash$,
\hbox{$\vslash F_v^\mu=F_v^\mu=-F_v^\mu\vslash$}, and
$F_v^\mu\,\gamma_\mu=F_v^\mu\,v_\mu=0$.  

To leading order in $\Lambda_{\rm QCD}/m_Q$ and $\alpha_s$, matrix
elements of the $b\to c$ flavor changing current between the states destroyed 
by the fields in $H_v$ and $F_{v'}^\sigma$ are
\begin{equation}\label{lo}
\bar c\, \Gamma\, b = \bar h^{(c)}_{v'}\, \Gamma\, h^{(b)}_v = \tau(w)\,
  {\rm Tr}\, \Big\{ v_\sigma \bar F^\sigma_{v'}\, \Gamma\, H_v \Big\} \,.
\end{equation}
Here $\tau(w)$ is a dimensionless function, and $h_v^{(Q)}$ is the heavy quark
field in the effective theory ($\tau$ is $\sqrt3$ times the function 
$\tau_{3/2}$ of Ref.~\cite{IWsr}).  This matrix
element vanishes at zero recoil for any Dirac structure $\Gamma$ and for any
value of $\tau(1)$, since the $B$ meson and the $(D_1,D_2^*)$ mesons are in
different heavy quark spin symmetry multiplets, and the current at zero recoil
is related to the conserved charges of heavy quark spin-flavor symmetry. 
Eq.~(\ref{lo}) leads to the $m_Q\to\infty$ predictions for the form factors
$f_i$ and $k_i$ given in Ref.~\cite{IWsr}.

At order $\Lambda_{\rm QCD}/m_Q$, there are corrections originating from the
matching of the $b\to c$ flavor changing current onto the effective theory, and
from order $\Lambda_{\rm QCD}/m_Q$ corrections to the effective Lagrangian. 
The current corrections modify the first equality in Eq.~(\ref{lo}) to
\begin{equation}\label{1/mcurrent}
\bar c\, \Gamma\, b = \bar h_{v'}^{(c)}\, 
  \bigg( \Gamma - \frac i{2m_c} \overleftarrow D\!\!\!\!\!\slash\, \Gamma  
  + \frac i{2m_b}\, \Gamma \overrightarrow D\!\!\!\!\!\slash\ 
  \bigg)\, h_v^{(b)} \,. 
\end{equation}
For matrix elements between the states destroyed by the fields in 
$F_{v'}^\sigma$ and $H_v$, the new order $\Lambda_{\rm QCD}/m_Q$ 
operators in Eq.~(\ref{1/mcurrent}) are
\begin{eqnarray}\label{1/mc1}
\bar h^{(c)}_{v'}\, i\overleftarrow D_{\!\lambda}\, \Gamma\, h^{(b)}_v &=&
  {\rm Tr}\, \Big\{ {\cal S}^{(c)}_{\sigma\lambda}\, 
  \bar F^\sigma_{v'}\, \Gamma\, H_v \Big\} \,, \nonumber\\*
\bar h^{(c)}_{v'}\, \Gamma\, i\overrightarrow D_{\!\lambda}\, h^{(b)}_v &=&
  {\rm Tr}\, \Big\{ {\cal S}^{(b)}_{\sigma\lambda}\, 
  \bar F^\sigma_{v'}\, \Gamma\, H_v \Big\} \,.  \label{curr}
\end{eqnarray}
The most general form for these quantities is
\begin{equation}\label{1/mc2}
{\cal S}^{(Q)}_{\sigma\lambda} = v_\sigma \Big[ \tau_1^{(Q)}\, v_\lambda 
  + \tau_2^{(Q)}\, v'_\lambda + \tau_3^{(Q)}\, \gamma_\lambda \Big] 
  + \tau_4^{(Q)}\, g_{\sigma\lambda} \,. \label{Sdef}
\end{equation}
The functions $\tau_i$ depend on $w$, and have mass dimension
one.\footnote{Order $\Lambda_{\rm QCD}/m_c$ corrections were also analyzed in
Ref.~\cite{Mannel}.  We find that $\tau_4$ (denoted $\xi_4$ in \cite{Mannel})
does contribute in Eq.~(\ref{curr}) for
$\Gamma=\gamma_\lambda\widetilde\Gamma$, and corrections to the Lagrangian are
parameterized by more functions than in \cite{Mannel}.}  They are not all
independent.  

The equation of motion for the heavy quarks, $(v\cdot D)\,h_v^{(Q)}=0$, implies
\begin{eqnarray}\label{const1}
w\,\tau_1^{(c)} + \tau_2^{(c)} - \tau_3^{(c)} &=& 0 \,, \nonumber\\*
  \tau_1^{(b)} + w\,\tau_2^{(b)} - \tau_3^{(b)} + \tau_4^{(b)} &=& 0 \,.
\end{eqnarray}
Four more relations can be derived using 
\begin{equation}\label{momcons}
i\partial_\nu\,(\bar h_{v'}^{(c)}\,\Gamma\,h_v^{(b)}) = 
  (\bar\Lambda v_\nu-\bar\Lambda'v'_\nu)\,
  \bar h_{v'}^{(c)}\,\Gamma\,h_v^{(b)} \,,
\end{equation}
which is valid between the states destroyed by the fields in $F_{v'}^\sigma$
and $H_v$.  This relation follows from translation invariance and the 
definition of the heavy quark fields $h_v^{(Q)}$.  It implies that
\begin{equation}\label{const2a}
{\cal S}^{(c)}_{\sigma\lambda} + {\cal S}^{(b)}_{\sigma\lambda}
  = (\bar\Lambda v_\lambda-\bar\Lambda'v'_\lambda)\, v_\sigma\, \tau \,.
\end{equation}
Eq.~(\ref{const2a}) gives the following relations\footnote{In Ref.~\cite{llsw}
two out of these four relations were obtained (only those two were needed to 
get Eq.~(\ref{tau41})).  We thank M.~Neubert for pointing out that there are 
two additional constraints.}
\begin{eqnarray}\label{const2}
\tau_1^{(c)} + \tau_1^{(b)} &=& \bar\Lambda\, \tau \,, \nonumber\\*
  \tau_2^{(c)} + \tau_2^{(b)} &=& -\bar\Lambda'\, \tau \,, \nonumber\\
  \tau_3^{(c)} + \tau_3^{(b)} &=& 0 \,, \nonumber\\*
  \tau_4^{(c)} + \tau_4^{(b)} &=& 0 \,.
\end{eqnarray}
These relations express the $\tau_j^{(b)}$'s in terms of the $\tau_j^{(c)}$'s.
Furthermore, combining Eqs.~(\ref{const1}) with (\ref{const2}) yields
\begin{eqnarray}\label{const2b}
\tau_3^{(c)} &=& w\, \tau_1^{(c)} + \tau_2^{(c)} \,, \nonumber\\*
\tau_4^{(c)} &=& (w-1)\, (\tau_1^{(c)} - \tau_2^{(c)})
  - (w \bar\Lambda' - \bar\Lambda)\, \tau \,.
\end{eqnarray}
All order $\Lambda_{\rm QCD}/m_Q$ corrections to the form factors coming from
the matching of the QCD currents onto those in the effective theory are
expressible in terms of $\bar\Lambda\,\tau$ and $\bar\Lambda'\,\tau$ and two
functions, which we take to be $\tau_1^{(c)}$ and $\tau_2^{(c)}$.  From
Eqs.~(\ref{curr}) and (\ref{Sdef}) it is evident that only $\tau_4^{(Q)}$ can
contribute at zero recoil.  Eq.~(\ref{const2b}) determines this contribution in
terms of $\tau(1)$ and measurable mass splittings given in
Eq.~(\ref{HQET_diff}),
\begin{equation}\label{tau41}
\tau^{(b)}_4(1) = -\tau^{(c)}_4(1) = (\bar\Lambda'-\bar\Lambda)\,\tau(1) \,.  
\end{equation}

Note that with our methods Eq.~(\ref{tau41}) cannot be derived working
exclusively at zero recoil.  At that kinematic point, matrix elements of the
operator $\bar h_v^{(c)}\,\Gamma\,h_v^{(b)}$ vanish between a $B$ meson and an
excited charmed meson, and so Eq.~(\ref{momcons}) only implies that
$\tau_4^{(c)}+\tau_4^{(b)}=0$.  Eq.~(\ref{tau41}) relies on the assumption that
the $\tau_j^{(Q)}(w)$ are continuous at $w=1$.

Next consider the terms originating from order $\Lambda_{\rm QCD}/m_Q$ 
corrections to the HQET Lagrangian, $\delta {\cal L}$ in Eq.~(\ref{lag}).
These corrections modify the heavy meson states compared to their infinite
heavy quark mass limit.  For example, they cause the mixing of the $D_1$ with
the $J^P=1^+$ member of the $s_l^{\pi_l}=\frac12^+$ doublet.  (This is a very
small effect, since the $D_1$ is not any broader than the $D_2^*$.)  For matrix
elements between the states destroyed by the fields in $F_{v'}^\sigma$ and
$H_v$, the time ordered products of the kinetic energy term in $\delta{\cal L}$
with the leading order currents are 
\begin{eqnarray}\label{kinetic}
i \int {\rm d}^4x\, T\,\Big\{ O_{{\rm kin},v'}^{(c)}(x)\, 
  \Big[ \bar h_{v'}^{(c)}\, \Gamma\, h_{v}^{(b)} \Big](0)\, \Big\} 
  &=& \eta^{(c)}_{\rm ke}\, {\rm Tr}\, \Big\{ v_\sigma 
  \bar F^\sigma_{v'}\, \Gamma\, H_v \Big\} \,, \nn\\*
i \int {\rm d}^4x\, T\,\Big\{ O_{{\rm kin},v}^{(b)}(x)\, 
  \Big[ \bar h_{v'}^{(c)}\, \Gamma\, h_{v}^{(b)} \Big](0)\, \Big\} 
  &=& \eta^{(b)}_{\rm ke}\, {\rm Tr}\, \Big\{ v_\sigma 
  \bar F^\sigma_{v'}\, \Gamma\, H_v \Big\} \,.
\end{eqnarray}
These corrections do not violate spin symmetry, so their contributions enter
the same way as the $m_Q\to\infty$ Isgur-Wise function, $\tau$.

For matrix elements between the states destroyed by the fields in
$F_{v'}^\sigma$ and $H_v$, the time ordered products of the chromomagnetic 
term in $\delta{\cal L}$ with the leading order currents are
\begin{eqnarray}\label{magnetic}
i \int {\rm d}^4x\, T\,\Big\{ O_{{\rm mag},v'}^{(c)}(x)\, 
  \Big[ \bar h_{v'}^{(c)}\, \Gamma\, h_{v}^{(b)} \Big](0)\, \Big\} 
  &=& {\rm Tr}\, \bigg\{ {\cal R}_{\sigma\alpha\beta}^{(c)}\,
  \bar F_{v'}^\sigma\, i\sigma^{\alpha\beta}\, \frac{1+\vslash'}2\, 
  \Gamma\, H_v \bigg\} \,, \nonumber\\*
i \int {\rm d}^4x\, T\,\Big\{ O_{{\rm mag},v}^{(b)}(x)\, 
  \Big[ \bar h_{v'}^{(c)}\, \Gamma\, h_{v}^{(b)} \Big](0)\, \Big\} 
  &=& {\rm Tr}\, \bigg\{ {\cal R}_{\sigma\alpha\beta}^{(b)}\,
  \bar F_{v'}^\sigma\, \Gamma\, \frac{1+\vslash}2\, i\sigma^{\alpha\beta}
  H_v \bigg\} \,. 
\end{eqnarray}
The most general parameterizations of ${\cal R}^{(Q)}$ are
\begin{eqnarray}\label{Rdef}
{\cal R}_{\sigma\alpha\beta}^{(c)} &=& 
  \eta_1^{(c)}\, v_\sigma \gamma_\alpha \gamma_\beta 
  + \eta_2^{(c)}\, v_\sigma v_\alpha \gamma_\beta 
  + \eta_3^{(c)}\, g_{\sigma\alpha} v_\beta \,, \nonumber\\*
{\cal R}_{\sigma\alpha\beta}^{(b)} &=& 
  \eta_1^{(b)}\, v_\sigma \gamma_\alpha \gamma_\beta 
  + \eta_2^{(b)}\, v_\sigma v'_\alpha \gamma_\beta 
  + \eta_3^{(b)}\, g_{\sigma\alpha} v'_\beta \,.
\end{eqnarray}
Only the part of ${\cal R}_{\sigma\alpha\beta}^{(Q)}$ antisymmetric in
$\alpha$ and $\beta$ contributes when inserted into Eq.~(\ref{magnetic}).  
The functions $\eta_i$ depend on $w$, and have mass dimension one.  Note that
$g_{\sigma\alpha}\gamma_\beta$ is dependent on the tensor structures included
in Eq.~(\ref{Rdef}) for matrix elements between these states.  For example, 
for the $\Lambda_{\rm QCD}/m_c$ corrections the following trace identity holds
\begin{equation}\label{magicTr}
{\rm Tr}\, \bigg\{ \Big[ v_\sigma \gamma_\alpha\gamma_\beta 
  + 2g_{\sigma\alpha} v_\beta + 2(1+w)\, g_{\sigma\alpha} \gamma_\beta 
  \Big] \bar F_{v'}^\sigma\, \sigma^{\alpha\beta}\, 
  \frac{1+\vslash'}2\, \Gamma H_v \bigg\} = 0 \,. 
\end{equation}
All contributions arising from the time ordered products in
Eq.~(\ref{magnetic}) vanish at zero recoil, since $v_\sigma\bar F_v^\sigma=0$
and \hbox{$v_\alpha(1+\vslash)\sigma^{\alpha\beta}(1+\vslash)=0$}.  Thus we
find that at zero recoil the only $\Lambda_{\rm QCD}/m_Q$ corrections that
contribute are determined by measured meson mass splittings and the value of
the leading order Isgur-Wise function at zero recoil.

The form factors in Eq.~(\ref{formf1}) depend on $\eta_i^{(b)}$ only through
the linear combination $\eta_b=\eta_{\rm
ke}^{(b)}+6\,\eta_1^{(b)}-2(w-1)\,\eta_2^{(b)}+\eta_3^{(b)}$.  Denoting
$\varepsilon_Q=1/(2m_Q)$ and dropping the superscript on $\tau_i^{(c)}$ and
$\eta_i^{(c)}$, the $B\to D_1\,e\,\bar\nu_e$ form factors are \cite{llsw}
\begin{eqnarray}\label{expf}
\sqrt6\, f_A &=& - (w+1)\tau 
  - \varepsilon_b \{ (w-1) [(\bar\Lambda'+\bar\Lambda)\tau 
  - (2w+1)\tau_1-\tau_2] + (w+1)\eta_b \} \nonumber\\*
&& - \varepsilon_c [ 4(w\bar\Lambda'-\bar\Lambda)\tau - 3(w-1) (\tau_1-\tau_2) 
  + (w+1) (\eta_{\rm ke}-2\eta_1-3\eta_3) ] \,,\nonumber\\*
\sqrt6\, f_{V_1} &=&  (1-w^2)\tau 
  - \varepsilon_b (w^2-1) [(\bar\Lambda'+\bar\Lambda)\tau 
  - (2w+1)\tau_1-\tau_2 + \eta_b] \nonumber\\*
&& - \varepsilon_c [ 4(w+1)(w\bar\Lambda'-\bar\Lambda)\tau
  - (w^2-1)(3\tau_1-3\tau_2-\eta_{\rm ke}+2\eta_1+3\eta_3) ] \,, \nonumber\\
\sqrt6\, f_{V_2} &=& -3\tau - 3\varepsilon_b [(\bar\Lambda'+\bar\Lambda)\tau 
  - (2w+1)\tau_1-\tau_2 + \eta_b] \nonumber\\* 
&& - \varepsilon_c [ (4w-1)\tau_1+5\tau_2 +3\eta_{\rm ke} +10\eta_1 
  + 4(w-1)\eta_2-5\eta_3 ] \,, \nonumber\\*
\sqrt6\, f_{V_3} &=&  (w-2)\tau 
  + \varepsilon_b \{ (2+w) [(\bar\Lambda'+\bar\Lambda)\tau 
  - (2w+1)\tau_1-\tau_2] - (2-w)\eta_b \} \nonumber\\*
&& + \varepsilon_c [ 4(w\bar\Lambda'-\bar\Lambda)\tau + 
  (2+w)\tau_1 + (2+3w)\tau_2 \nonumber\\*
&& \phantom{\varepsilon_c [}
  + (w-2)\eta_{\rm ke} - 2(6+w)\eta_1 - 4(w-1)\eta_2 - (3w-2)\eta_3 ] \,. 
\end{eqnarray}
The analogous formulae for $B\to D_2^*\,e\,\bar\nu_e$ are
\begin{eqnarray}\label{expk}
k_V &=& - \tau - \varepsilon_b [(\bar\Lambda'+\bar\Lambda)\tau 
  - (2w+1)\tau_1-\tau_2 + \eta_b] 
  - \varepsilon_c (\tau_1-\tau_2+\eta_{\rm ke}-2\eta_1+\eta_3) , \nonumber\\*
k_{A_1} &=& - (1+w)\tau - \varepsilon_b \{ (w-1)
  [(\bar\Lambda'+\bar\Lambda)\tau - (2w+1)\tau_1-\tau_2] + (1+w)\eta_b \} 
  \nonumber\\*
&& - \varepsilon_c [ (w-1)(\tau_1-\tau_2)
  + (w+1)(\eta_{\rm ke}-2\eta_1+\eta_3) ] , \nonumber\\
k_{A_2} &=& - 2\varepsilon_c (\tau_1+\eta_2) , \\*
k_{A_3} &=& \tau + \varepsilon_b [(\bar\Lambda'+\bar\Lambda)\tau 
  - (2w+1)\tau_1-\tau_2 + \eta_b] 
  - \varepsilon_c (\tau_1+\tau_2-\eta_{\rm ke}+2\eta_1-2\eta_2-\eta_3) . 
  \nonumber
\end{eqnarray}
Recall that $f_{V_1}$ determines the zero recoil matrix elements of the weak
currents.  From Eqs.~(\ref{expf}) it follows that
\begin{equation}
\sqrt6\, f_{V_1}(1) = -8\, \varepsilon_c\, (\bar\Lambda'-\bar\Lambda)\, 
  \tau(1)\,.
\end{equation}

The allowed kinematic range for $B\to D_1\,e\,\bar\nu_e$ decay is $1<w<1.32$,
while for $B\to D_2^*\,e\,\bar\nu_e$ decay it is $1<w<1.31$.  Since these
ranges are fairly small, and at zero recoil there are some constraints on the
$\Lambda_{\rm QCD}/m_Q$ corrections, it is useful to consider the decay rates
given in Eq.~(\ref{rate1}) expanded in powers of $w-1$.  The general structure
of the expansion of ${\rm d}\Gamma/{\rm d}w$ is elucidated schematically below,
\begin{eqnarray}\label{schematic}
{{\rm d}\Gamma_{D_1}^{(\lambda=0)}\over {\rm d}w} &\sim&
  \sqrt{w^2-1}\, \Big[ (w-1)^0\, \Big( 0
  + 0\,\varepsilon + \varepsilon^2 + \varepsilon^3 + \ldots \Big) \nonumber\\*
&& \phantom{\sqrt{w^2-1}}
  + (w-1)^1\, \Big(0 + \varepsilon + \varepsilon^2 + \ldots\Big) 
  + (w-1)^2\, \Big(1 + \varepsilon + \ldots\Big) + \ldots \Big] \,, 
  \nonumber\\*
{{\rm d}\Gamma_{D_1}^{(|\lambda|=1)}\over {\rm d}w} &\sim&
  \sqrt{w^2-1}\, \Big[ (w-1)^0\, \Big( 0
  + 0\,\varepsilon + \varepsilon^2 + \varepsilon^3 + \ldots \Big) 
  \nonumber\\*
&& \phantom{\sqrt{w^2-1}}
  + (w-1)^1\, \Big(1 + \varepsilon + \ldots\Big) 
  + (w-1)^2\, \Big(1 + \varepsilon + \ldots\Big) + \ldots \Big] \,,
  \nonumber\\*
{{\rm d}\Gamma_{D_2}^{(|\lambda|=0,1)}\over {\rm d}w} &\sim&
  (w^2-1)^{3/2}\, \Big[ (w-1)^0\, \Big( 1 + \varepsilon + \ldots \Big) 
  + (w-1)^1\, \Big(1 + \varepsilon + \ldots\Big) + \ldots \Big] \,. 
\end{eqnarray}
Here $\varepsilon^n$ denotes a term of order $(\Lambda_{\rm QCD}/m_Q)^n$.  The
zeros in Eq.~(\ref{schematic}) are consequences of heavy quark symmetry, as
the leading contribution to the matrix elements of the weak currents at zero
recoil is of order $\Lambda_{\rm QCD}/m_Q$.  Thus, the $D_1$ decay rate at
$w=1$ starts out at order $\Lambda_{\rm QCD}^2/m_Q^2$.  Similarly, from
Eq.~(\ref{rate1}) it is evident that the vanishing of $f_{V_1}(1)$ in the
$m_Q\to\infty$ limit implies that at order $w-1$ the $D_1^{(\lambda=0)}$ rate
starts out at order $\Lambda_{\rm QCD}/m_Q$.  The $D_2^*$ decay rate is
suppressed by an additional power of $w^2-1$, so there is no further
restriction on its structure.

In this paper we present predictions using two different approximations to the
decay rates.  In approximation~A we treat $w-1$ as order $\Lambda_{\rm
QCD}/m_Q$ and expand the decay rates in these parameters.  In approximation~B
the known order $\Lambda_{\rm QCD}/m_Q$ contributions to the form factors are
kept, as well as the full $w$-dependence of the decay rates.

Expanding the terms in the square brackets in Eq.~(\ref{rate1}) in powers of
$w-1$ gives
\begin{eqnarray}\label{rt_expn1}
{{\rm d}^2\Gamma_{D_1}\over {\rm d}w\,{\rm d}\!\cos\theta} &=& 
  \Gamma_0\, \tau^2(1)\, r_1^3\, \sqrt{w^2-1}\, \sum_n\, (w-1)^n\,
  \bigg\{ \sin^2\theta\, s_1^{(n)}  \\*
&& + (1-2r_1w+r_1^2)\, \Big[ (1+\cos^2\theta)\, t_1^{(n)} 
  - 4\cos\theta\, \sqrt{w^2-1}\, u_1^{(n)} \Big] \bigg\} \,,\nonumber\\*
{{\rm d}^2\Gamma_{D_2^*}\over {\rm d}w\,{\rm d}\!\cos\theta} &=& 
   \frac32\,\Gamma_0\, \tau^2(1)\, r_2^3\, (w^2-1)^{3/2}\, \sum_n\, (w-1)^n\,
  \bigg\{ \frac43\,\sin^2\theta\, s_2^{(n)}  \nn\\*
&& + (1-2r_2w+r_2^2)\, \Big[ (1+\cos^2\theta)\, t_2^{(n)} 
  - 4\cos\theta\, \sqrt{w^2-1}\, u_2^{(n)} \Big] \bigg\} \,.\nonumber
\end{eqnarray}
(We do not expand the factors of $\sqrt{w^2-1}$ that multiply $\cos\theta$).
The subscripts of the coefficients $s,t,u$ denote the spin of the excited $D$
meson, while the superscripts refer to the order in the $w-1$ expansion.   
The $u_i^{(n)}$ terms proportional to $\cos\theta$ only affect the
lepton spectrum, since they vanish when integrated over $\theta$.

Eqs.~(\ref{rate1}), (\ref{expf}), and (\ref{expk}) yield the following 
expressions for the coefficients in the $D_1$ decay rate in 
Eq.~(\ref{rt_expn1}),
\begin{eqnarray}\label{stuD1}
s_1^{(0)} &=& 32\varepsilon_c^2\, (1-r_1)^2\,
  (\bar\Lambda'-\bar\Lambda)^2 + \ldots \,, \nn\\*
s_1^{(1)} &=& 32\varepsilon_c\, (1-r_1^2)\,
  (\bar\Lambda'-\bar\Lambda) + \ldots \,, \nn\\*
s_1^{(2)} &=& 8\, (1+r_1)^2 + \ldots \,, \nonumber\\
t_1^{(0)} &=& 32\varepsilon_c^2\, (\bar\Lambda'-\bar\Lambda)^2 + \ldots \,, 
  \nn\\*
t_1^{(1)} &=& 4 + 8\varepsilon_c\, \Big[ 4(\bar\Lambda'-\bar\Lambda) 
  + \hat\eta_{\rm ke} - 2\hat\eta_1 - 3\hat\eta_3 \Big] 
  + 8\varepsilon_b\, \hat\eta_b + \ldots \,, \nn\\*
t_1^{(2)} &=& 8\,(1+\hat\tau') + \ldots \,, \nn\\
u_1^{(0)} &=& 8\varepsilon_c\, (\bar\Lambda'-\bar\Lambda) + \ldots \,, \nn\\*
u_1^{(1)} &=& 2 + \ldots \,. 
\end{eqnarray}
For the decay rate into $D_2^*$ the first two terms in the $w-1$ expansion are
\begin{eqnarray}\label{stuD2}
s_2^{(0)} &=& 4\, (1-r_2)^2\, \Big[1 + 2\varepsilon_b\,\hat\eta_b
  + 2\varepsilon_c\,(\hat\eta_{\rm ke}-2\hat\eta_1+\hat\eta_3) \Big] + \ldots
  \,,\nn\\*
s_2^{(1)} &=& 4\,(1-r_2)^2\, (1+2\hat\tau') + \ldots \,, \nn\\
t_2^{(0)} &=& 4 + 8\varepsilon_b\,\hat\eta_b
  + 8\varepsilon_c\, (\hat\eta_{\rm ke}-2\hat\eta_1+\hat\eta_3) + \ldots \,, 
  \nn\\*
t_2^{(1)} &=& 2(3+4\hat\tau') + \ldots \,, \nn\\
u_2^{(0)} &=& 2 + \ldots \,. 
\end{eqnarray}
In Eqs.~(\ref{stuD1}) and (\ref{stuD2}) the functions $\tau$, $\tau'={\rm
d}\tau/{\rm d}w$, and $\eta_i$ are all evaluated at $w=1$, and the functions
with a hat are normalized to $\tau(1)$ (e.g., $\hat\eta_i=\eta_i/\tau(1)$,
$\hat\tau'=\tau'/\tau(1)$, etc.).  The ellipses denote higher order terms in
the $\Lambda_{\rm QCD}/m_Q$ expansion.  The $u_i^{(n)}$ terms are suppressed by
$\sqrt{w^2-1}$ compared to $s_i^{(n)}$ and $t_i^{(n)}$, therefore we displayed
the $u$'s to one lower order than the $s$ and $t$ coefficients.  (Note that
$u_1^{(0)}$ also starts out at order $\Lambda_{\rm QCD}/m_Q$ as a consequence
of the vanishing of $f_{V_1}(1)$ in the $m_Q\to\infty$ limit, as it was shown
for $s_1^{(1)}$ after Eq.~(\ref{schematic}).)

The order $\Lambda_{\rm QCD}/m_Q$ terms proportional to
$\bar\Lambda'-\bar\Lambda$ are very significant for the $D_1$ decay rate.  The
decay rate into $D_2^*$ does not receive a similarly large enhancement from
order $\Lambda_{\rm QCD}/m_Q$ terms proportional to $\bar\Lambda'-\bar\Lambda$.
The coefficients $s_2^{(n)}$ and $t_2^{(n)}$ are independent of $\bar\Lambda'$
and $\bar\Lambda$ to the order displayed in Eq.~(\ref{stuD2}).  

The values of $s_1^{(0)}$ and $t_1^{(0)}$ are known to order $\Lambda_{\rm
QCD}^2/m_Q^2$, and $s_1^{(1)}$ and $u_1^{(0)}$ are known to order $\Lambda_{\rm
QCD}/m_Q$.  At order $\Lambda_{\rm QCD}/m_Q$, the only unknowns in $t_1^{(1)}$,
$s_2^{(0)}$, $t_2^{(0)}$ are the $\hat\eta_i$ functions that parameterize
corrections to the HQET Lagrangian.  The remaining coefficients in
Eqs.~(\ref{stuD1}) and (\ref{stuD2}) (i.e., $s_1^{(2)}$, $t_1^{(2)}$,
$u_1^{(1)}$, $s_2^{(1)}$, $t_2^{(1)}$, $u_2^{(0)}$) are known in the infinite
mass limit in terms of $\hat\tau'(1)$, the slope of the $m_Q\to\infty$
Isgur-Wise function at zero recoil.  At order $\Lambda_{\rm QCD}/m_Q$, these
six coefficients depend on the unknown subleading $\tau_i$ and $\eta_i$
functions.

The values of $\tau'$, $\eta_i^{(Q)}$ and $\tau_{1,2}$ that occur in
Eqs.~(\ref{stuD1}) and (\ref{stuD2}) are not known ($\tau_i$ only appears in
the terms replaced by ellipses).  $\eta_{1,2,3}^{(Q)}$, which parameterize time
ordered products of the chromomagnetic operator, are expected to be small
(compared to $\Lambda_{\rm QCD}$), and we neglect them hereafter.  This is
supported by the very small $D_2^*-D_1$ mass splitting, and the fact that model
calculations indicate that the analogous functions parameterizing time ordered
products of the chromomagnetic operator for $B\to D^{(*)}\,e\,\bar\nu_e$ decays
are small \cite{qcdsr}.  On the other hand, there is no reason to expect
$\tau_{1,2}$ and $\eta_{\rm ke}^{(Q)}$ to be much smaller than about
$500\,$MeV.  Note that the large value for $\lambda_1'$ is probably a
consequence of the $D_1$ and $D_2^*$ being $P$-waves in the quark model, and
does not necessarily imply that $O_{\rm kin}^{(Q)}$ significantly distorts the
overlap of wave functions that yield $\eta_{\rm ke}^{(Q)}$.  

Even though $\varepsilon_c(\bar\Lambda'-\bar\Lambda)\simeq0.14$ is quite small,
the order $\Lambda_{\rm QCD}/m_Q$ correction to $t_1^{(1)}$ proportional to
$\varepsilon_c(\bar\Lambda'-\bar\Lambda)$ is as large as the leading
$m_Q\to\infty$ contribution.  This occurs because it has an anomalously large
coefficient and does not necessarily mean that the $\Lambda_{\rm QCD}/m_Q$
expansion has broken down.  For example, the part of the $\Lambda_{\rm
QCD}^2/m_c^2$ corrections that involve $\bar\Lambda'$, $\bar\Lambda$, and
$\tau'(1)$ affect $s_1^{(1)}$ by $(21+10\hat\tau')\%$, and $t_1^{(1)}$ by
$(44+15\hat\tau')\%$ (using $\bar\Lambda=0.4\,$GeV \cite{gklw}).  These
corrections follow from Eq.~(\ref{expf}), but they are neglected in
Eq.~(\ref{stuD1}) (i.e., approximation~A), because there are other order
$\Lambda_{\rm QCD}^2/m_Q^2$ effects we have not calculated.

As the kinetic energy operator does not violate spin symmetry, effects of
$\eta_{\rm ke}^{(Q)}$ can be absorbed into $\tau$ by the replacement of $\tau$
by $\widetilde\tau=\tau+\varepsilon_c\,\eta_{\rm
ke}^{(c)}+\varepsilon_b\,\eta_{\rm ke}^{(b)}$.  This replacement introduces an
error of order $\Lambda_{\rm QCD}^2/m_Q^2$, in $t_1^{(1)}$, etc.  But due to
the presence of large $\Lambda_{\rm QCD}/m_Q$ corrections, the resulting
$\Lambda_{\rm QCD}^2/m_Q^2$ error is also sizable, and is expected to be more
like an order $\Lambda_{\rm QCD}/m_Q$ correction.  Hereafter, unless explicitly
stated otherwise, it is understood that the replacement $\tau\to\widetilde\tau$
is made.  But we shall examine the sensitivity of our results to~$\eta_{\rm
ke}$ (assuming it has the same shape as $\tau$).

In approximation~A we treat $w-1$ as order $\Lambda_{\rm QCD}/m_Q$ \cite{llsw},
and keep terms up to order $(\Lambda_{\rm QCD}/m_Q)^{2-n}$ in $s_1^{(n)}$ and
$t_1^{(n)}$ ($n=0,1,2$) in Eq.~(\ref{stuD1}), and up to order $(\Lambda_{\rm
QCD}/m_Q)^{1-n}$ in $s_2^{(n)}$ and $t_2^{(n)}$ ($n=0,1$) in Eq.~(\ref{stuD2}).
Since the $u_i^{(n)}$ are suppressed by $\sqrt{w^2-1}$ compared to $s_i^{(n)}$
and $t_i^{(n)}$, we keep $u_i^{(n)}$ to one lower order than the $s$ and $t$
coefficients, i.e., to order $(\Lambda_{\rm QCD}/m_Q)^{1-n}$ ($n=0,1$) for
$B\to D_1$ decay and order $(\Lambda_{\rm QCD}/m_Q)^{n}$ ($n=0$) for $B\to
D_2^*$ decay.  The terms included in approximation~A are precisely the ones
explicitly shown in Eqs.~(\ref{stuD1}) and (\ref{stuD2}).  This power counting
has the advantage that the unknown functions, $\tau_1$ and $\tau_2$, do not
enter the predictions.\footnote{Approximation~A differs from our discussion in
Ref.~\cite{llsw} only in the separation of the different helicity states of the
excited charmed mesons, and keeping the $1-2rw+r^2$ factors for the helicity
one states as well as the $(w^2-1)^{3/2}$ terms for the $D_2^*$ rates
unexpanded.}  Neglecting higher order terms in the $w-1$ expansion in
approximation~A gives rise to a sizable error for the $B\to D_1\,e\,\bar\nu_e$
decay\footnote{We thank A.\ Le Yaouanc for pointing out the importance of these
terms.}.  The order $(w-1)^3$ term is important for the decay into helicity
zero $D_1$ in the $m_Q\to\infty$ limit, since the helicity zero rate (which, as
we shall see, dominates over the helicity one rate) starts out at order
$(w-1)^2$ as shown in Eq.~(\ref{schematic}).

In approximation~B we do not expand the decay rates in powers of $w-1$.  We
keep the $\Lambda_{\rm QCD}/m_Q$ corrections to the form factors that involve
$\bar\Lambda'$ and $\bar\Lambda$ and examine the sensitivity of our results to
the corrections involving $\tau_1$ and $\tau_2$ (assuming that they have the 
same shape as $\tau$, which is not a strong assumption). This approximation 
retains some order $\Lambda_{\rm QCD}^2/m_Q^2$ terms away from zero recoil in 
the differential decay rates.  Furthermore, a linear form for the Isgur-Wise
function is assumed, $\tau(w)=\tau(1)\,[1+\hat\tau'(w-1)]$.  The uncertainty in
the $\Lambda_{\rm QCD}/m_Q$ corrections is parameterized by the functions
$\tau_{1,2}(w)$.  A different choice of $\tau_{1,2}(w)$ changes what is
retained by terms involving $\bar\Lambda/m_Q$ and $\bar\Lambda'/m_Q$.  In an
approximation, which we shall refer to as B$_1$, we set $\tau_1=\tau_2=0$ in
Eqs.~(\ref{expf}) and (\ref{expk}).  (This is identical to saturating the first
two relations in Eq.~(\ref{const2}) by $\tau_{1,2}^{(b)}$, i.e., setting
$\tau_1^{(b)}=\bar\Lambda\,\tau$ and $\tau_2^{(b)}=-\bar\Lambda'\,\tau$.)  An
equally reasonable approximation, which we refer to as B$_2$, is given by
setting $\tau_1=\bar\Lambda\,\tau$ and $\tau_2=-\bar\Lambda'\,\tau$ in
Eqs.~(\ref{expf}) and (\ref{expk}).  (This is identical to setting
$\tau_{1,2}^{(b)}=0$.)  If the first two relations in Eq.~(\ref{const2}) are
taken as hints to the signs of $\tau_1$ and $\tau_2$, then the difference
between approximations~B$_1$ and B$_2$ gives a rough estimate of the
uncertainty related to the unknown $\Lambda_{\rm QCD}/m_Q$ corrections.  When
our predictions are sensitive to $\tau_1$ and $\tau_2$, we shall vary these in
a range larger than that spanned by approximations~B$_1$ and B$_2$.  Note that
the infinite mass limits of B$_1$ and B$_2$ coincide.  Predictions of
approximation~A are within the spread of the approximation~B results, except
for those that depend on the helicity zero $D_1$ rate.  In that case, including
the order $(w-1)^3$ term in the infinite mass limit alone,
$s_1^{(3)}=8\,(1+r_1)^2\,(1+2\hat\tau')$, would bring the approximation~A
results close to approximation~B.

Eqs.~(\ref{stuD1}) and (\ref{stuD2}) show that the heavy quark expansion for
$B$ decays into excited charmed mesons is controlled by the excitation energies
of the hadrons, $\bar\Lambda'$ and $\bar\Lambda$.  For highly excited mesons
that have $\bar\Lambda'$ comparable to $m_c$, the $1/m_Q$ expansion is not
useful.  For the $s_\ell^{\pi_\ell}=\frac32^+$ doublet
$\varepsilon_c\,\bar\Lambda'\sim0.3$.  However, near zero recoil only
$\varepsilon_c\,(\bar\Lambda'-\bar\Lambda)\sim0.14$ occurs at order
$\Lambda_{\rm QCD}/m_Q$.

The expressions for the decay rates in terms of form factors in
Eq.~(\ref{rate1}) imply that one form factor dominates each decay rate near
zero recoil, independent of the helicity of the $D_1$ or $D_2^*$ ($f_{V_1}$ for
$D_1$ and $k_{A_1}$ for $D_2^*$).  Thus, to all orders in the $\Lambda_{\rm
QCD}/m_Q$ expansion, $s_1^{(0)}/t_1^{(0)}=(1-r_1)^2$, and
$s_2^{(0)}/t_2^{(0)}=(1-r_2)^2$.  This implies that for $B\to D_1$ decay
$\lim_{w\to1}\Big[({\rm d}\Gamma_{D_1}^{(\lambda=0)}/{\rm d}w)\Big/({\rm
d}\Gamma_{D_1}^{(|\lambda|=1)}/{\rm d}w)\Big]=1/2$, and for $B\to D_2^*$ decay
$\lim_{w\to1}\Big[({\rm d}\Gamma_{D_2^*}^{(\lambda=0)}/{\rm d}w)\Big/({\rm
d}\Gamma_{D_2^*}^{(|\lambda|=1)}/{\rm d}w)\Big]=2/3$.  Note that the first of
these ratios would vanish if the rates were calculated in the $m_Q\to\infty$
limit.  In that case $f_{V_1}(1)=0$, so the ratio of helicity zero and helicity
one $B\to D_1$ rates is determined by the other form factors at zero recoil.

\subsection{Predictions}

The relationship between $s_1^{(0)}$ and $s_1^{(1)}$ implies a model
independent prediction for the slope parameter of semileptonic $B$ decay into
helicity zero $D_1$.  This holds independent of the subleading Isgur-Wise
functions that arise at order $\Lambda_{\rm QCD}/m_Q$.  The semileptonic decay
rate to a helicity zero $D_1$ meson is
\begin{equation}\label{mir1}
{{\rm d}\Gamma_{D_1}^{(\lambda=0)} \over {\rm d}w} = \frac{128}3\,
  \Gamma_0\, r_1^3\, (1-r_1)^2\, \sqrt{w^2-1}\, \tau^2(1)\, 
  \varepsilon_c^2\, (\bar\Lambda'-\bar\Lambda)^2\,
  \Big[ 1 - \rho_{D_1}^2\, (w-1) + \ldots \Big] \,,
\end{equation}
where the slope parameter $\rho_{D_1}^2$ for helicity zero $D_1$ has the value
\begin{equation}
\rho_{D_1}^2 = - {1+r_1\over1-r_1}\, {2m_c\over\bar\Lambda'-\bar\Lambda} 
  + {\cal O}(1) \,.
\end{equation}
Since the decay rate at zero recoil is suppressed, $\rho_{D_1}^2$ is of order
$m_Q/\Lambda_{\rm QCD}$.  Note that this slope parameter is negative.

Recently the ALEPH \cite{ALEPH} and CLEO \cite{CLEO} Collaborations measured,
with some assumptions, the $B\to D_1\,e\,\bar\nu_e$ branching ratio.  The 
average of their results is
\begin{equation}\label{data}
  {\cal B}(B\to D_1\,e\,\bar\nu_e) = (6.0\pm1.1) \times 10^{-3} \,.  
\end{equation}
The $B\to D_2^*\,e\,\bar\nu_e$ branching ratio has not yet been measured;
CLEO set the limit ${\cal B}(B\to~D_2^*\,e\,\bar\nu_e)<1\%$ \cite{CLEO}, 
while ALEPH found ${\cal B}(B\to D_2^*\,e\,\bar\nu_e)<0.2\%$ \cite{ALEPH}.

Predictions for various quantities of experimental interest are made in
Table~\ref{tab:sec2res} using $\bar\Lambda'-\bar\Lambda=0.39\,$GeV,
$\bar\Lambda=0.4\,$GeV, $\tau_B=1.6\,$ps, $|V_{cb}|=0.04$, $m_c=1.4\,$GeV,
$m_b=4.8\,$GeV.  Keeping $m_b-m_c$ fixed and varying $m_c$ by $\pm0.1\,$GeV
only affects our results at the few percent level.  These predictions depend on
the shape of the Isgur-Wise function.  In our approximations this enters
through the slope parameter, $\hat\tau'=\tau'(1)/\tau(1)$, which is expected to
be of order $-1$.  We shall quote results for the ``central value"
$\hat\tau'=-1.5$, motivated by model predictions \cite{ISGW,Cola,VeOl,More},
and discuss the sensitivity to this assumption.  For $B\to D_1\,e\,\bar\nu_e$
decay we use $r_1=0.459$ and $1<w<1.319$, whereas for $B\to
D_2^*\,e\,\bar\nu_e$ decay $r_2=0.466$ and $1<w<1.306$.  

\begin{table}[t]
\begin{tabular}{c|cccc}  
Approximation  &  $R=\Gamma_{D_2^*}\Big/\Gamma_{D_1}$  &  
  $\Gamma_{D_1}^{(\lambda=0)}\Big/\Gamma_{D_1}$  &
  $\Gamma_{D_2^*}^{(\lambda=0)}\Big/\Gamma_{D_2^*}$  &
  $\tau(1)\, \bigg[\displaystyle {6.0\times10^{-3} \over 
    {\cal B}(B\to D_1\,e\,\bar\nu_e)} \bigg]^{1/2}$  \\[6pt] \hline 
A$_\infty$  &  $0.93$  &  $0.88$  &  $0.64$  &  $0.92$  \\
B$_\infty$  &  $1.65$  &  $0.80$  &  $0.66$  &  $1.24$  \\ \hline
A  &  $0.40$  &  $0.81$  &  $0.64$  &  $0.60$  \\
B$_1$  &  $0.52$  &  $0.72$  &  $0.63$  &  $0.71$  \\
B$_2$  &  $0.67$  &  $0.77$  &  $0.64$  &  $0.75$  \\
\end{tabular} \vspace{6pt}
\tighten{
\caption[2]{Predictions for various ratios of $B\to D_1\,e\,\bar\nu_e$ and 
$B\to D_2^*\,e\,\bar\nu_e$ decay rates, as described in the text.  
The extracted value of $\tau(1)$ is also shown.  A$_\infty$ and B$_\infty$
denote the $m_Q\to\infty$ limits of approximations~A and B.  These results 
correspond to $\hat\tau'=\tau'(1)/\tau(1)=-1.5$.} \label{tab:sec2res} }
\end{table}

\begin{figure}[t]
\centerline{\epsfysize=8truecm \epsfbox{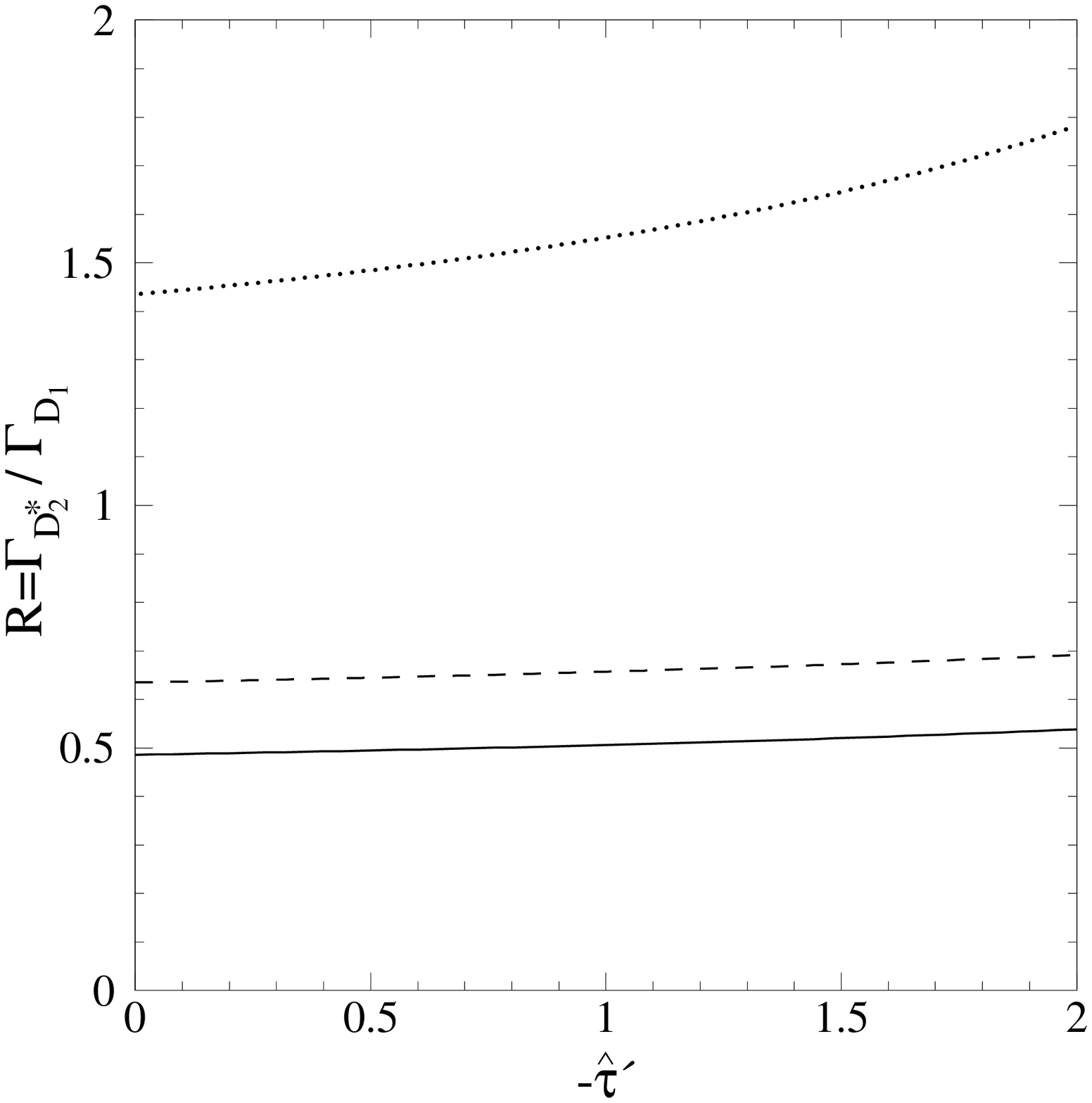}
  \epsfysize=8truecm \epsfbox{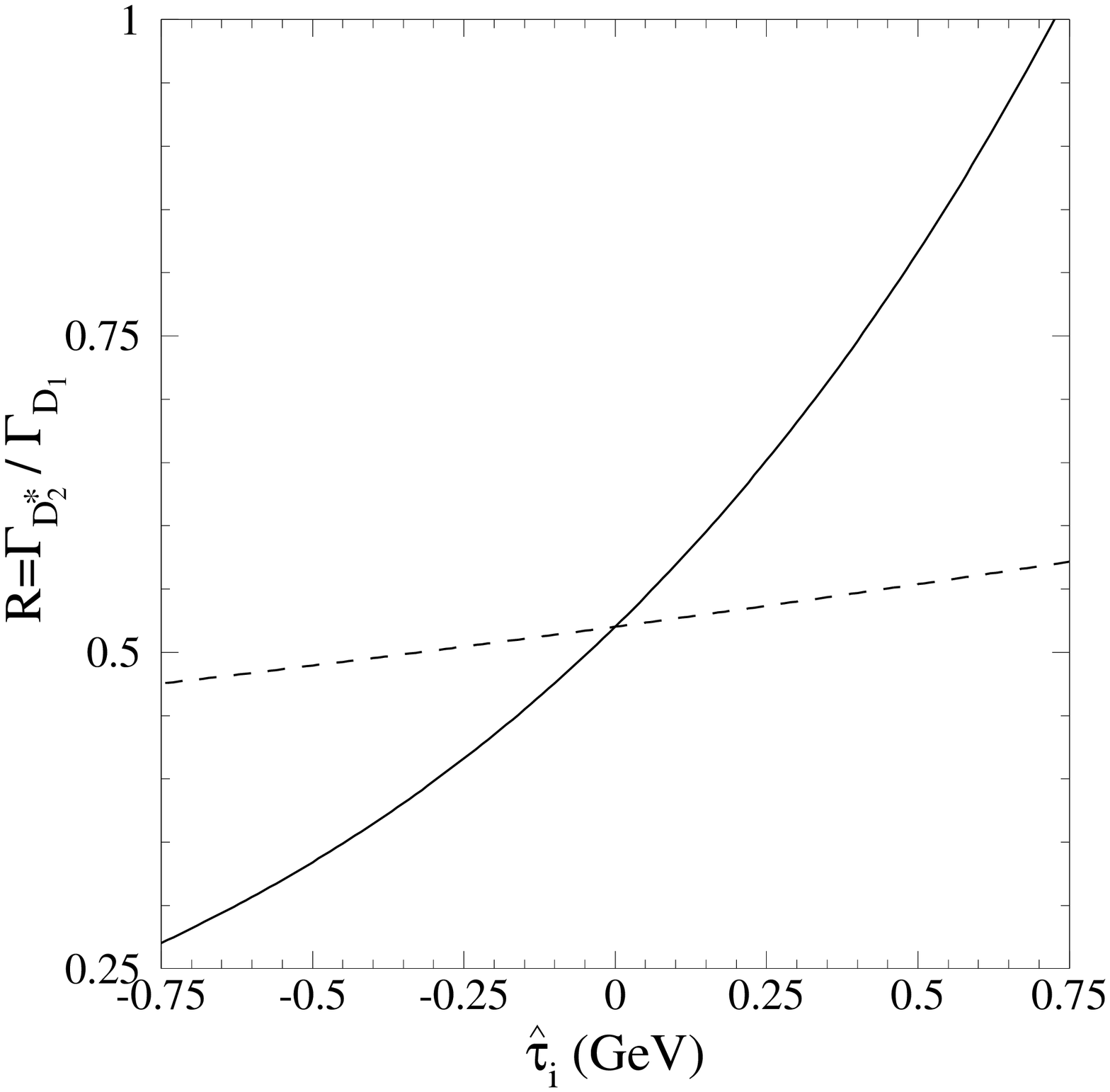}}
\tighten{
\caption[tau1]{Fig.~\ref{fig:Rtau}a shows
$R={\cal B}(B\to D_2^*\,e\,\bar\nu_e)/{\cal B}(B\to D_1\,e\,\bar\nu_e)$
as a function of $\hat\tau'$.  The dotted curve is the $m_Q\to\infty$ limit 
(B$_\infty$), solid curve is approximation~B$_1$, dashed curve is B$_2$.
Fig.~\ref{fig:Rtau}b shows $R$ as a function of $\hat\tau_1(=\tau_1/\tau)$ 
for $\hat\tau_2=0$ (solid curve), and as a function of $\hat\tau_2$ for
$\hat\tau_1=0$ (dashed curve).  Note that the scales in 
Fig.~\ref{fig:Rtau}a and \ref{fig:Rtau}b are different.} 
\label{fig:Rtau} }
\end{figure}

The order $\Lambda_{\rm QCD}/m_Q$ corrections are important for predicting
\begin{equation}
R \equiv {{\cal B}(B\to D_2^*\,e\,\bar\nu_e) \over 
  {\cal B}(B\to D_1\,e\,\bar\nu_e) } \,.
\end{equation}
In the $m_Q\to\infty$ limit $R\simeq1.65$ for $\hat\tau'=-1.5$ (this is the
B$_\infty$ result in Table~\ref{tab:sec2res}).  The sizable difference between
approximations~A and B is mainly due to the order $(w-1)^3$ contribution to the
helicity zero $D_1$ rate.  For $\hat\tau'=-1.5$ this term by itself would shift
the approximation~A result for $R$ from 0.40 to 0.49 and the A$_\infty$
prediction from 0.93 to 1.65.  The $\Lambda_{\rm QCD}/m_Q$ correction to the
form factors yield a large suppression of $R$ as shown in
Table~\ref{tab:sec2res} and Fig.~\ref{fig:Rtau}a.  Fig.~\ref{fig:Rtau}a also
shows that $R$ is fairly insensitive to $\hat\tau'$.  The difference of the
B$_1$ and B$_2$ results in Table~\ref{tab:sec2res} and Fig.~\ref{fig:Rtau}a
shows that $R$ is sensitive to the unknown $\Lambda_{\rm QCD}/m_Q$ corrections,
$\tau_1$ and $\tau_2$.  In Fig.~\ref{fig:Rtau}b we plot $R$ in approximation~B
as a function of $\hat\tau_1$ setting $\hat\tau_2=0$ (solid curve), and as a
function of $\hat\tau_2$ setting $\hat\tau_1=0$ (dashed curve). 
Fig.~\ref{fig:Rtau}b shows that $R$ is fairly insensitive to $\tau_2$, whereas
it depends sensitively on $\tau_1$.  In the range $-0.75\,{\rm
GeV}<\hat\tau_1<0.75\,{\rm GeV}$, $R$ goes over $0.27<R<1.03$.  This
suppression of $R$ compared to the infinite mass limit is supported by the
experimental data.  (It is possible that part of the reason for the strong
ALEPH bound ${\cal B}(B\to D_2^*\,e\,\bar\nu_e\,X)\times{\cal B}(D_2^*\to
D^{(*)}\pi)\lesssim(1.5-2.0)\times10^{-3}$ \cite{ALEPH} is a suppression of
${\cal B}(D_2^*\to D^{(*)}\pi)$ compared to ${\cal B}(D_1\to D^*\pi)$.) 

The prediction for the fraction of helicity zero $D_1$'s in semileptonic $B\to
D_1$ decay, $\Gamma_{D_1}^{(\lambda=0)}/\Gamma_{D_1}$, is surprisingly stable
in the different approximations (see Table~\ref{tab:sec2res}).  The weak
dependence of this ratio on $\hat\tau'$ is well described in approximation~B
for $|1.5+\hat\tau'|<1$ by adding $0.05(1.5+\hat\tau')$.  The dependence on
$\tau_1$ is at the 0.01 level, while the $\tau_2$-dependence is
$-0.07\,\hat\tau_2/{\rm GeV}$.  This is why the B$_2$ result for this quantity
is 0.05 larger than the B$_1$ prediction.  A linear dependence of $({\rm
d}\Gamma_{D_1}^{(\lambda=0)}/{\rm d}w)\Big/({\rm d}\Gamma_{D_1}/{\rm d}w)$ on
$w$ between $\lim_{w\to1}\Big[({\rm d}\Gamma_{D_1}^{(\lambda=0)}/{\rm
d}w)\Big/({\rm d}\Gamma_{D_1}/{\rm d}w)\Big]=1/3$ and $\Big[({\rm
d}\Gamma_{D_1}^{(\lambda=0)}/{\rm d}w)\Big/({\rm d}\Gamma_{D_1}/{\rm
d}w)\Big]=1$ at $w=w_{\rm max}$ is consistent with our result.

A similar prediction exists for the fraction of helicity zero $D_2^*$'s in
semileptonic $B\to D_2^*$ decay.  As can be seen from Table~\ref{tab:sec2res},
it is again quite stable.  The dependence on $\hat\tau'$ in approximation~B is
given by adding $0.04(1.5+\hat\tau')$.  However,
$\Gamma_{D_2^*}^{(\lambda=0)}/\Gamma_{D_2^*}$ is sensitive to both $\tau_1$ and
$\tau_2$ at the $(10-20)\%$ level, and the small difference between the B$_1$
and B$_2$ predictions for this quantity in Table~\ref{tab:sec2res} is due to an
accidental cancellation.  The prediction for the $w$ dependence of $({\rm
d}\Gamma_{D_2^*}^{(\lambda=0)}/{\rm d}w)\Big/({\rm d}\Gamma_{D_2^*}/{\rm d}w)$
between $\lim_{w\to1}\Big[({\rm 
d}\Gamma_{D_2^*}^{(\lambda=0)}/{\rm d}w)\Big/({\rm d
}\Gamma_{D_2^*}/{\rm d}w)\Big]=2/5$ and $\Big[({\rm 
d}\Gamma_{D_2^*}^{(\lambda=0)}/{\rm d}w)\Big/({\rm d}\Gamma_{D_2^*}/{\rm 
d}w)\Big]=1$ at $w=w_{\rm max}$ in this case is not linear.

\begin{figure}[t]
\centerline{\epsfysize=8truecm \epsfbox{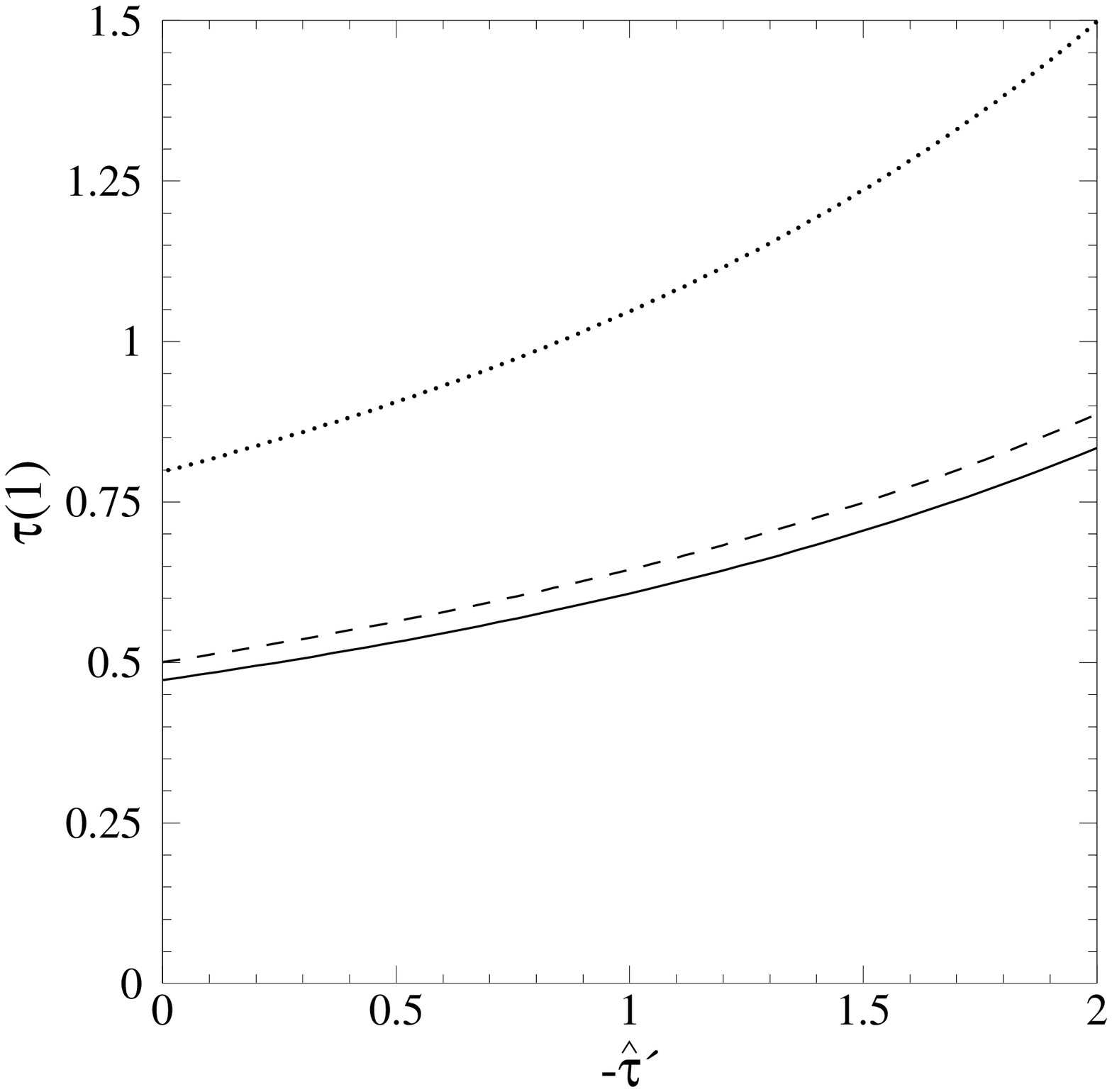}
  \epsfysize=8truecm \epsfbox{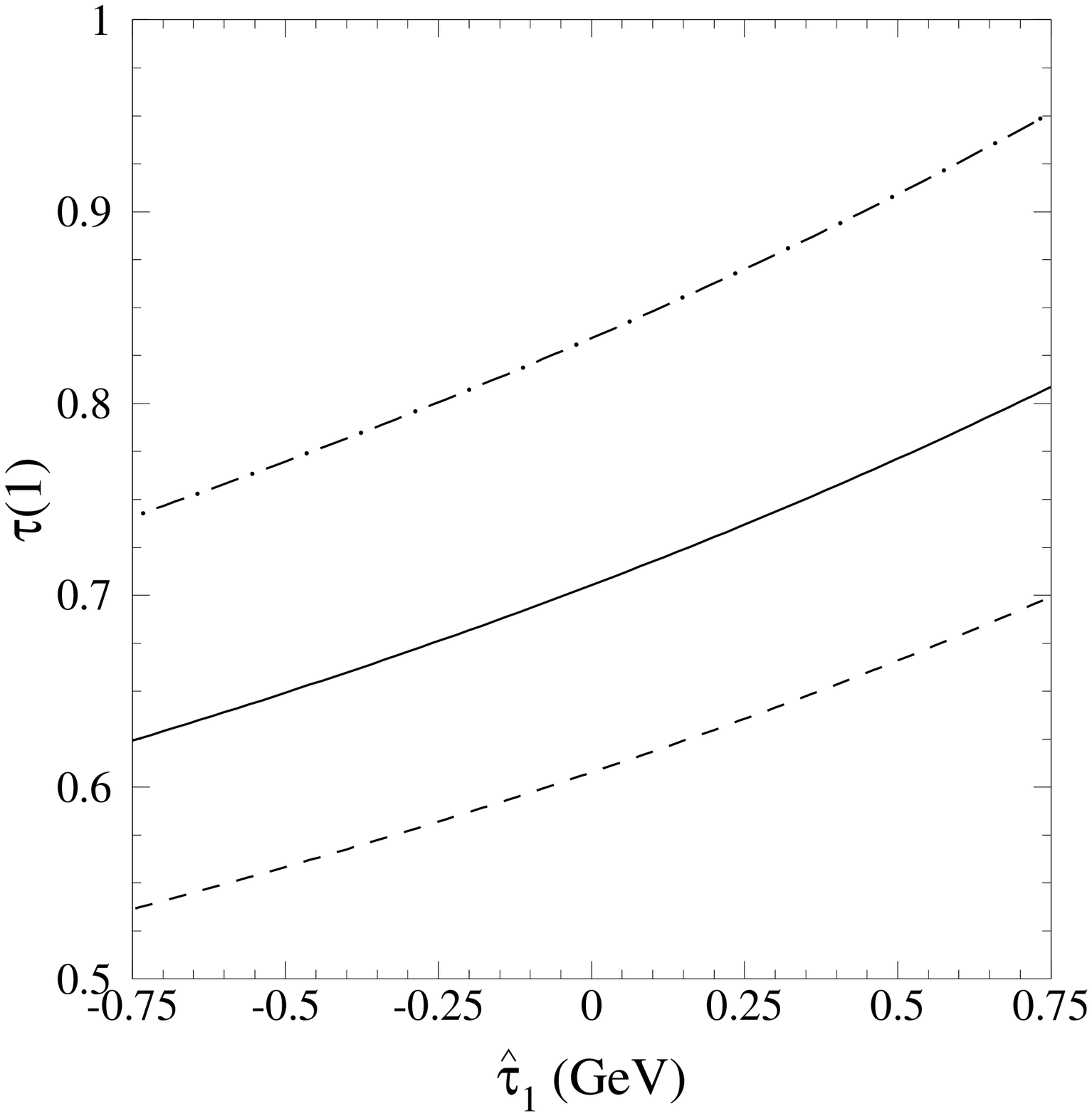}}
\tighten{
\caption[tau1]{Fig.~\ref{fig:tau1}a shows the extracted value of $\tau(1)$ 
as a function of $\hat\tau'$ in approximations~B$_\infty$, B$_1$, and B$_2$.  
The notation is the same as in Fig.~\ref{fig:Rtau}. 
Fig.~\ref{fig:tau1}b shows the dependence of $\tau(1)$ on $\hat\tau_1$ for 
$\hat\tau'=-1$ (dashed curve), $\hat\tau'=-1.5$ (solid curve), 
$\hat\tau'=-2$ (dash-dotted curve).} \label{fig:tau1} }
\end{figure}

The predictions considered so far do not depend on the value of $\tau(1)$, but
$\tau(1)$ affects some results that we discuss later.  $\tau(1)$ can be
determined from the measured $B\to D_1\,e\,\bar\nu_e$ branching ratio
using the expressions in Eq.~(\ref{rt_expn1}) and (\ref{stuD1}).  Using 
approximation~B$_1$ and $\hat\tau'=-1.5$, we obtain 
\begin{equation}\label{tau1}
\tau(1)\, \bigg[ {6.0\times10^{-3} \over {\cal B}(B\to D_1\,e\,\bar\nu_e)} 
  \bigg]^{1/2} = 0.71 \,.
\end{equation}
The extracted value of $\tau(1)$ is plotted in Fig.~\ref{fig:tau1}a in
approximations~B$_\infty$, B$_1$, and B$_2$ as functions of $\hat\tau'$.  The
suppression of $\tau(1)$ compared to the infinite mass limit indicates that the
order $\Lambda_{\rm QCD}/m_Q$ corrections enhance the semileptonic $B\to D_1$
width by about a factor of three.  In approximation~B the value of $\tau(1)$
changes by less than 0.01 as $\tau_2$ is varied in the range $-0.75\,{\rm
GeV}<\hat\tau_2<0.75\,{\rm GeV}$, but $\tau(1)$ is sensitive to $\tau_1$ at the
15\% level.  In Fig.~\ref{fig:tau1}b we plot $\tau(1)$ as a function of
$\hat\tau_1$ for $\hat\tau'=-1$ (dashed curve), $\hat\tau'=-1.5$ (solid curve),
and $\hat\tau'=-2$ (dash-dotted curve).  For $\tau_1>0$ (such as approximation
B$_2$) $\tau(1)$ is enhanced compared to the B$_1$ value of 0.71.  

The value of $\tau(1)$ in approximation~B is larger than that in
approximation~A.  Most of the difference arises from the inclusion of the order
$(w-1)^3$ term, $s_1^{(3)}$, which reduces the theoretical expression for the
helicity zero $B\to D_1\,e\,\bar\nu_e$ rate (for $\hat\tau'<-0.5$), resulting
in an increase in the value of $\tau(1)$ needed to accommodate the measured
rate.  For $\hat\tau'=-1.5$ this term by itself would shift the approximation~A
result from 0.60 to 0.66, and the A$_\infty$ prediction from 0.92 to 1.22.  The
ISGW nonrelativistic constituent quark model predicts $\tau(1)=0.54$, in rough
agreement with Eq.~(\ref{tau1}) \cite{ISGW,IWsr}.  (For some other quark model
predictions, see, e.g., Ref.~\cite{VeOl,More}.  QCD sum rules can also be used
to estimate $\tau$, see, e.g., Ref~\cite{Cola}.)

The ALEPH and CLEO analyses that yield Eq.~(\ref{data}) assume that $B\to
D_1\,e\,\bar\nu_e\,X$ is dominated by $B\to D_1\,e\,\bar\nu_e$, and that
$D_1$ decays only into $D^*\,\pi$.  If the first assumption turns out to be
false then $\tau(1)$ will decrease, if the second assumption is false then
$\tau(1)$ will increase compared to Eq.~(\ref{tau1}).  

The predictions discussed above would change if we had not absorbed into $\tau$
the time ordered product involving the kinetic energy operator.  As discussed
earlier (in the paragraph preceding the description of approximation~A), the
replacement of $\tau$ by $\widetilde\tau=\tau+\varepsilon_c\,\eta_{\rm
ke}^{(c)}+\varepsilon_b\,\eta_{\rm ke}^{(b)}$ introduces an error, which is
formally of order $\Lambda_{\rm QCD}^2/m_Q^2$.  Absorbing $\eta_{\rm ke}$ into
$\tau$ almost fully eliminates the $\eta_{\rm ke}$ dependence of the $D_2^*$
rate.  For the $D_1$ rate, however, absorbing $\eta_{\rm ke}$ into $\tau$
generates at order $\Lambda_{\rm QCD}^2/m_Q^2$ a formally suppressed but
numerically sizable $\eta_{\rm ke}$ dependence.  This $\eta_{\rm ke}$
dependence is more like a typical $\Lambda_{\rm QCD}/m_Q$ correction, since the
$\Lambda_{\rm QCD}/m_Q$ current corrections are as important as the infinite
mass limit for the $D_1$ rate.  Keeping $\hat\eta_{\rm ke}^{(Q)}=\eta_{\rm
ke}^{(Q)}/\tau$ explicit in the results, the total $B\to D_1$ semileptonic rate
in units of $\Gamma_0\,\tau^2(1)$ is
$0.033\,(1+1.1\,\varepsilon_c\,\hat\eta_{\rm ke}^{(c)}+\ldots)$, while the
$B\to D_2^*$ rate is $0.017\,(1+2.0\,\varepsilon_c\,\hat\eta_{\rm
ke}^{(c)}+\ldots)$.  From these expressions it is evident that, for
$-0.75\,{\rm GeV}<\hat\eta_{\rm ke}<0.75\,{\rm GeV}$, $\tau(1)$ changes only by
$\pm15\%$, while $R$ has a larger variation.  In the future this uncertainty
will be reduced if differential spectra can also be measured besides total
rates in $B\to D_1,D_2^*$ decays.  Note that $\eta_{\rm ke}$ does not enter
into predictions for the $B\to D_1\,e\,\bar\nu_e$ decay rate near zero recoil.  

Order $\alpha_s$ corrections to the results of this section can be calculated
in a straightforward way, using well-known methods.  Details of this
calculation are given in Appendix~A.  The order $\alpha_s$ corrections to the
results shown in Table~\ref{tab:sec2res} are given in Table~\ref{tab:alphas}. 
These are smaller than the uncertainty in our results from higher order terms
in the $\Lambda_{\rm QCD}/m_Q$ expansion that have been neglected.  The
corrections are most significant for $R=\Gamma_{D_2^*}\Big/\Gamma_{D_1}$ and
$\tau(1)$ in approximation~B; the central values of these quantities are
reduced by about $9\%$ and $4\%$, respectively.  Some of these $\alpha_s$
corrections depend sensitively on $\hat\tau'$, but they remain small for
$0>\hat\tau'>-2$.  For the remainder of this paper, we neglect the small
$\alpha_s$ corrections.

\begin{table}[t]
\begin{tabular}{c|cccc}  
Approximation  &  $\delta\Big(\Gamma_{D_2^*}\Big/\Gamma_{D_1}\Big)$  &  
  $\delta\Big(\Gamma_{D_1}^{(\lambda=0)}\Big/\Gamma_{D_1}\Big)$  &
  $\delta\Big(\Gamma_{D_2^*}^{(\lambda=0)}\Big/\Gamma_{D_2^*}\Big)$  &
  $\delta\tau(1)\, \bigg[\displaystyle {6.0\times10^{-3} \over 
    {\cal B}(B\to D_1\,e\,\bar\nu_e)} \bigg]^{1/2}$  \\[6pt] \hline 
A$_\infty$  &  $-0.68$  &  $0.10$  &  $0.02$  &  $-0.26$  \\
B$_\infty$  &  $-1.63$  &  $0.19$  &  $-0.003$  &  $-0.32$  \\ \hline
A  &  $-0.22$  &  $0.04$  &  $0.05$  &  $-0.24$  \\
B$_1$  &  $-0.55$  &  $0.06$  &  $-0.02$  &  $-0.32$  \\
B$_2$  &  $-0.68$  &  $0.07$  &  $-0.05$  &  $-0.33$  \\
\end{tabular} \vspace{6pt}
\tighten{
\caption[3]{Order $\alpha_s$ and $\alpha_s(\Lambda_{\rm QCD}/m_Q)$ corrections 
to the predictions in Table~\ref{tab:sec2res} for $\hat\tau'=-1.5$.  
These numbers should be multiplied by $\alpha_s(\sqrt{m_cm_b})/\pi$ to get 
the corrections to Table~\ref{tab:sec2res}.} \label{tab:alphas} }
\end{table}

Our predictions for the single differential $B\to(D_1,D_2^*)\,e\,\bar\nu_e$
spectra follow from Eqs.~(\ref{rt_expn1}), (\ref{stuD1}), and (\ref{stuD2}). 
${\rm d}\Gamma/{\rm d}w$ is given by integrating Eqs.~(\ref{rt_expn1}) over
${\rm d}\!\cos\theta$.  This amounts to the replacements $\sin^2\theta\to4/3$,
$(1+\cos^2\theta)\to8/3$, and $\cos\theta\to0$.  Thus ${\rm d}\Gamma/{\rm d}w$
is trivial to obtain using either approximations~A or B.  The electron energy
spectra are obtained by expressing $\cos\theta$ in terms of $y$ (where
$y=2E_e/m_B$ is the rescaled electron energy) using Eq.~(\ref{yct}), and
integrating $w$ over $[(1-y)^2+r^2]/[(2r(1-y)]<w<(1+r^2)/(2r)$.  They depend on
the coefficients $u_i^{(n)}$ which did not enter our results so far.  

\begin{figure}[pthb]
\centerline{\epsfysize=8truecm \epsfbox{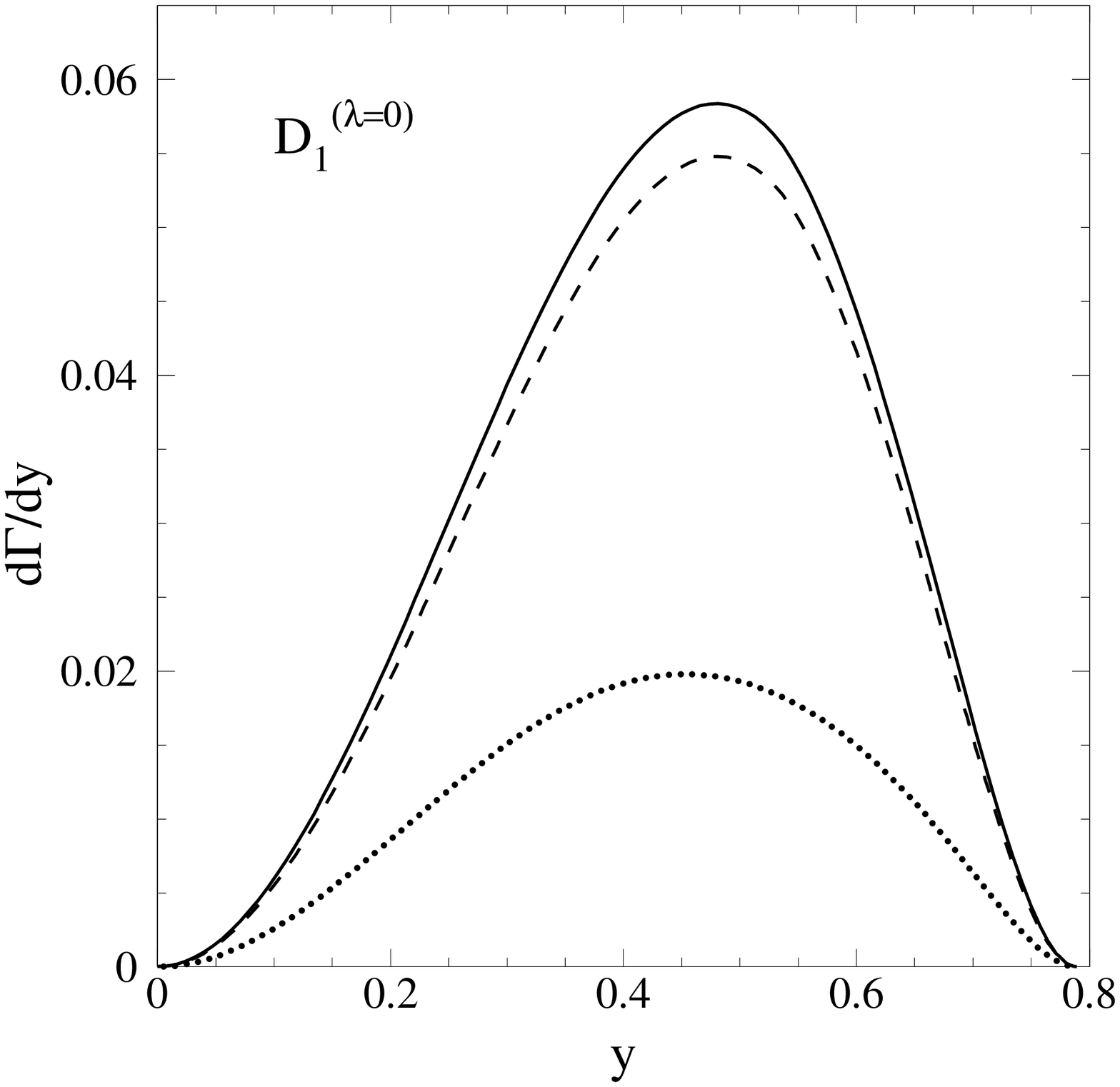}
  \epsfysize=8truecm \epsfbox{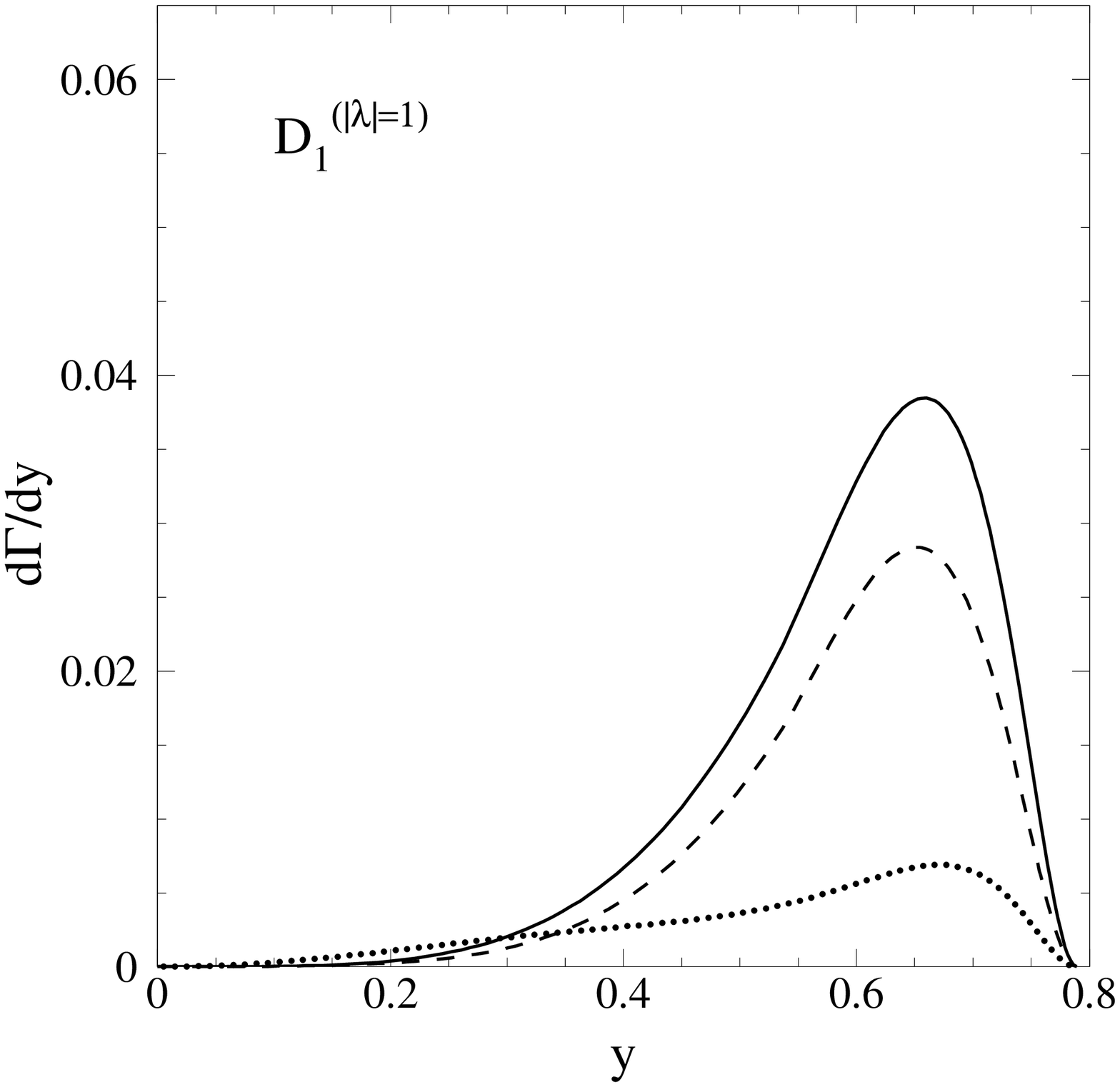}}
\tighten{
\caption[1]{Electron spectrum for $B\to D_1\,e\,\bar\nu_e$ in units of 
$\Gamma_0\,\tau^2(1)$ for $\hat\tau'=-1.5$.  Figs.~\ref{fig:D1spect}a and 
\ref{fig:D1spect}b are the spectra for helicity zero and helicity one $D_1$, 
respectively.  Dotted curves show the $m_Q\to\infty$ limit (B$_\infty$), 
solid curves are approximation~B$_1$, dashed curves are B$_2$.} 
\label{fig:D1spect} }
\vspace{1truecm}
\centerline{\epsfysize=8truecm \epsfbox{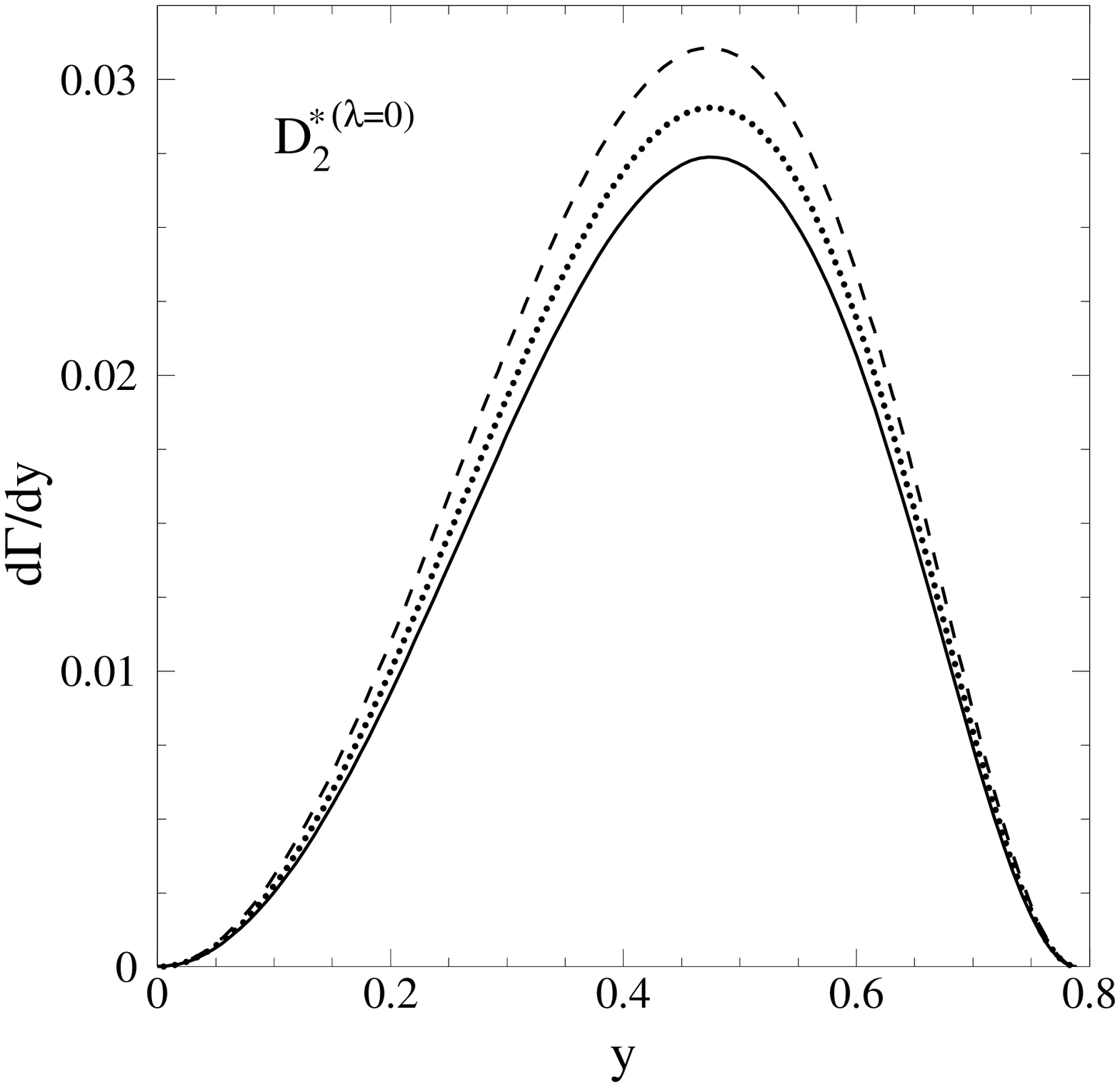}
  \epsfysize=8truecm \epsfbox{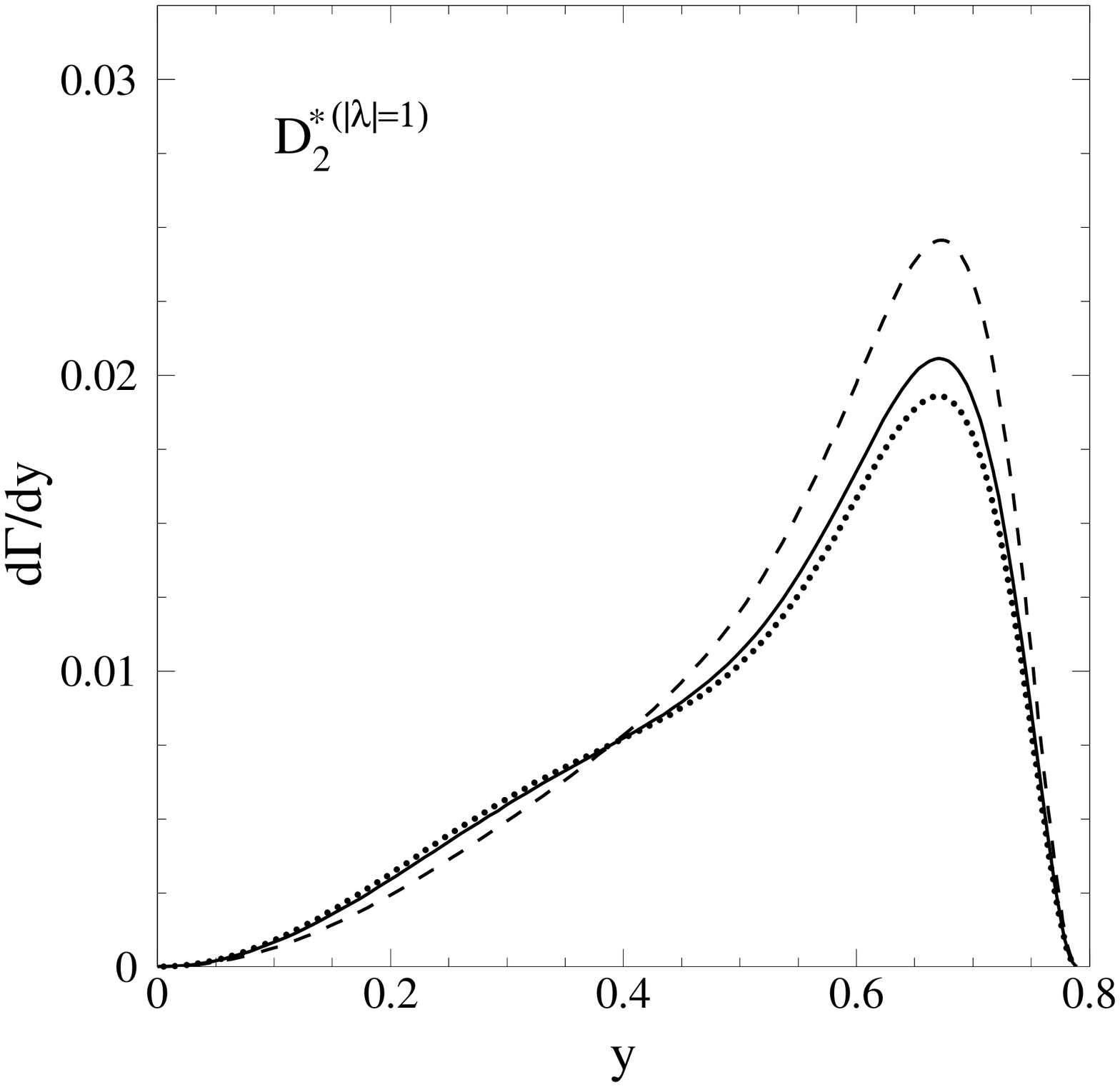}}
\tighten{
\caption[1]{Electron spectrum for $B\to D_2^*\,e\,\bar\nu_e$ in units of 
$\Gamma_0\,\tau^2(1)$ for $\hat\tau'=-1.5$.  The notations are the same as 
in Fig.~\ref{fig:D1spect}, but the scales are different.} 
\label{fig:D2spect} }
\end{figure}

In Fig.~\ref{fig:D1spect} the electron spectrum for $B\to D_1\,e\,\bar\nu_e$ is
plotted in units of $\Gamma_0\,\tau^2(1)$.  Figs.~\ref{fig:D1spect}a and
\ref{fig:D1spect}b are the spectra for helicity zero and helicity one $D_1$,
respectively.  In these plots $\hat\tau'=-1.5$.  The dotted curve shows the
$m_Q\to\infty$ limit (B$_\infty$), the solid curve is approximation~B$_1$, the
dashed curve is B$_2$.  Note that the kinematic range for $y$ is $0<y<1-r^2$. 
Near $y=0$ and $y=1-r^2$ the spectrum is dominated by contributions from $w$
near $w_{\rm max}$.  In this case, we expect sizable uncertainties in our
results, for example, from unknown terms that occur in the $u_i^{(n)}$ terms in
Eq.~(\ref{stuD1}) at a lower order than in the $s$ and $t$ coefficients. 
Fig.~\ref{fig:D1spect} shows the large enhancement of the $D_1$ rate due to
order $\Lambda_{\rm QCD}/m_Q$ corrections, and that the difference between
approximations~B$_1$ and B$_2$ is small compared to this enhancement.  In
Figs.~\ref{fig:D2spect}a and \ref{fig:D2spect}b we plot the electron spectrum
for $B\to D_2^*\,e\,\bar\nu_e$ for helicity zero and helicity one $D_2^*$,
respectively.  In this case the $\Lambda_{\rm QCD}/m_Q$ corrections are less
important.

\section{$B\to D_0^*\,\lowercase{e}\,\bar\nu_{\lowercase{e}}$ and 
  $B\to D_1^*\,\lowercase{e}\,\bar\nu_{\lowercase{e}}$ decays}

The other low lying states above the $D^{(*)}$ ground states occur in a doublet
with $s_l^{\pi_l}=\frac12^+$.  These states are expected to be broad since they
can decay into $D^{(*)}\,\pi$ in an $S$-wave, unlike the $D_1$ and $D_2^*$
which can only decay in a $D$-wave.  (An $S$-wave decay amplitude for the $D_1$
is forbidden by heavy quark spin symmetry \cite{IWprl}.)  This section repeats
the analysis of the previous section for these states.  Since the notation,
methods, and results are similar to those used in Sec.~II, the discussion here
will be briefer.

The matrix elements of the vector and axial currents between $B$ mesons and 
$D_0^*$ or $D_1^*$ mesons can be parameterized by
\begin{eqnarray}\label{formf2}
\langle D_0^*(v')|\, V^\mu\, |B(v)\rangle 
  &=& 0, \nn\\*
{\langle D_0^*(v')|\, A^\mu\, |B(v)\rangle \over\sqrt{m_{D_0^*}\,m_B}}
  &=& g_+\, (v^\mu+v'^\mu) + g_-\, (v^\mu-v'^\mu) \,, \nn\\*
{\langle D_1^*(v',\epsilon)|\, V^\mu\, |B(v)\rangle \over\sqrt{m_{D_1^*}\,m_B}}
  &=& g_{V_1}\, \epsilon^{*\, \mu} 
  + (g_{V_2} v^\mu + g_{V_3} v'^\mu)\, (\epsilon^*\cdot v) \,, \nn \\*
{\langle D_1^*(v',\epsilon)|\, A^\mu\, |B(v)\rangle \over\sqrt{m_{D_1^*}\,m_B}}
  &=& i\, g_A\, \varepsilon^{\mu\alpha\beta\gamma}\, 
  \epsilon^*_\alpha v_\beta\, v'_\gamma \,,  
\end{eqnarray}
where $g_i$ are functions of $w$.  At zero recoil the matrix elements are
determined by $g_+(1)$ and $g_{V_1}(1)$.  In terms of these form factors 
the double differential decay rates for $B\to D_0^*\,e\,\bar\nu_e$ and 
$B\to D_1^*\,e\,\bar\nu_e$ decays are
\begin{eqnarray}\label{rate2}
{{\rm d}^2\Gamma_{D_0^*}\over {\rm d}w\,{\rm d}\!\cos\theta} &=& 
  3\Gamma_0\, r_0^{*3}\, (w^2-1)^{3/2}\, \sin^2\theta\, 
  \Big[ (1+r_0^*)\,g_+ - (1-r_0^*)\, g_- \Big]^2 \,,\\*
{{\rm d}^2\Gamma_{D_1^*}\over {\rm d}w\,{\rm d}\!\cos\theta} &=& 
  3\Gamma_0\, r_1^{*3}\, \sqrt{w^2-1}\, \bigg\{ \sin^2\theta\,
  \Big[ (w-r_1^*) g_{V_1}+(w^2-1) (g_{V_3}+r_1^* g_{V_2}) \Big]^2 \nonumber\\*
&& + (1-2r_1^*w+r_1^{*2})\, \Big[(1+\cos^2\theta)\, [g_{V_1}^2 + (w^2-1) g_A^2] 
  - 4\cos\theta\, \sqrt{w^2-1}\, g_{V_1}\, g_A \Big] \bigg\} \,.\nonumber
\end{eqnarray}
where $\Gamma_0 = {G_F^2\,|V_{cb}|^2\,m_B^5 /(192\pi^3)}$, 
$r_0^*=m_{D_0^*}/m_B$ and $r_1^*=m_{D_1^*}/m_B$.

We follow the previous section to obtain expressions for the form factors $g_i$
in terms of Isgur-Wise functions to order $\Lambda_{\rm QCD}/m_Q$.  The fields
$P_v$ and $P_v^{*\mu}$ that destroy members of the $s_l^{\pi_l}=\frac12^+$
doublet with four-velocity $v$ are in the $4\times4$ matrix
\begin{equation}
K_v = \frac{1+\vslash}2\, \Big[ P_v^{*\mu}\, \gamma_5\gamma_\mu 
  + P_{v}\, \Big] \,.
\end{equation} 
This matrix $K$ satisfies $\vslash K_v=K_v=K_v\vslash$.  In the infinite mass 
limit matrix elements of the leading order current operator are \cite{IWsr}
\begin{equation}
\bar h^{(c)}_{v'}\, \Gamma\, h^{(b)}_v = \zeta(w)\,
  {\rm Tr}\, \Big\{ \bar K_{v'}\, \Gamma\, H_v \Big\} 
\end{equation}
Here $\zeta(w)$ is the leading order Isgur-Wise function ($\zeta$ is twice the
function $\tau_{1/2}$ of Ref.~\cite{IWsr}).  Since the $(D_0^*,D_1^*)$ states
are in a different spin multiplet than the ground state, $g_+(1)=g_{V_1}(1)=0$
in the infinite mass limit, independent of $\zeta(1)$.

The order $\Lambda_{\rm QCD}/m_Q$ corrections to the current can be 
parameterized as
\begin{eqnarray}\label{1/mc21}
\bar h^{(c)}_{v'}\, i\overleftarrow D_{\!\lambda}\, \Gamma\, h^{(b)}_v &=&
  {\rm Tr}\, \Big\{ {\cal S}^{(c)}_{\lambda}\, 
  \bar K_{v'}\, \Gamma\, H_v \Big\} \,, \nonumber\\*
\bar h^{(c)}_{v'}\, \Gamma\, i\overrightarrow D_{\!\lambda}\, h^{(b)}_v &=&
  {\rm Tr}\, \Big\{ {\cal S}^{(b)}_{\lambda}\, 
  \bar K_{v'}\, \Gamma\, H_v \Big\} \,.
\end{eqnarray}
This is the analogue of Eq.~(\ref{1/mc1}), except that in the present case
\begin{equation}\label{1/mc22}
{\cal S}^{(Q)}_{\lambda} = \zeta_1^{(Q)} v_\lambda + \zeta_2^{(Q)} 
  v'_\lambda + \zeta_3^{(Q)} \gamma_\lambda \,.
\end{equation}
The functions $\zeta_i^{(Q)}(w)$ have mass dimension one.  The heavy quark 
equation of motion yield
\begin{eqnarray}\label{constbr1}
w\, \zeta_1^{(c)} + \zeta_2^{(c)} + \zeta_3^{(c)} &=& 0 \,,\nn\\*
\zeta_1^{(b)} + w\, \zeta_2^{(b)} - \zeta_3^{(b)} &=& 0 \,.
\end{eqnarray}
Eq.~(\ref{momcons}) implies 
${\cal S}^{(c)}_\lambda+{\cal S}^{(b)}_\lambda = 
(\bar\Lambda v_\lambda-\bar\Lambda^*v'_\lambda)\, \zeta$, which gives three
more relations
\begin{eqnarray}\label{constbr2}
\zeta_1^{(c)} + \zeta_1^{(b)} &=& \bar\Lambda\, \zeta \,, \nonumber\\*
  \zeta_2^{(c)} + \zeta_2^{(b)} &=& -\bar\Lambda^*\, \zeta \,, \nonumber\\
  \zeta_3^{(c)} + \zeta_3^{(b)} &=& 0 \,.
\end{eqnarray}
These relations express the $\zeta_j^{(b)}$'s in terms of the 
$\zeta_j^{(c)}$'s.  Combining Eqs.~(\ref{constbr1}) with (\ref{constbr2}) 
yields
\begin{eqnarray}\label{constbr3}
\zeta_2^{(c)} &=& -{w\bar\Lambda^*-\bar\Lambda\over w+1}\, \zeta 
  - \zeta_1^{(c)} \,, \nonumber\\*
\zeta_3^{(c)} &=& {w\bar\Lambda^*-\bar\Lambda\over w+1}\, \zeta 
  - (w-1)\, \zeta_1^{(c)} \,.
\end{eqnarray}
At zero recoil, only $\zeta_3^{(Q)}$ can give a non-vanishing contribution to
the matrix elements of the weak currents in Eq.~(\ref{formf2}).  It is 
determined in terms of $\bar\Lambda^*-\bar\Lambda$ and $\zeta(1)$, since
Eqs.~(\ref{constbr2}) and (\ref{constbr3}) imply that
\begin{equation}
\zeta_3^{(c)}(1) = -\zeta_3^{(b)}(1) = 
  {\bar\Lambda^*-\bar\Lambda\over2}\, \zeta(1) \,.
\end{equation}
We use Eq.~(\ref{constbr3}) to eliminate $\zeta_2^{(c)}$ and 
$\zeta_3^{(c)}$ in favor of $\zeta_1^{(c)}$ and $\zeta$.

There are also order $\Lambda_{\rm QCD}/m_Q$ corrections to the effective
Lagrangian, given in Eq.~(\ref{lag}).  Time ordered products involving
$O_{\rm kin}$ can be parameterized as
\begin{eqnarray} 
i \int {\rm d}^4x\, T\,\Big\{ O_{{\rm kin},v'}^{(c)}(x)\, 
  \Big[ \bar h_{v'}^{(c)}\, \Gamma\, h_{v}^{(b)} \Big](0)\, \Big\} 
  &=& \chi_{\rm ke}^{(c)}\, {\rm Tr}\, 
  \Big\{ \bar K_{v'}\, \Gamma\, H_v \Big\} \,, \nn\\*
i \int {\rm d}^4x\, T\,\Big\{ O_{{\rm kin},v}^{(b)}(x)\, 
  \Big[ \bar h_{v'}^{(c)}\, \Gamma\, h_{v}^{(b)} \Big](0)\, \Big\} 
  &=& \chi_{\rm ke}^{(b)}\, {\rm Tr}\, 
  \Big\{ \bar K_{v'}\, \Gamma\, H_v \Big\} \,.
\end{eqnarray}
These corrections do not contribute at zero recoil.  The chromomagnetic 
corrections have the form
\begin{eqnarray}\label{timeo2}
i \int {\rm d}^4x\, T\,\Big\{ O_{{\rm mag},v'}^{(c)}(x)\, 
  \Big[ \bar h_{v'}^{(c)}\, \Gamma\, h_{v}^{(b)} \Big](0)\, \Big\} 
  &=& {\rm Tr}\, \bigg\{ {\cal R}_{\alpha\beta}^{(c)}\,
  \bar K_{v'}\, i\sigma^{\alpha\beta}\, \frac{1+\vslash'}2\, 
  \Gamma\, H_v \bigg\} \,, \nonumber\\*
i \int {\rm d}^4x\, T\,\Big\{ O_{{\rm mag},v}^{(b)}(x)\, 
  \Big[ \bar h_{v'}^{(c)}\, \Gamma\, h_{v}^{(b)} \Big](0)\, \Big\} 
  &=& {\rm Tr}\, \bigg\{ {\cal R}_{\alpha\beta}^{(b)}\,
  \bar K_{v'}\, \Gamma\, \frac{1+\vslash}2\, i\sigma^{\alpha\beta}
  H_v \bigg\} , 
\end{eqnarray} 
In this case the most general form of ${\cal R}_{\alpha\beta}^{(Q)}$ is
\begin{equation}\label{Rdef2}
{\cal R}_{\alpha\beta}^{(c)} = \chi_1^{(c)} \gamma_\alpha \gamma_\beta +
  \chi_2^{(c)} v_\alpha \gamma_\beta \,, \qquad 
{\cal R}_{\alpha\beta}^{(b)} = \chi_1^{(b)} \gamma_\alpha \gamma_\beta +
  \chi_2^{(b)} v'_\alpha \gamma_\beta \,. 
\end{equation} 
At zero recoil the contribution of $\chi_2^{(Q)}$ vanish because
$v_\alpha(1+\vslash)\sigma^{\alpha\beta}(1+\vslash)=0$, while that of
$\chi_1^{(Q)}$ vanish because
$(1-\vslash)\gamma_\alpha\gamma_\beta(1+\vslash)=(1-\vslash)(\gamma_\alpha
v_\beta-\gamma_\beta v_\alpha)(1+\vslash)$.

Using Eqs.~(\ref{1/mc21})--(\ref{timeo2}), it is straightforward to express the
form factors $g_i$ parameterizing $B\to D_0^*\,e\,\bar\nu_e$ and $B\to
D_1^*\,e\,\bar\nu_e$ semileptonic decays in terms of Isgur-Wise functions.  The
order $\Lambda_{\rm QCD}/m_b$ Lagrangian corrections arise only in the
combination $\chi_b=\chi_{\rm ke}^{(b)}+6\chi_1^{(b)}-2(w+1)\chi_2^{(b)}$. 
Dropping the $c$ superscript from $\zeta_1^{(c)}$ and $\chi_i^{(c)}$, we obtain
\begin{eqnarray}\label{expg0}
g_+ &=& \varepsilon_c\, \bigg[ 2(w-1)\zeta_1
  - 3\zeta\, {w\bar\Lambda^*-\bar\Lambda\over w+1} \bigg] 
  - \varepsilon_b\, \bigg[ {\bar\Lambda^*(2w+1)-\bar\Lambda(w+2)\over w+1}\,
  \zeta - 2(w-1)\,\zeta_1 \bigg] \,, \nonumber\\*
g_- &=& \zeta + \varepsilon_c\, \Big[ \chi_{\rm ke}+6\chi_1-2(w+1)\chi_2 \Big] 
  + \varepsilon_b\, \chi_b \,. 
\end{eqnarray}
The analogous formulae for $B\to D_1^*\,e\,\bar\nu_e$ are
\begin{eqnarray}\label{expg1}
g_A &=& \zeta 
  + \varepsilon_c\, \bigg[ {w\bar\Lambda^*-\bar\Lambda \over w+1} \zeta
  +\chi_{\rm ke}-2\chi_1 \bigg] - 
  \varepsilon_b\, \bigg[ {\bar\Lambda^*(2w+1)-\bar\Lambda(w+2)\over w+1}\,
  \zeta - 2(w-1)\,\zeta_1 - \chi_b \bigg] \,,\nonumber\\*
g_{V_1} &=&  (w-1)\,\zeta + \varepsilon_c\, 
  \Big[(w\bar\Lambda^*-\bar\Lambda)\zeta + (w-1)(\chi_{\rm ke}-2\chi_1) \Big] 
  \nonumber\\*
&& - \varepsilon_b\, \Big\{ [\bar\Lambda^*(2w+1)-\bar\Lambda(w+2)]\, \zeta 
  - 2(w^2-1)\,\zeta_1 - (w-1)\chi_b \Big\} \,, \nn\\*
g_{V_2} &=& 2\varepsilon_c\, (\zeta_1-\chi_2) \,, \nn\\*
g_{V_3} &=& - \zeta 
  - \varepsilon_c\, \bigg[ {w\bar\Lambda^*-\bar\Lambda \over w+1}\zeta 
  + 2\zeta_1 + \chi_{\rm ke} - 2\chi_1 +2\chi_2 \bigg] \nonumber\\*
&& + \varepsilon_b\, \bigg[ {\bar\Lambda^*(2w+1)-\bar\Lambda(w+2)\over w+1}\,
  \zeta - 2(w-1)\,\zeta_1 - \chi_b \bigg] \,. 
\end{eqnarray}
These equations show that at zero recoil the leading contributions to $g_{V_1}$
and $g_+$ of order $\Lambda_{\rm QCD}/m_Q$ are determined in terms of
$\bar\Lambda^*-\bar\Lambda$ and $\zeta(1)$.  Explicitly,
\begin{eqnarray}
g_+(1) &=& -\frac32\, (\varepsilon_c+\varepsilon_b)\, 
  (\bar\Lambda^*-\bar\Lambda)\, \zeta(1) \,, \nonumber\\*
g_{V_1}(1) &=& (\varepsilon_c-3\,\varepsilon_b)\, 
  (\bar\Lambda^*-\bar\Lambda)\, \zeta(1) \,.
\end{eqnarray}

For approximation~A we shall again expand the double differential decay rates 
in Eq.~(\ref{rate2}) in powers of $w-1$,
\begin{eqnarray}
{{\rm d}^2\Gamma_{D_0^*}\over {\rm d}w\,{\rm d}\!\cos\theta} &=& 
  3\Gamma_0\, \zeta^2(1)\, r_0^{*3}\, (w^2-1)^{3/2}\, \sin^2\theta\, 
  \sum_n\, (w-1)^n\, s_0^{(n)} \,, \\*
{{\rm d}^2\Gamma_{D_1^*}\over {\rm d}w\,{\rm d}\!\cos\theta} &=& 
  3\Gamma_0\, \zeta^2(1)\, r_1^{*3}\, \sqrt{w^2-1}\, \sum_n\, (w-1)^n\,
  \bigg\{ \sin^2\theta\, s_1^{(n)}\, \nonumber\\*
&& + (1-2r_1^*w+r_1^{*2})\, \Big[(1+\cos^2\theta)\, t_1^{(n)}
  - 4\cos\theta\, \sqrt{w^2-1}\, u_1^{(n)} \Big] \bigg\} \,.\nonumber
\end{eqnarray}
The coefficients for the decay rate into $D_0^*$ are
\begin{eqnarray}
s_0^{(0)} &=& (1-r_0)^{*2}\, \Big[ 1 + 
  2 \varepsilon_c (\hat\chi_{\rm ke} + 6\hat\chi_1 - 4\hat\chi_2) 
  + 4 \varepsilon_b \hat\chi_b \Big] +
  3(\varepsilon_c+\varepsilon_b)\, (1-r_0^{*2})\, (\bar\Lambda^*-\bar\Lambda) 
  + \ldots \,,\nn\\*
s_0^{(1)} &=& 2(1-r_0^*)^2\, \hat\zeta' + \ldots \,.
\end{eqnarray}
For the decay into $D_1^*$ the coefficients are
\begin{eqnarray}
s_1^{(0)} &=& (\varepsilon_c-3\varepsilon_b)^2\, (1-r_1^*)^2\,
  (\bar\Lambda^*-\bar\Lambda)^2 + \ldots \,, \nn\\*
s_1^{(1)} &=& - 2(\varepsilon_c-3\varepsilon_b)\, (1-r_1^{*2})\,
  (\bar\Lambda^*-\bar\Lambda) + \ldots \,, \nn\\*
s_1^{(2)} &=& (1+r_1^*)^2 + \ldots  \,,\nn\\
t_1^{(0)} &=& (\varepsilon_c-3\varepsilon_b)^2\, 
  (\bar\Lambda^*-\bar\Lambda)^2 + \ldots \,, \nn\\*
t_1^{(1)} &=& 2 + 4(\varepsilon_c-3\varepsilon_b)\, (\bar\Lambda^*-\bar\Lambda)
  + 4\varepsilon_c (\hat\chi_{\rm ke}-2\hat\chi_1) 
  + 4\varepsilon_b \hat\chi_b + \ldots \,, \nn\\*
t_1^{(2)} &=& 2(1 + 2\hat\zeta') + \ldots \,,\nn\\
u_1^{(0)} &=& (\varepsilon_c-3\varepsilon_b)\, (\bar\Lambda^*-\bar\Lambda) 
  + \ldots \,, \nonumber \\*
u_1^{(1)} &=& 1 + \ldots \,.  
\end{eqnarray}
Note that at zero recoil and at order $w-1$ the contributions to $D_1^*$ decay
proportional to $\bar\Lambda^*-\bar\Lambda$ depend on the anomalously small
combination $\varepsilon_c-3\varepsilon_b\sim0.05{\rm GeV}^{-1}$.  Thus
$\Lambda_{\rm QCD}/m_Q$ corrections enhance $B\to D_1^*$ by a much smaller
amount than they enhance $B\to D_1$ decay.  On the other hand, the $B\to D_0^*$
decay rate receives a large enhancement from $\Lambda_{\rm QCD}/m_Q$
corrections, similar to $B\to D_1$.

In approximation~A, $B\to D_1^*$ is treated the same way as $B\to D_1$ in
Sec.~II.  $B\to D_0^*$ is treated as $B\to D_2^*$ in Sec.~II, since these rates
contain an additional factor of $w^2-1$.  Approximation~B is also very similar
to that in Sec.~II, except that in the present case there is only one unknown
$\Lambda_{\rm QCD}/m_Q$ Isgur-Wise function, $\zeta_1$ (once time ordered
products involving the chromomagnetic operator are neglected, and the matrix
elements of the time ordered products involving the kinetic energy operator are
absorbed into the $m_Q\to\infty$ Isgur-Wise function, $\zeta$).  In
approximation~B$_1$ we set $\zeta_1=0$ in Eqs.~(\ref{expg0}) and (\ref{expg1}).
This is identical to saturating the first relation in Eq.~(\ref{constbr2}) by
$\zeta_1^{(b)}$, i.e., setting $\zeta_1^{(b)}=\bar\Lambda\,\zeta$.  In
approximation B$_2$ we set $\zeta_1=\bar\Lambda\,\zeta$ in Eqs.~(\ref{expg0})
and (\ref{expg1}), which is identical to setting $\zeta_1^{(b)}=0$.  To the
extent the first relation in Eq.~(\ref{constbr2}) can be taken as a hint to the
sign of $\zeta_1$, the difference between approximations~B$_1$ and B$_2$ gives
a crude estimate of the uncertainty related to the unknown $\Lambda_{\rm
QCD}/m_Q$ corrections.  

As in the previous section, the expression for the decay rate in terms of form
factors in Eq.~(\ref{rate2}) implies that $s_1^{(0)}/t_1^{(0)}=(1-r_1^*)^2$ to
all orders in the $\Lambda_{\rm QCD}/m_Q$ expansion.  Thus the ratio of
helicity zero and helicity one $B\to D_1^*$ decay rates at zero recoil is
$\lim_{w\to1}\Big[({\rm d}\Gamma_{D_1^*}^{(\lambda=0)}/{\rm d}w)\Big/({\rm
d}\Gamma_{D_1^*}^{(|\lambda|=1)}/{\rm d}w)\Big]=1/2$.

\subsection{Predictions}

A model independent prediction similar to that in Sec.~II can be made for the
slope parameter of semileptonic $B$ decay into the helicity zero $D_1^*$.  We
write the semileptonic decay rate into the helicity zero $D_1^*$ as
\begin{equation}\label{mir2}
{{\rm d}\Gamma_{D_1^*}^{(\lambda=0)} \over {\rm d}w} = 4 \Gamma_0\, 
  r_1^{*3}\, (1-r_1^*)^2\, \sqrt{w^2-1}\, \zeta^2(1)\, 
  (\varepsilon_c-3\varepsilon_b)^2\, (\bar\Lambda^*-\bar\Lambda)^2\,
  \Big[ 1 - \rho_{D_1^*}^2\, (w-1) + \ldots \Big] \,.
\end{equation}
The relationship between $s_1^{(0)}$ and $s_1^{(1)}$ implies that the slope
parameter $\rho_{D_1^*}^2$ for helicity zero $D_1^*$ is 
\begin{equation}
\rho_{D_1^*}^2 = {1+r_1^*\over1-r_1^*}\, 
  {2\over(\varepsilon_c-3\varepsilon_b)\, (\bar\Lambda^*-\bar\Lambda)} 
  + {\cal O}(1) \,.
\end{equation}
As in Sec.~II, this slope parameter is of order $m_Q/\Lambda_{\rm QCD}$.  It
would be very hard experimentally to test this model independent prediction,
since the $D_1^*$ is expected to be of order $100\,$MeV broad, and also because 
$\varepsilon_c-3\varepsilon_b$ is so small.

Predictions for the $B\to D_0^*\,e\,\bar\nu_e$ and $B\to D_1^*\,e\,\bar\nu_e$
rates are shown in the first two columns of Table~\ref{tab:sec3res}, normalized
to $\zeta^2(1)$ times the measured $B\to D_1\,e\,\bar\nu_e$ rate.  These
results are obtained using $\hat\zeta'=-1$, and
$\bar\Lambda^*-\bar\Lambda\simeq0.35\,$GeV corresponding to $1<w<1.33$.  This
value of $\bar\Lambda^*-\bar\Lambda$ has at least a $50\,$MeV uncertainty at
present, as it follows from model predictions for the masses of the
$s_\ell^{\pi_\ell}=\frac12^+$ charmed mesons,
$\overline{m}_D^{\,*}\simeq2.40\,$GeV \cite{GI}, and from the fact that
$\lambda_1^*=\lambda_1'$ in nonrelativistic quark models with spin-orbit
independent potentials.  Although the $D_1^*$ state is expected to be somewhat
heavier than the $D_0^*$, we use the kinematic range $1<w<1.33$ for both
decays.  The results in the first two columns of Table~\ref{tab:sec3res} are
quite sensitive to the value of $\hat\zeta'$ and $\zeta_1$.  In
approximation~B$_1$, for example, ${\cal B}(B\to
D_0^*\,e\,\bar\nu_e)/[\zeta^2(1)\times0.006]$ changes from 1.92 at
$\hat\zeta'=0$ to 0.54 at $\hat\zeta'=-2$.  In the same range of $\hat\zeta'$,
${\cal B}(B\to D_1^*\,e\,\bar\nu_e)/[\zeta^2(1)\times0.006]$ changes from 0.72
to 0.24.  The effect of $\zeta_1$ is also important; in the range $-0.75\,{\rm
GeV}<\hat\zeta_1<0.75\,{\rm GeV}$, the $D_0^*$ and $D_1^*$ branching ratios
change from 1.68 to 0.66 and 0.30 to 0.63, respectively.  Therefore, even if
$\zeta$ were known from models or lattice calculations, there would still be a
factor of two uncertainty in the theoretical predictions for the semileptonic
$B\to D_0^*$ and $D_1^*$ rates; but the uncertainty in the sum of these two
rates is smaller.

\begin{table}[t]
\begin{tabular}{c|ccc} 
Approximation  &  ${\displaystyle{\cal B}(B\to D_0^*\,e\,\bar\nu_e)\over
  \displaystyle\zeta^2(1)\times0.006}$
  &  ${\displaystyle{\cal B}(B\to D_1^*\,e\,\bar\nu_e)\over
  \displaystyle\zeta^2(1)\times0.006}$
  &  $\Gamma_{D_0^*+D_1^*}\Big/\Gamma_{D_1}$  \\[6pt] \hline 
A$_\infty$  &  $0.30$  &  $0.66$  &  $1.07$  \\
B$_\infty$  &  $0.33$  &  $0.46$  &  $1.61$  \\ \hline
A  &  $1.03$  &  $0.65$  &  $0.80$  \\
B$_1$  &  $1.11$  &  $0.44$  &  $1.03$  \\
B$_2$  &  $0.85$  &  $0.53$  &  $1.05$  \\
\end{tabular} \vspace{6pt}
\tighten{
\caption[4]{The first two columns show semileptonic $B$ branching ratios into 
$D_0^*$ and $D_1^*$ normalized to $\zeta^2(1)$ times the measured
branching ratio ${\cal B}(B\to D_1\,e\,\bar\nu_e)=0.6\%$, assuming 
$\hat\zeta'=\zeta'(1)/\zeta(1)=-1$.  The sum of $D_0^*+D_1^*$ rates relative 
to $B\to D_1$ is in the third column, using the nonrelativistic constituent
quark model prediction in Eq.~(\ref{ztau}) and $\hat\tau'=-1.5$.} 
\label{tab:sec3res} }
\end{table}

To obtain even a crude absolute prediction for the $B\to D_1^*,D_0^*$ rates, a
relation between the $s_\ell^{\pi_\ell}=\frac12^+$ and $\frac32^+$ Isgur-Wise
functions is needed.  In any nonrelativistic constituent quark model with
spin-orbit independent potential, $\zeta$ and $\tau$ are related by
\cite{VeOl,IWsr}
\begin{equation}\label{ztau}
  \zeta(w) = {w+1\over\sqrt3}\, \tau(w) \,,
\end{equation}
since both of these spin symmetry doublets correspond to $L=1$ orbital
excitations.  This implies
\begin{equation}\label{zetarel}
\zeta(1) = {2\over\sqrt3}\, \tau(1) \,, \qquad 
  \hat\zeta' = \frac12 + \hat\tau' \,. 
\end{equation}
In the same approximation, $\hat\eta_{\rm ke}=\hat\chi_{\rm ke}$.\footnote{A
relation between $\tau_{1,2}$ and $\zeta_1$ may also hold in this model.}

Predictions for the $B$ semileptonic decay rate into the states in the
$s_\ell^{\pi_\ell}=\frac12^+$ doublet that follow from Eq.~(\ref{zetarel}) are
shown in the last column of Table~\ref{tab:sec3res}.  (For this quantity,
approximations~B$_i$ ($i=1,2$) contain a somewhat ad hoc input of combining the
B$_i$ prediction in Sec.~II with the B$_i$ prediction for $B\to D_0^*,D_1^*$.) 
For $\hat\tau'=-1.5$, the $\frac12^+$ doublet contributes about $1.0\times{\cal
B}(B\to D_1\,e\,\bar\nu_e)\sim0.6\%$ to the total $B$ decay rate.  Varying
$\tau_{1,2}$ and $\zeta_1$ in approximation~B results in the range
$(0.6-1.7)\times{\cal B}(B\to D_1\,e\,\bar\nu_e)$ for the sum of the $D_0^*$
and $D_1^*$ rates.  This combined with our results for
$R=\Gamma_{D_2^*}/\Gamma_{D_1}$ in Sec.~II is consistent with the ALEPH
measurement \cite{ALEPH} of the branching ratio for the sum of all semileptonic
decays containing a $D^{(*)}\,\pi$ in the final state to be $(2.26\pm0.44)\%$.  

The semileptonic decay rate into $D$ and $D^*$ is about $6.6\%$ of the total
$B$ decay rate \cite{PDG}.  Our results then suggest that the six lightest
charmed mesons contribute about 8.2\% of the $B$ decay rate.  Therefore,
semileptonic decays into higher excited states and non-resonant multi-body
channels should be at least 2\% of the $B$ decay rate, and probably around 3\%
if the semileptonic $B$ branching ratio is closer to the LEP result of about
11.5\%.  Such a sizable contribution to the semileptonic rate from higher mass
excited charmed mesons and non-resonant modes would soften the lepton spectrum,
and may make the agreement with data on the inclusive lepton spectrum worse. 
Of course, the decay rates to the broad $\frac12^+$ states would change
substantially if the nonrelativistic quark model prediction in Eq.~(\ref{ztau})
is wrong.  Semileptonic $B$ decay rate to the six lightest charmed mesons could
add up to close to $10\%$ if $\zeta$ were enhanced by a factor of two compared
to the prediction of Eq.~(\ref{ztau}).  However, model calculations \cite{More}
seem to obtain a suppression rather than an enhancement of $\zeta$ compared to
Eq.~(\ref{ztau}).  Thus, taking the measurements for the $B\to D$, $D^*$, and
$D_1$ semileptonic branching ratios on face value, a decomposition of the
semileptonic rate as a sum of exclusive channels seems problematic both in
light of our results and the above ALEPH measurement for the sum of all
semileptonic decays containing a $D^{(*)}\,\pi$ in the final state.

\section{Other excited charmed mesons at zero recoil}

In the previous two sections matrix elements of the weak vector current and
axial-vector current between a $B$ meson and an excited charmed mesons with
$s_\ell^{\pi_\ell}=\frac32^+$ and $\frac12^+$ quantum numbers were considered. 
Here we consider such matrix elements at zero recoil for excited charmed mesons
with other $s_\ell^{\pi_\ell}$ quantum numbers.  Only charmed mesons with spin
zero or spin one can contribute at this kinematic point.  The polarization
tensor of a spin $n$ state is rank $n$, traceless and symmetric in its indices,
and vanishes if it is contracted with the 4-velocity of the state.  For matrix
elements of the axial-vector or vector current, at least $n-1$ indices of the
charmed meson polarization tensor are contracted with $v^\mu$, the four
velocity of the $B$ meson.  Consequently, for $n>1$ these matrix elements
vanish at zero recoil, where $v=v'$.  In this section we work in the rest
frame, $v=v'=(1,\vec0\,)$, and four-velocity labels on the fields and states
are suppressed.

For spin zero and spin one excited charmed mesons, the possible spin parities
for the light degrees of freedom are $s_\ell^{\pi_\ell}=\frac12^+$,
$\frac32^+$, which we have already considered in the previous sections, and
$s_\ell^{\pi_\ell}=\frac12^-$, $\frac32^-$.  In the nonrelativistic constituent
quark model, the $\frac12^-$ states are interpreted as radial excitations of
the ground state $(D,D^*)$ doublet and the $\frac32^-$ states are $L=2$ orbital
excitations.  In the quark model, these states are typically expected to be
broad.  The mass of the lightest $s_l^{\pi_l}=\frac32^-$ doublet is expected
around $2.8\,$GeV, while the lightest excited states with
$s_l^{\pi_l}=\frac12^-$ are expected around $2.6\,$GeV \cite{GI}.\footnote{The
lightest $\frac12^-$ states may be narrow since decays to the
$s_\ell^{\pi_\ell}=\frac12^-$ and $\frac32^-$ multiplets are suppressed by the
available phase space, and decays to $D^{(*)}\,\pi$ in an $S$-wave are
forbidden by parity.}  ($B$ decays into radial excitations of the
$s_l^{\pi_l}\neq\frac12^-$ states have similar properties as the decay into the
lightest state with the same quantum numbers.)

In the $m_Q\to\infty$ limit, the zero recoil matrix elements vanish by heavy
quark symmetry.  For the excited $s_\ell^{\pi_\ell}=\frac12^-$ states, the
$m_Q\to\infty$ Isgur-Wise functions vanish at zero recoil due to the
orthogonality of the states.  The matrix elements for the
$s_\ell^{\pi_\ell}\neq\frac12^-$ states vanish at zero recoil due to spin
symmetry alone, and therefore the corresponding $m_Q\to\infty$ Isgur-Wise
functions need not vanish at zero recoil.  

Using the same methods as in Sections II and III, it is straightforward to show
that $\Lambda_{\rm QCD}/m_Q$ corrections to the current do not contribute at
zero recoil.  For the $s_\ell^{\pi_\ell}=\frac12^-$ states, this follows from
the heavy quark equation of motion.  For the $s_\ell^{\pi_\ell}=\frac32^-$
states, the $\Lambda_{\rm QCD}/m_Q$ corrections to the current can be
parameterized similar to Eqs.~(\ref{1/mc1}) and (\ref{1/mc2}).  In this case
the analogue of $F_v^\mu$ in Eq.~(\ref{Fdef}) satisfies $\vslash
F_v^\mu=F_v^\mu=F_v^\mu\,\vslash$.  Recall that the
$\tau_4^{(Q)}g_{\sigma\lambda}$ in Eq.~(\ref{1/mc2}) was the only term whose
contribution at zero recoil did not vanish due to the $v_\mu F_v^\mu=0$
property of the Rarita-Schwinger spinors.  Here, the analogous term is placed
between $1-\vslash$ and $1+\vslash'$, and therefore also disappears at $v=v'$.

It remains to consider the $\Lambda_{\rm QCD}/m_Q$ contributions to the
$\frac12^-$ and $\frac32^-$ matrix elements coming from corrections to the
Lagrangian in Eq.~(\ref{lag}).  These are written as time ordered products of
$O_{\rm kin}^{(Q)}(x)$ and $O_{\rm mag}^{(Q)}(x)$ with the leading order
$m_Q\to\infty$ currents (e.g., Eq~(\ref{kinetic})).  At zero recoil it is
useful to insert a complete set of states between these operators.  Since the
zero recoil weak currents are charge densities of heavy quark spin-flavor
symmetry, only one state from this sum contributes.  For the
$s_\ell^{\pi_\ell}=\frac12^-$ multiplet this procedure gives
\begin{eqnarray}\label{css1}
\frac{\langle D^{*(n)}(\varepsilon)|\, \vec A\, |B\rangle}
  {\sqrt{m_{D^{*(n)}}\,m_B}} &=& 
  \frac{-\vec\varepsilon}{(\bar\Lambda^{(n)}-\bar\Lambda)}\,
  \Bigg\{ \left(\frac1{2m_c} + \frac3{2m_b}\right)
  \frac{\langle D^{*(n)}(\varepsilon)|\, O_{\rm mag}^{(c)}(0)\, 
  |D^*(\varepsilon)\rangle}{\sqrt{m_{D^{*(n)}}\,m_{D^*}}} \nonumber\\*
&& \phantom{\frac{-\vec\varepsilon}{(\bar\Lambda^{(n)}-\bar\Lambda)}}
  + \left(\frac1{2m_c} - \frac1{2m_b}\right)
  \frac{\langle D^{*(n)}(\varepsilon)|\, O_{\rm kin}^{(c)}(0)\, 
  |D^*(\varepsilon)\rangle}{\sqrt{m_{D^{*(n)}}\,m_{D^*}}} \Bigg\} \,.
\end{eqnarray}
and 
\begin{equation}\label{css2}
\frac{\langle D^{(n)}|\, V^0\, |B\rangle}
  {\sqrt{m_{D^{(n)}}\,m_B}} = \frac{1}{(\bar\Lambda^{(n)}-\bar\Lambda)}
  \left(-\frac1{2m_c} + \frac1{2m_b}\right)
  \frac{\langle D^{(n)}|\, O_{\rm mag}^{(c)}(0)+O_{\rm kin}^{(c)}(0)\,
  |D\rangle}{\sqrt{m_{D^{(n)}}\,m_D}} \,.
\end{equation}
Here we have denoted spin zero and spin one members of the excited
$s_\ell^{\pi_\ell}=\frac12^-$ multiplet by $D^{(n)}$ and $D^{*(n)}$
respectively, and the analogues of $\bar\Lambda$ by $\bar\Lambda^{(n)}$.  Heavy
quark spin-flavor symmetry was used to write the effects of $O_{\rm kin}^{(b)}$
and $O_{\rm mag}^{(b)}$ in terms of matrix elements of $O_{\rm kin}^{(c)}$ and
$O_{\rm mag}^{(c)}$.  This neglects the weak logarithmic dependence on the
heavy quark mass in the matrix elements of $O_{\rm mag}$.
For the spin one member of the $s_\ell^{\pi_\ell} = \frac32^-$
multiplet, which we denote by $D_1^{**}$,
\begin{equation}\label{css3}
\frac{\langle D_1^{**}(\varepsilon)|\, \vec A\, |B\rangle}
  {\sqrt{m_{D_1^{**}}\,m_B}} = 
  \frac{-\vec\varepsilon}{(\bar\Lambda^{**}-\bar\Lambda)}
  \left(\frac1{2m_c}\right) \frac{\langle D_1^{**}(\varepsilon)|\, 
  O_{\rm mag}^{(c)}(0)\, |D^*(\varepsilon)\rangle}{\sqrt{m_{D_1^{**}}\,m_D}}\,.
\end{equation}

For the $s_\ell^{\pi_\ell}=\frac12^-$ and $\frac32^-$ excited charmed mesons,
the correction to the Lagrangian, $\delta{\cal L}$ in Eq.~(\ref{lag}), 
gives rise to an order $\Lambda_{\rm QCD}/m_c$ contribution to the matrix
elements of the weak currents at zero recoil.  Formulae similar to those in
Eqs.~(\ref{css1})--(\ref{css3}) hold in the $s_\ell^{\pi_\ell}=\frac12^+$,
$\frac32^+$ cases, but the corresponding matrix elements vanish due to the
parity invariance of the strong interaction.

\section{Applications}

\subsection{Factorization}

Factorization should be a good approximation for $B$ decay into charmed mesons
and a charged pion.  Contributions that violate factorization are suppressed by
$\Lambda_{\rm QCD}$ divided by the energy of the pion in the $B$ rest frame
\cite{GrDu} or by $\alpha_s(m_Q)$.  Furthermore for these decays, factorization
also holds in the limit of large number of colors.  Neglecting the pion mass,
the two-body decay rate, $\Gamma_\pi$, is related to the differential decay
rate ${\rm d}\Gamma_{\rm sl}/{\rm d}w$ at maximal recoil for the analogous
semileptonic decay (with the $\pi$ replaced by the $e\,\bar\nu_e$ pair).  This
relation is independent of the identity of the charmed meson in the final
state,
\begin{equation}\label{factor}
\Gamma_\pi = {3\pi^2\, |V_{ud}|^2\, C^2\, f_\pi^2 \over m_B^2\, r} \times
  \left( {{\rm d} \Gamma_{\rm sl}\over {\rm d}w} \right)_{w_{\rm max}} .
\end{equation}
Here $r$ is the mass of the charmed meson divided by $m_B$, $w_{\rm
max}=(1+r^2)/(2r)$, and $f_\pi\simeq132\,$MeV is the pion decay constant.  
$C$ is a combination of Wilson coefficients of four-quark operators, and
numerically $C\,|V_{ud}|$ is very close to unity.

These nonleptonic decay rates can therefore be predicted from a measurement of
${\rm d}\Gamma_{\rm sl}/{\rm d}w$ at maximal recoil.  The semileptonic decay
rate near maximal recoil is only measured for $B\to D^{(*)}\,e\,\bar\nu_e$ at
present.  The measured $B\to D^{(*)}\,\pi$ rate is consistent with
Eq.~(\ref{factor}) at the level of the 10\% experimental uncertainties.  In the
absence of a measurement of the $B\to(D_1,D_2^*)\,e\,\bar\nu_e$ differential
decay rates, we can use our results for the shape of ${\rm d}\Gamma_{\rm
sl}/{\rm d}w$ to predict the $B\to D_1\,\pi$ and $B\to D_2^*\,\pi$ decay rates.
These predictions depend on the semileptonic differential decay rates at
$w_{\rm max}$, where we are the least confident that $\Lambda_{\rm QCD}/m_Q$
terms involving $\bar\Lambda$ and $\bar\Lambda'$ are the most important.  With
this caveat in mind, we find the results shown in Table~\ref{tab:factor}.

\begin{table}[t] 
\begin{tabular}{c|ccc} 
Approximation  
  &  ${\displaystyle{\cal B}(B\to D_1\,\pi)\over 
  \displaystyle{\cal B}(B\to D_1\,e\,\bar\nu_e)}$
  &  ${\displaystyle{\cal B}(B\to D_2^*\,\pi)\over
  \displaystyle{\cal B}(B\to D_1\,\pi)}$  \\[6pt] \hline 
A$_\infty$  &  $0.39$  &  $0.36$  \\
B$_\infty$  &  $0.26$  &  $1.00$  \\ \hline
A  &  $0.29$  &  $0.21$  \\
B$_1$  &  $0.19$  &  $0.41$  \\
B$_2$  &  $0.20$  &  $0.56$  \\
\end{tabular} \vspace{6pt}
\tighten{
\caption[5]{Predictions for the ratios of branching ratios,
${\cal B}(B\to D_1\,\pi)/{\cal B}(B\to D_1\,e\,\bar\nu_e)$ and 
${\cal B}(B\to D_2^*\,\pi)/{\cal B}(B\to D_1\,\pi)$, using factorization and
assuming $\hat\tau'=\tau'(1)/\tau(1)=-1.5$.} \label{tab:factor} }
\end{table}

At present there are only crude measurements of the ${\cal B}(B\to D_1\,\pi)$
and ${\cal B}(B\to D_2^*\,\pi)$ branching ratios.  Assuming ${\cal
B}(D_1(2420)^0\to D^{*+}\,\pi^-)=2/3$ and ${\cal B}(D_2^*(2460)^0\to
D^{*+}\,\pi^-)=0.2$, the measured rates are \cite{CLEOfact}
\begin{eqnarray}\label{factordata}
{\cal B}(B^-\to D_1(2420)^0\,\pi^-) &=& (1.17\pm0.29) \times 10^{-3} \,, 
  \nonumber\\*
{\cal B}(B^-\to D_2^*(2460)^0\,\pi^-) &=& (2.1\pm0.9) \times 10^{-3} \,.
\end{eqnarray}
A reduction of the experimental uncertainty in ${\cal B}(B\to D_2^*\,\pi)$ is
needed to test the prediction in the second column of Table~\ref{tab:factor}.

\begin{figure}[t]
\centerline{\epsfysize=9truecm \epsfbox{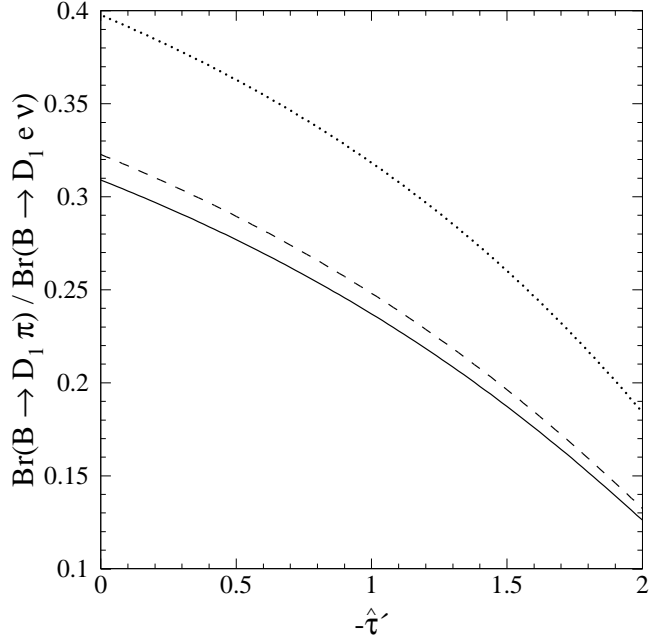}}
\tighten{
\caption[tau1]{Factorization prediction for 
${\cal B}(B\to D_1\,\pi)/{\cal B}(B\to D_1\,e\,\bar\nu_e)$ as a function of
$\hat\tau'=\tau'(1)/\tau(1)$.  The dotted curve shows the $m_Q\to\infty$ limit 
(B$_\infty$), solid curve is approximation~B$_1$, dashed curve is B$_2$.} 
\label{fig:factor1} }
\end{figure}

The prediction for ${\cal B}(B\to D_1\,\pi)/{\cal B}(B\to D_1\,e\,\bar\nu_e)$
in approximation~B is fairly independent of $\tau_{1,2}$, but more sensitive to
$\hat\tau'$.  The latter dependence is plotted in Fig.~\ref{fig:factor1} for
$0>\hat\tau'>-2$.  Not absorbing $\eta_{\rm ke}$ into $\tau$ results in the
following weak dependence: ${\cal B}(B\to D_1\,\pi)/{\cal B}(B\to
D_1\,e\,\bar\nu_e)\propto1+0.27\,\varepsilon_c\,\hat\eta_{\rm ke}+\ldots$. 
Assuming that factorization works at the 10\% level, a precise measurement of
the ${\cal B}(B\to D_1\,\pi)$ rate may provide a determination of $\hat\tau'$. 
The present experimental data, ${\cal B}(B\to D_1\,\pi)/{\cal B}(B\to
D_1\,e\,\bar\nu_e)\simeq0.2$, does in fact support $\hat\tau'\sim-1.5$, which
we took as the ``central value" in this paper, motivated by model calculations.

\begin{figure}[t]
\centerline{\epsfysize=9truecm \epsfbox{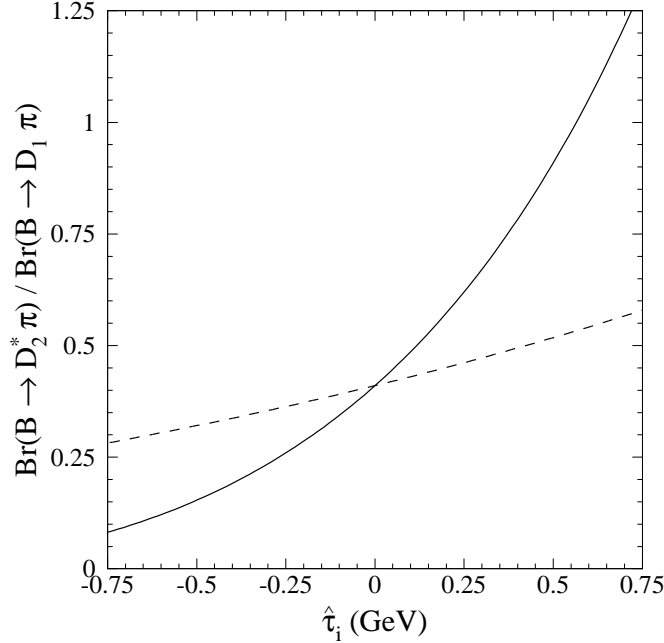}}
\tighten{
\caption[tau1]{Factorization prediction for 
${\cal B}(B\to D_2^*\,\pi)/{\cal B}(B\to D_1\,\pi)$ as a function of 
$\hat\tau_1(=\tau_1/\tau)$ for $\hat\tau_2=0$ (solid curve), and as a 
function of $\hat\tau_2$ for $\hat\tau_1=0$ (dashed curve).} 
\label{fig:factor2} }
\end{figure}

The prediction for ${\cal B}(B\to D_2^*\,\pi)/{\cal B}(B\to D_1\,\pi)$, on the
other hand, only weakly depends on $\hat\tau'$, but it is more sensitive to
$\tau_{1,2}$.  Varying $\tau_{1,2}$ in the range $-0.75\,{\rm
GeV}<\hat\tau_{1,2}<0.75\,{\rm GeV}$, we can accommodate almost any value of
${\cal B}(B\to D_2^*\,\pi)/{\cal B}(B\to D_1\,\pi)$ between 0 and 1.5.  This
quantity depends more sensitively on $\tau_1$ than on $\tau_2$.
In Fig.~\ref{fig:factor2} we plot ${\cal B}(B\to D_2^*\,\pi)/{\cal B}(B\to
D_1\,\pi)$ in approximation~B as a function of $\hat\tau_1$
setting $\hat\tau_2=0$ (solid curve), and as a function of $\hat\tau_2$ setting
$\hat\tau_1=0$ (dashed curve).  Not absorbing $\eta_{\rm ke}$ into $\tau$
results in the following dependence: ${\cal B}(B\to D_2^*\,\pi)/{\cal B}(B\to
D_1\,\pi)\propto1+0.75\,\varepsilon_c\,\hat\eta_{\rm ke}+\ldots$.  This ratio
and $R$ depend on $\hat\eta_{\rm ke}$ and $\hat\tau_1$.  In the future
experimental data on these ratios may lead to a determination of 
$\hat\eta_{\rm ke}$ and $\hat\tau_1$.

If the experimental central value on ${\cal B}(B\to D_2^*\,\pi)$ does not
decrease compared to Eq.~(\ref{factordata}), then it would suggest a huge value
for $\hat\tau_1$, leading to a violation of the ALEPH bound on
$R$ (see Fig.~\ref{fig:Rtau}).  The approximation B results in
Tables~\ref{tab:sec2res} and \ref{tab:factor} can be combined to give
${\cal B}(B\to D_2^*\,\pi)/{\cal B}(B\to D_2^*\,e\,\bar\nu_e) = 0.15$.
Varying $\hat\tau_i$, $\hat\eta_{\rm ke}$ and $\hat\tau'$ does not
bring this quantity close to the current experimental limit.
Therefore, if the branching ratio for $B\to D_2^*\,e\,\bar\nu_e$ is
below the ALEPH bound, then ${\cal B}(B\to D_2^*\,\pi)$ should be
smaller than the central value in Eq.~(\ref{factordata}).

\subsection{Sum Rules}

Our results are important for sum rules that relate inclusive $B\to
X_c\,e\,\bar\nu_e$ decays to the sum of exclusive channels.  The Bjorken sum
rule bounds the slope of the $B\to D^{(*)}\,e\,\bar\nu_e$ Isgur-Wise function,
defined by the expansion $\xi(w)=1-\rho^2\,(w-1)+\ldots\,$.  Knowing $\rho^2$
would reduce the uncertainty in the determination of $|V_{cb}|$ from the
extrapolation of the $B\to D^{(*)}\,e\,\bar\nu_e$ spectrum to zero recoil.  
The Bjorken sum rule \cite{Bjsr,IWsr} is
\begin{equation}\label{bjorken}
\rho^2 = \frac14 + \sum_m\, {|\zeta^{(m)}(1)|^2 \over 4}
  + 2 \sum_p\, {|\tau^{(p)}(1)|^2 \over 3} + \ldots \,.
\end{equation}
Throughout this section the ellipses denote contributions from non-resonant
channels.  $\zeta^{(m)}$ and $\tau^{(p)}$ are the Isgur-Wise functions for the
exited $s_\ell^{\pi_\ell}=\frac12^+$ and $\frac32^+$ states, respectively (for
$m=p=0$ these are the orbitally excited states discussed in Sec.~II and III,
and $m,p\geq1$ are radial excitations of these).\footnote{In Ref.~\cite{IWsr}
$|\zeta^{(m)}(1)|^2/4$ was denoted by $|\tau_{1/2}^{(m)}(1)|^2$, and
$|\tau^{(p)}(1)|^2/3$ was denoted by $|\tau_{3/2}^{(p)}(1)|^2$.}  Since all
terms in the sums, as well as the contributions replaced by ellipses, are
non-negative, a lower bound on $\rho^2$ can be obtained by keeping only the
first few terms on the right-hand-side of Eq.~(\ref{bjorken}).  Using
Eqs.~(\ref{tau1}) and (\ref{ztau}), we find that the contribution of the
lowest lying $s_\ell^{\pi_\ell}=\frac12^+$ and $\frac32^+$ states implies the
bound
\begin{equation}\label{bjnew}
\rho^2 > \frac14 + {|\zeta(1)|^2\over4} + 2\, {|\tau(1)|^2\over3} 
  \simeq 0.75 \,.
\end{equation}
The contribution of the $\frac12^+$ states through $\zeta(1)$ to this bound,
which relies on the quark model result in Eq.~(\ref{ztau}), is only 0.17.

An upper bound on $\rho^2$ follows from an upper bound on the excited states
contribution to the right-hand-side of Eq.~(\ref{bjorken}).  This sum rule was
first derived by Voloshin \cite{Volo}
\begin{equation}\label{voloshin}
\frac12\, \bar\Lambda = \sum_m\, (\bar\Lambda^{*\,(m)}-\bar\Lambda)\, 
  {|\zeta^{(m)}(1)|^2\over4} + 2 \sum_p\, (\bar\Lambda'^{\,(p)}-\bar\Lambda)\, 
  {|\tau^{(p)}(1)|^2\over3} + \ldots \,.
\end{equation}

Here $\bar\Lambda^{*\,(m)}$ and $\bar\Lambda'^{\,(p)}$ are the analogues of
$\bar\Lambda^*$ and $\bar\Lambda'$ for the exited $s_\ell^{\pi_\ell}=\frac12^+$
and $\frac32^+$ states, respectively.  Eq.~(\ref{voloshin}) combined with
Eq.~(\ref{bjorken}) implies that $\rho^2<1/4+\bar\Lambda/(2\varepsilon_1)$,
where $\varepsilon_1$ is the excitation energy of the lightest excited charmed
meson state.  However, knowing $\zeta(1)$ and $\tau(1)$ does not strengthen
this bound on $\rho^2$ significantly.  On the other hand, Eq.~(\ref{voloshin})
implies the bound $\bar\Lambda>0.38\,$GeV (neglecting perturbative QCD
corrections).  The model dependent contribution of the $\frac12^+$ states to
this bound is only $0.12\,$GeV; while the bound $\bar\Lambda>0.26\,{\rm GeV}$
from only the $\frac32^+$ states is fairly model independent.

A class of zero recoil sum rules were considered in Ref.~\cite{VVsr}.  The
axial sum rule, which bounds the $B\to D^*$ form factor (that is used to
determine $|V_{cb}|$) only receives contributions from
$s_\ell^{\pi_\ell}=\frac12^-$ and $\frac32^-$ states, which were discussed in
Sec.~IV.  It has the form
\begin{equation}\label{aasr}
|F_{B\to D^*}(1)|^2 + \sum_{X_c} 
  {|\langle X_c(\varepsilon)|\, \vec A\, |B\rangle|^2 \over 4m_{X_c}\,m_B }
  = \eta_A^2 - {\lambda_2\over m_c^2} + {\lambda_1+3\lambda_2\over4}\, 
  \bigg( {1\over m_c^2} + {1\over m_b^2} + {2\over 3m_c\,m_b} \bigg) \,,
\end{equation}
where $\eta_A$ is the perturbative matching coefficient of the full QCD
axial-vector current onto the HQET current, $X_c$ denotes spin one states 
(continuum or resonant) with $s_\ell^{\pi_\ell}=\frac12^-$ and $\frac32^-$, 
and $F_{B\to D^*}(1)$ is defined by
\begin{equation}
{\langle D^*(\varepsilon)|\, \vec A\, |B\rangle\over 2\sqrt{m_{D^*}\,m_B} }
  = F_{B\to D^*}(1)\, \vec\varepsilon \,.
\end{equation}
Neglecting the contributions of the excited states $X_c$ to the left-hand-side,
gives an upper bound on $|F_{B\to D^*}(1)|^2$.  Using the nonrelativistic
constituent quark model, we estimate using Eq.~(\ref{css1}) that the
contribution of the first radial excitation of the $D^*$ to the sum over $X_c$
in Eq.~(\ref{aasr}) is about 0.1.  For this estimate we took
$\bar\Lambda^{(1)}-\bar\Lambda=450\,$MeV, $O_{\rm
mag}^{(c)}=C\,\delta^3(r)\,\vec s_c\cdot\vec s_{\bar q}$ (fixing the constant
$C$ by the measured $D^*-D$ mass splitting), $O_{\rm kin}^{(c)}=\vec\nabla^2$,
and used the harmonic oscillator quark model wave functions of
Ref.~\cite{ISGW}.  A 0.1 correction would significantly strengthen the upper
bound on $F_{B\to D^*}(1)$ and have important consequences for the extraction
of the magnitude of $V_{cb}$ from exclusive $B\to D^*e\bar\nu_e$ decay.  Note
that $s_\ell^{\pi_\ell}=\frac32^-$ states do not contribute to the zero recoil
axial sum rule in the quark model, because their spatial wave functions vanish
at the origin.  

The $J^P=1^+$ members of the $s_\ell^{\pi_\ell}=\frac12^+$ and
$s_\ell^{\pi_\ell}=\frac32^+$ doublets contribute to the vector sum rule, which
is used to bound $\lambda_1$.  This sum rule reads \cite{VVsr,llsw}
\begin{eqnarray}\label{vvsr}
&& {(m_b-3m_c)^2 \over 4m_b^2\,m_c^2}\, 
  \sum_m\, (\bar\Lambda^{*\,(m)}-\bar\Lambda)^2\, {|\zeta^{(m)}(1)|^2\over4}
  + {2\over m_c^2}\, \sum_p\, (\bar\Lambda'^{\,(p)}-\bar\Lambda)^2\, 
  {|\tau^{(p)}(1)|^2\over3} + \ldots \nonumber\\*
&& \qquad\qquad = {\lambda_2\over m_c^2} - {\lambda_1+3\lambda_2\over4}\, 
  \bigg( {1\over m_c^2} + {1\over m_b^2} - {2\over3m_c\,m_b} \bigg) \,.
\end{eqnarray}
This relation can be simplified by setting $m_b/m_c$ to different values.
Taking $m_b=m_c$ yields
\begin{equation}\label{l1sr}
\lambda_1 = -3\, \sum_m\, (\bar\Lambda^{*\,(m)}-\bar\Lambda)^2\,
  {|\zeta^{(m)}(1)|^2\over4} - 6 \sum_p\, 
  (\bar\Lambda'^{\,(p)}-\bar\Lambda)^2\, {|\tau^{(p)}(1)|^2\over3} + \ldots \,,
\end{equation}
whereas $m_c\gg m_b\gg\Lambda_{\rm QCD}$ gives \cite{llsw}
\begin{equation}\label{vvnew}
\lambda_1 + 3\lambda_2 = - 9 \sum_m\, (\bar\Lambda^{*\,(m)}-\bar\Lambda)^2\, 
  {|\zeta^{(m)}(1)|^2\over4} + \ldots \,.
\end{equation}
These relations can be combined to obtain a sum rule for $\lambda_2$,
\begin{equation}\label{l2sr}
\lambda_2 = -2\, \sum_m\, (\bar\Lambda^{*\,(m)}-\bar\Lambda)^2\,
  {|\zeta^{(m)}(1)|^2\over4} + 2 \sum_p\, 
  (\bar\Lambda'^{\,(p)}-\bar\Lambda)^2\, {|\tau^{(p)}(1)|^2\over3} + \ldots \,.
\end{equation}
Eqs.~(\ref{l1sr}) and (\ref{l2sr}) were previously obtained in
Ref.~\cite{srreview} using different methods.  The strongest constraint on
$\lambda_1$ is given by Eq.~(\ref{vvnew}) (the sum rule in Eq.~(\ref{l1sr})
only implies $-\lambda_1>(0.06+0.15)\,{\rm GeV}^2$).  Including the
contribution of the lightest $s_\ell^{\pi_\ell}=\frac12^+$ doublet to
Eq.~(\ref{vvnew}) yields
\begin{equation}\label{l1new}
\lambda_1 < - 3\lambda_2 
  - 9\, (\bar\Lambda^*-\bar\Lambda)^2\, {|\zeta(1)|^2\over4}
  \simeq - 3\lambda_2 - 0.18\,{\rm GeV}^2 \,,
\end{equation}
neglecting perturbative QCD corrections.  Note that only the broad $D_1^*$
state (and its radial excitations) contribute to this sum rule, so the result
in Eq~(\ref{l1new}) is sensitive to the relation between $\tau(1)$ and
$\zeta(1)$ in Eq.~(\ref{ztau}).

Perturbative corrections to the sum rules in this section can be found in 
Ref.~\cite{pert}.

\section{Summary and conclusions}

The branching ratios for $B\to D\,e\,\bar\nu_e$ and $B\to D^*\,e\,\bar\nu_e$
are $(1.8\pm0.4)\%$ and $(4.6\pm0.3)\%$, respectively \cite{PDG}.  This implies
that about $40\%$ of semileptonic $B$ decays are to excited charmed mesons and
non-resonant final states.  An excited charmed meson doublet
$(D_1(2420),D_2^*(2460))$ with $s_\ell^{\pi_\ell}={3\over2}^+$ has been
observed.  These states are narrow and have widths around $20\,$MeV.  With some
assumptions, the CLEO and ALEPH collaborations have measured about a
$(0.6\pm0.1)\%$ branching ratio for $B\to D_1\,e\,\bar\nu_e$.  The decay $B\to
D_2^*\,e\,\bar\nu_e$ has not been observed, and CLEO and ALEPH respectively
report limits of $1\%$ and $0.2\%$ on its branching ratio.  A detailed
experimental study of semileptonic $B$ decays to these states should be
possible in the future.

The semileptonic $B$ decay rate to an excited charmed meson is determined by
the corresponding matrix elements of the weak axial-vector and vector currents.
At zero recoil (where the final excited charmed meson is at rest in the rest
frame of the initial $B$ meson), these currents correspond to charges of the
heavy quark spin-flavor symmetry.  Consequently, in the $m_Q\to\infty$ limit,
the zero recoil matrix elements of the weak currents between a $B$ meson and
any excited charmed meson vanish.  However, at order $\Lambda_{\rm QCD}/m_Q$
these matrix elements are not necessarily zero.  Since for $B$ semileptonic
decay to excited charmed mesons most of the available phase space is near zero
recoil, the $\Lambda_{\rm QCD}/m_Q$ corrections can play a very important role.
In this paper we studied the predictions of HQET for the $B\to
D_1\,e\,\bar\nu_e$ and $B\to D_2\,e\,\bar\nu_e$ differential decay rates
including the effects of $\Lambda_{\rm QCD}/m_Q$ corrections to the matrix
elements of the weak currents.  Since the matrix elements of the weak currents
between a $B$ meson and any excited charmed meson can only be nonzero for spin
zero or spin one charmed mesons at zero recoil, the $\Lambda_{\rm QCD}/m_Q$
corrections are more important for the spin one member of the
$s_\ell^{\pi_\ell}={3\over2}^+$ doublet.

The $\Lambda_{\rm QCD}/m_Q$ corrections to the matrix elements of the weak
axial-vector and vector currents can be divided into two classes; corrections
to the currents themselves and corrections to the states.  For $B$ semileptonic
decays to the $D_1$, parity invariance of the strong interactions forces the
corrections to the states to vanish at zero recoil.  Furthermore, the
corrections to the current give a contribution which at zero recoil is
expressible in terms of the leading, $m_Q\to\infty$, Isgur-Wise function and
known meson mass splittings.  This correction leads to an enhancement of the
$B$ semileptonic decay rate to the $D_1$ over that to the $D_2$.  With some
model dependent assumptions, we made predictions for the differential decay
rates for $B\to D_1\,e\,\bar\nu_e$ and $B\to D_2^*\,e\,\bar\nu_e$ and
determined the zero recoil value of the leading $m_Q \to\infty$ Isgur-Wise
function from the measured $B$ to $D_1$ semileptonic decay rate.  The influence
of perturbative QCD corrections on these decay rates were also considered but
these are quite small.

Factorization was used to predict the rates for the nonleptonic decays $B\to
D_1\,\pi$ and $B\to D_2^*\,\pi$.  The ALEPH limit on the semileptonic decay
rate to $D_2^*$ implies a small branching ratio for $B\to D_2^*\,\pi$.  The
ratio ${\cal B}(B\to D_1\,\pi)/{\cal B}(B\to D_1\,e\,\bar\nu_e)$ can be used to
determine $\hat\tau'$.  The present experimental value for this quantity
favors $\hat\tau'$ near $-1.5$.

The most significant uncertainty at order $\Lambda_{\rm QCD}/m_Q$ arises from
$\hat\tau_1$ and $\hat\eta_{\rm ke}$.  It may be possible to determine these
quantities from measurements of $R=\Gamma_{D_2^*}/\Gamma_{D_1}$ and ${\cal
B}(B\to D_2^*\,\pi)/{\cal B}(B\to D_1\,\pi)$.  The $w$-dependence of the
semileptonic decay rates can provide important similar information.  

A broad multiplet of excited charmed mesons with masses near those of the $D_1$
and $D_2^*$ is expected.  It has spin of the light degrees of freedom
$s_\ell^{\pi_\ell}={1\over2}^+$, giving spin zero and spin one states that are
usually denoted by $D_0^*$ and $D_1^*$.  We studied the predictions of HQET for
the $B\to D_0^*\,e\,\bar\nu_e$ and $B\to D_1^*\,e\,\bar\nu_e$ differential
decay rates including the effects of $\Lambda_{\rm QCD}/m_Q$ corrections to the
matrix elements of the weak current.  The situation here is similar to that in
the case of the  $s_\ell^{\pi_\ell}={3\over2}^+$ doublet.  Using a relation
between the leading, $m_Q\to\infty$, Isgur-Wise functions for these two excited
charmed meson doublets that is valid in the nonrelativistic constituent quark
model with any spin-orbit independent potential (and a few other assumptions),
we determined the rates for $B$ semileptonic decays to these excited charmed
mesons.  We find that branching ratio for $B$ semileptonic decays into the four
states in the $s_\ell^{\pi_{\ell}}={1\over2}^+$ and ${3\over2}^+$ doublets is
about $1.6\%$.  Combining this with the measured rates to the ground state $D$
and $D^*$ implies that more than $2\%$ of the $B$ meson decays must be
semileptonic decays to higher mass excited charmed states or nonresonant modes.
Some of the more important results in Tables~\ref{tab:sec2res} and
\ref{tab:sec3res} are summarized in Table~\ref{tab:concl}.  

\begin{table}[t] 
\begin{tabular}{c|ccc}  
Approximation  &  $R=\Gamma_{D_2^*}\Big/\Gamma_{D_1}$  &  
  $\tau(1)\, \bigg[\displaystyle {6.0\times10^{-3} \over 
    {\cal B}(B\to D_1\,e\,\bar\nu_e)} \bigg]^{1/2}$  &
  $\Gamma_{D_1+D_2^*+D_1^*+D_0^*}\Big/\Gamma_{D_1}$  \\[6pt] \hline 
B$_\infty$  &  $1.65$  &  $1.24$  &  $4.26$  \\
B$_1$  &  $0.52$  &  $0.71$  &  $2.55$  \\
B$_2$  &  $0.67$  &  $0.75$  &  $2.71$  \\
\end{tabular} \vspace{6pt}
\tighten{
\caption[6]{Predictions for $\Gamma_{D_2^*}/\Gamma_{D_1}$, $\tau(1)$,
and $\Gamma_{D_1+D_2^*+D_1^*+D_0^*}/\Gamma_{D_1}$ using $\hat\tau'=-1.5$.  
The results in the last column assume the nonrelativistic quark model 
prediction in Eq.~(\ref{ztau}).} \label{tab:concl} }
\end{table}

We considered the zero recoil matrix elements of the weak currents between a
$B$ meson and other excited charmed mesons at order $\Lambda_{\rm QCD}/m_Q$. 
Only the corrections to the states contribute and these were expressed in terms
of matrix elements of local operators.

Our results have implications for $B$ decay sum rules, where including the
contributions of the excited charmed meson states strengthens the bounds on
$\rho^2$ (the slope of the Isgur-Wise function for $B\to
D^{(*)}\,e\,\bar\nu_e$), on $\lambda_1$, and on the zero recoil matrix element
of the axial-vector current between $B$ and $D^*$ mesons.  The latter bound has
implications for the extraction of $|V_{cb}|$ from exclusive $B\to
D^*\,e\,\bar\nu_e$ decay.

\acknowledgments
We thank A. Le Yaouanc, M. Neubert, A. Vainshtein for useful discussions,
and A. Falk for keeping us from being disingenuous.  
This work was supported in part by the Department of Energy under grant 
no.~DE-FG03-92-ER 40701.

\appendix
\section{Perturbative order $\alpha_{\lowercase{s}}$ corrections}

In this Appendix we compute order $\alpha_s$ and order $\alpha_s\,\Lambda_{\rm
QCD}/m_Q$ corrections to the $B\to(D_1,D_2^*)\,e\,\bar\nu_e$ form factors.  At
this order both the current in Eq.~(\ref{1/mcurrent}) and the order
$\Lambda_{\rm QCD}/m_Q$ corrections to the Lagrangian in Eq.~(\ref{lag})
receive corrections.  Matrix elements of the kinetic energy operator,
$\eta_{\rm ke}^{(Q)}$, enter proportional to $\tau$ to all orders in $\alpha_s$
due to reparameterization invariance \cite{LuMa}.  The matrix elements
involving the chromomagnetic operator are probably very small and have been
neglected.  Order $\alpha_s$ corrections to the $b\to c$ flavor changing
current in the effective theory introduce a set of new operators at each order
in $\Lambda_{\rm QCD}/m_Q$, with the appropriate dimensions and quantum
numbers.  The Wilson coefficients for these operators are known $w$-dependent
functions \cite{falk,qcdcorr}, which we take from \cite{physrep}.

The vector and axial-vector currents can be written at order $\alpha_s$ as
\begin{eqnarray}\label{A1}
V^\mu &=& \bar h^{(c)}_{v'}\, \Bigg[ \gamma^\mu-{i\overleftarrow 
  D\!\!\!\!\!\slash\, \gamma^\mu \over 2\,m_c} + 
  {i\gamma^\mu \overrightarrow D\!\!\!\!\!\slash \over 2\,m_b}\,   
  \Bigg]\, h^{(b)}_v 
  + {\alpha_s\over\pi}\, \Big[ V^{\mu\,(1)} + V^{\mu\,(2)} \Big] +\ldots \,,  
  \nonumber\\*
A^\mu &=& \bar h^{(c)}_{v'}\, \Bigg[ \gamma^\mu\gamma_5-{i\overleftarrow 
  D\!\!\!\!\!\slash\, \gamma^\mu\gamma_5 \over 2\,m_c} + 
  {i\gamma^\mu\gamma_5 \overrightarrow D\!\!\!\!\!\slash \over 2\,m_b}\,   
  \Bigg]\, h^{(b)}_v 
  + {\alpha_s\over\pi}\, \Big[ A^{\mu\,(1)} + A^{\mu\,(2)} \Big] +\ldots \,,
\end{eqnarray}
where the ellipses denote terms higher order in $\alpha_s$ and 
$\Lambda_{\rm QCD}/m_Q$.  Superscripts $(1)$ denote corrections 
proportional to $\alpha_s$,
\begin{eqnarray}\label{A2}
V^{\mu\,(1)} &=& \bar h^{(c)}_{v'}\, \Big[   
  c_{V1}\gamma^\mu+c_{V2}v^\mu+c_{V3}v'^\mu \Big]\, h^{(b)}_v \,, \nn\\*
A^{\mu\,(1)} &=& \bar h^{(c)}_{v'}\, \Big[ 
  c_{A1}\gamma^\mu+c_{A2}v^\mu+c_{A3}v'^\mu \Big]\,\gamma_5\, h^{(b)}_v \,.
\end{eqnarray}
The terms with superscript $(2)$ in Eq.~(\ref{A1}) denote corrections
proportional to $\alpha_s\,\Lambda_{\rm QCD}/m_Q$,
\begin{eqnarray}\label{A3}
V^{\mu\,(2)} = \bar h^{(c)}_{v'}\, &\Bigg\{&
  {i\overrightarrow D_{\!\lambda}\over 2\,m_b}\,  
  \bigg[ \Big(c_{V1}\gamma^\mu+c_{V2}v^\mu+c_{V3}v'^\mu\Big)
  \bigg(\gamma^\lambda + 2 v'^\lambda\,{\frac{\overleftarrow\partial}
  {\partial w}}\bigg) 
  + 2 c_{V2} g^{\mu\,\lambda} \bigg] \\*
&-& {i\overleftarrow D_{\!\lambda}\over 2\,m_c}\, 
  \bigg[ 2 c_{V3} g^{\mu\,\lambda} + \bigg(\gamma^\lambda + 2v^\lambda\,
  {\frac{\overrightarrow\partial}{\partial w}}\bigg) 
  \Big(c_{V1}\gamma^\mu+c_{V2}v^\mu+c_{V3}v'^\mu\Big) \bigg] 
  \Bigg\}\, h^{(b)}_v, \nn\\*
A^{\mu\,(2)} = \bar h^{(c)}_{v'}\, &\Bigg\{&
  {i\overrightarrow D_{\!\lambda}\over 2\,m_b}\,  
  \bigg[ \Big(c_{A1}\gamma^\mu+c_{A2}v^\mu + c_{A3}v'^\mu\Big)\,\gamma_5\,
  \bigg(\gamma^\lambda + 2v'^\lambda\, 
  {\frac{\overleftarrow\partial}{\partial w}}\bigg) 
  + 2 c_{A2} g^{\mu\,\lambda}\,\gamma_5\, \bigg] \nn\\*
&-& {i\overleftarrow D_{\!\lambda}\over 2\,m_c}\, 
  \bigg[ 2 c_{A3} g^{\mu\,\lambda} + \bigg(\gamma^\lambda + 2v^\lambda\,
  {\frac{\overrightarrow\partial }{\partial w}}\bigg) 
  \Big(c_{A1}\gamma^\mu+c_{A2}v^\mu+c_{A3}v'^\mu\Big) 
  \bigg]\, \gamma_5\, \Bigg\}\, h^{(b)}_v \,.\nn
\end{eqnarray}
In these expressions the covariant derivatives, $D_\lambda$, act on the fields
$h_v^{(b)}$ or $h_{v'}^{(c)}$, and partial derivatives with respect to $w$,
$\partial/\partial w$, act on the coefficient functions $c_{Vi}(w)$ and
$c_{Ai}(w)$.  Using Eqs.~(\ref{A2}) and (\ref{A3}) it is straightforward to
include the order $\alpha_s$ and $\alpha_s\,\Lambda_{\rm QCD}/m_Q$ corrections
using trace formalism presented in Sec.~II.  The corrections with superscript
(1) simply change the form of $\Gamma$ in Eq.~(\ref{lo}), while those with
superscript (2) change $\Gamma$ in Eq.~(\ref{1/mc1}).

The $B\to D_1\,e\,\bar\nu_e$ form factors were defined in Eq.~(\ref{formf1}),
and their expansions in terms of Isgur-Wise functions at leading order in
$\alpha_s$ were given in Eq.~(\ref{expf}).  The order $\alpha_s$ and order
$\alpha_s\,\Lambda_{\rm QCD}/m_Q$ corrections modify the results for $f_i$ in
Eq.~(\ref{expf}) to $f_i+(\alpha_s/\pi)\,\delta f_i$.  The functions $\delta
f_i$ are given by
\begin{eqnarray}
\sqrt6\,\delta f_A &=&  -(w+1)c_{A1}\tau 
  -2\varepsilon_c\,(w\bar\Lambda'-\bar\Lambda)\Big[ 
  2c_{A1} +(w+1)c_{A1}' +c_{A3} \Big] \tau \nn\\*
&+& \varepsilon_c\,(w-1)\, \Big\{ [ 3c_{A1}-2(w-1)c_{A3} ] \tau_1-
   (3c_{A1}+4c_{A3})\tau_2 \Big\} \nn\\*
&-& \varepsilon_b\,\Big[(\bar\Lambda'+\bar\Lambda)(w-1)c_{A1}-
   2(\bar\Lambda'-w\bar\Lambda)(w+1)c_{A1}'+2\,(w\bar\Lambda'-\bar\Lambda)
   c_{A2}\Big] \tau \nn\\*
&+& \varepsilon_b\,(w-1)\, \Big\{ [(2w+1)c_{A1}-2(w-1)c_{A2} ] \tau_1+
   (c_{A1}-4c_{A2})\tau_2 \Big\} , 
\end{eqnarray}
\begin{eqnarray}
\sqrt6\,\delta f_{V_1} &=& (1-w^2)c_{V1}\tau 
  - 2\varepsilon_c\, (w\bar\Lambda'-\bar\Lambda)(w+1)\Big[ 
  2c_{V1}+(w-1)c_{V1}'+2c_{V3} \Big] \tau \nn\\*
&+& \varepsilon_c\, (w^2-1) \Big\{ [ 3c_{V1}+2(w+2)c_{V3} ] \tau_1 - 
   (3c_{V1}+2c_{V3})\tau_2 \Big\} \nn\\*
&-& \varepsilon_b\, (w+1) \Big[(\bar\Lambda'+\bar\Lambda)(w-1)c_{V1} -
   2(\bar\Lambda'-w\bar\Lambda)(w-1)c_{V1}'+4(w\bar\Lambda'-\bar\Lambda)c_{V2}
   \Big] \tau \nn\\*
&+& \varepsilon_b\,(w^2-1) \Big\{ [ (2w+1)c_{V1}+2(w+2)c_{V2} ] \tau_1 + 
   (c_{V1}-2c_{V2})\tau_2 \Big\} , 
\end{eqnarray}
\begin{eqnarray}
\sqrt6\, \delta f_{V_2} &=& - [3c_{V1}+2(w+1)c_{V2}] \tau 
  - 2\varepsilon_c\, (w\bar\Lambda'-\bar\Lambda) 
  \Big[ 3c_{V1}'+2c_{V2}+2(w+1)c_{V2}' \Big] \tau \nn\\*
&-& \varepsilon_c \Big\{ [ (4w-1)c_{V1}-2(2w+1)(w-1)c_{V2}-2(w+2)c_{V3} ]
    \tau_1 \nn\\*
&&\phantom{\varepsilon_c} + [5c_{V1}+2(1-w)c_{V2}+2c_{V3}]\tau_2 \Big\} \nn\\*
&-& \varepsilon_b\, \Big\{ 3 (\bar\Lambda'+\bar\Lambda)c_{V1}-
  6(\bar\Lambda'-w\bar\Lambda)c_{V1}'+2[(w-1)\bar\Lambda'+(3w+1)\bar\Lambda]
  c_{V2}\nn\\*
&&\phantom{\varepsilon_b} - 4(\bar\Lambda'-w\bar\Lambda)(w+1)c_{V2}' 
  \Big\}\tau \nn\\*
&+& \varepsilon_b \Big\{ [ 3(2w+1)c_{V1}+2(2w^2+1)c_{V2} ] \tau_1
  + [ 3c_{V1}+ 2(w-2)c_{V2} ] \tau_2 \Big\} ,
\end{eqnarray}
\begin{eqnarray}
\sqrt6\, \delta f_{V_3} &=& [ (w-2)c_{V1}-2(w+1)c_{V3} ] \tau \nn\\*
&+& 2\varepsilon_c\, (w\bar\Lambda'-\bar\Lambda) \Big\{ 2c_{V1} 
    +(w-2)c_{V1}'- 2 [ c_{V3}+(w+1)c_{V3}' ] \Big\} \tau \nn\\*
&+& \varepsilon_c \Big\{ [ (2+w)c_{V1}+2(w^2-3w-1)c_{V3} ] \tau_1 
  + [ (3w+2)c_{V1}+(4w-2)c_{V3} ] \tau_2 \Big\} \nn\\*
&+& \varepsilon_b\, \Big[ (\bar\Lambda'+\bar\Lambda)(w+2)c_{V1}+
  2(\bar\Lambda'-w\bar\Lambda) (2-w)c_{V1}'+4\bar\Lambda'(w+1)c_{V2}\nn\\*
&&\phantom{\varepsilon_b} - 2(\bar\Lambda'+\bar\Lambda)(w-1)c_{V3}
  +4(\bar\Lambda'-w\bar\Lambda)(w+1)c_{V3}' \Big] \tau \nn\\*
&-& \varepsilon_b\, \Big\{ 
  [ (2w^2+5w+2)c_{V1}+2w(2+w)c_{V2}+2(1+w-2w^2)c_{V3} ] \tau_1 \nn\\*
&&\phantom{\varepsilon_b} + [(2+w)c_{V1}-2wc_{V2}-2(w-1)c_{V3}]\tau_2 \Big\} .
\end{eqnarray}
Here $c_{Vi}$ and $c_{Ai}$ are functions of $w$, and prime denotes a derivative
with respect to $w$.  Note that at zero recoil $\delta f_{V_1}$ is known in
terms of $\bar\Lambda'-\bar\Lambda$ and $\tau(1)$, as expected from our results
in Sec.~II.

For $B\to D_2^*\,e\,\bar\nu_e$ decay, the $\alpha_s$ and order
$\alpha_s\,\Lambda_{\rm QCD}/m_Q$ corrections modify the leading order form
factors in Eq.~(\ref{expk}) to $k_i\to k_i+(\alpha_s/\pi)\,\delta k_i$.  The
functions $\delta k_i$ are 
\begin{eqnarray}
\delta k_V &=&  -c_{V1}\tau 
  -\varepsilon_c\,\Big[ 2c_{V1}'(w\bar\Lambda'-\bar\Lambda) \tau +
  (c_{V1}-2wc_{V3})\tau_1-(c_{V1}+2c_{V3})\tau_2 \Big] \\*
&-& \varepsilon_b\,\Big\{ [(\bar\Lambda'+\bar\Lambda)c_{V1}-2
  (\bar\Lambda'-w\bar\Lambda) c_{V1}']\tau - [(2w+1)c_{V1}+2wc_{V2}]
  \tau_1 - (c_{V1}+2c_{V2})\tau_2 \Big\}, \nn
\end{eqnarray}
\begin{eqnarray}
\delta k_{A_1} &=&  -(w+1)c_{A1}\tau 
  -\varepsilon_c\,\Big[ 2 (c_{A1}'+wc_{A1}'-c_{A3}) (w\bar\Lambda'-\bar\Lambda)
  \tau + (w-1)c_{A1} (\tau_1-\tau_2) \nn\\*
&& +2(w^2-1)c_{A3}\tau_1 \Big]
  - \varepsilon_b\, \Big\{ [(\bar\Lambda'+\bar\Lambda)(w-1)c_{A1}-
  2(\bar\Lambda'-w\bar\Lambda)(w+1)c_{A1}'\nn\\*
&& -2(w\bar\Lambda'-\bar\Lambda)c_{A2}]\tau - (w-1)[c_{A1}(\tau_1+\tau_2)
  +2(wc_{A1}-wc_{A2} -c_{A2})\tau_1] \Big\},
\end{eqnarray}
\begin{eqnarray}
\delta k_{A_2} &=& c_{A2}\tau 
  + \varepsilon_c\,\Big\{ 2c_{A2}' (w\bar\Lambda'-\bar\Lambda) \tau 
  - [ 2c_{A1}-(2w+1)c_{A2}+2c_{A3} ] \tau_1 + c_{A2}\tau_2 
  \Big\} \nn\\*
&+& \varepsilon_b\,\Big\{ [(\bar\Lambda'+3\bar\Lambda)c_{A2} 
  - 2(\bar\Lambda'-w\bar\Lambda)c_{A2}' ]\tau 
  - (2w+3)c_{A2}\tau_1 - c_{A2}\tau_2  \Big\}, 
\end{eqnarray}
\begin{eqnarray}
\delta k_{A_3} &=&  (c_{A1}+c_{A3})\tau 
  +\varepsilon_c\,\Big[ 2(c_{A1}'+c_{A3}') (w\bar\Lambda'-\bar\Lambda) \tau 
  -(c_{A1}-c_{A3})(\tau_1+\tau_2)+4wc_{A3}\tau_1\Big] \nn\\*
&+& \varepsilon_b\,\Big\{ [ (\bar\Lambda'+\bar\Lambda)(c_{A1}+c_{A3}) 
  -2 \bar\Lambda'c_{A2} - 2 (\bar\Lambda'-w\bar\Lambda) (c_{A1}'+c_{A3}')]\tau 
  \nn\\*
&&\phantom{\varepsilon_b} -(c_{A1}+c_{A3})(\tau_1+\tau_2)
  - 2w(c_{A1}-c_{A2}+c_{A3})\tau_1 \Big\} .
\end{eqnarray}

To compute the corrections to the results obtained in Sec.~II, it is sufficient
to expand the Wilson coefficients $c_{Vi}$ and $c_{Ai}$ to linear order in $w$.
We take $c_{Vi}$ and $c_{Ai}$ and their first derivatives at zero recoil from
Ref.~\cite{physrep}.  To evaluate these, we choose to integrate out the $c$ and
$b$ quarks at a common scale $\mu=\sqrt{m_c\,m_b}$, giving for $c_{Vi}$ and
$c_{Ai}$ 
\begin{eqnarray}
c_{V1}(1) &=&  -\frac43 - {1+z\over1-z}\, \ln z  
  \simeq 0.91 \,, \nn\\*
c_{V2}(1) &=& -  {2\,(1-z+z \ln z) \over 3 (1-z)^2}
  \simeq -0.46 \,, \nn\\*
c_{V3}(1) &=& {2z\, (1-z+\ln z) \over 3 (1-z)^2}
  \simeq -0.20 \,, \nn\\
c_{A1}(1) &=& -\frac83 - {1+z\over1-z}\, \ln z 
  \simeq -0.42 \,, \nn\\* 
c_{A2}(1) &=& - {2\,[3-2z-z^2+(5-z)z\ln z] \over 3(1-z)^3} 
  \simeq -1.20 \,, \nn\\*
c_{A3}(1) &=&  {2z\, [1+2z-3z^2+(5z-1)\ln z] \over 3 (1-z)^3} 
  \simeq 0.42 \,.
\end{eqnarray}
The derivatives $c_{Vi}'$ and $c_{Ai}'$ at zero recoil are
\begin{eqnarray}
c_{V1}'(1) &=& - {2 [ 13-9z+9z^2-13z^3+3(2+3z+3z^2+2z^3)\,\ln z] 
  \over 27\,(1-z)^3 } \simeq 0.20 \,, \nn\\*
c_{V2}'(1) &=& {2 [ 2+3z-6z^2+z^3+6z\,\ln z ] \over 9\,(1-z)^4 }
  \simeq 0.21 \,,  \nn\\*
c_{V3}'(1) &=& {2z [1-6z+3z^2+2z^3-6z^2\,\ln z] \over 9\,(1-z)^4 }
  \simeq 0.05 \,, \nn\\
c_{A1}'(1) &=& - {2 [7+9z-9z^2-7z^3+3(2+3z+3z^2+2z^3)\,\ln z ]\over 
  27\,(1-z)^3} \simeq 0.64 \,, \nn\\* 
c_{A2}'(1) &=& {2 [2-33z+9z^2+25z^3-3z^4-6z(1+7z)\,\ln z]\over 9\,(1-z)^5 }
  \simeq 0.37 \,, \nn\\*
c_{A3}'(1) &=& - {2z [3-25z-9z^2+33z^3-2z^4-6z^2(7+z)\,\ln z]\over 9\,(1-z)^5}
  \simeq -0.12 \,. 
\end{eqnarray}
Here $z=m_c/m_b$, and the numbers quoted are for $z=1.4/4.8$.  

Using these values and the $\alpha_s$ corrections for the form factors above,
we find the corrections given in Table~\ref{tab:alphas} to the leading order
results summarized in Table~\ref{tab:sec2res}.

{\tighten

} 

\end{document}